\documentclass[pra,a4paper,reprint,superscriptaddress,preprintnumbers]{revtex4-2}

\usepackage{mymacros}
\definecolor{change}{rgb}{0.0, 0.0, 0.0} 
\newcommand{\change}[1]{{\textcolor{change}{#1}}}

\begin{document}
\title{Relativistic meson spectra on ion-trap quantum simulators}

\author{Johannes Knaute} 
\email{johannes.knaute@aei.mpg.de}
\affiliation{Max  Planck  Institute  for  Gravitational  Physics (Albert Einstein Institute), 
14476 Potsdam-Golm, Germany}
\affiliation{Department of Physics, Freie Universit\"{a}t Berlin, 
14195 Berlin, Germany}

\author{Philipp Hauke} 
\email{philipp.hauke@unitn.it}
\affiliation{INO-CNR BEC Center and Physics Department of Trento University, Via Sommarive 14, I-38123 Povo (Trento), Italy}


\begin{abstract}
The recent rapid experimental advancement in the engineering of quantum many-body systems opens the avenue to controlled studies of fundamental physics problems via digital or analog quantum simulations. Here, we systematically analyze the capability of analog ion traps to explore relativistic meson spectra on current devices. We focus on the E$_8$ quantum field theory regime, which arises due to longitudinal perturbations at the critical point of the transverse-field Ising model. As we show through exact numerics, for sufficiently strong long-range suppression in experimentally accessible spin chain models, absorption spectroscopy allows for the identification of the low-lying meson excitations with a good degree of accuracy even for small system sizes. Our proposal thus opens a way for probing salient features of quantum many-body systems reminiscent of meson properties in high-energy physics.
\end{abstract}
\maketitle

\section{Introduction}

Emergent phenomena of quantum many-body (QMB) systems play a major role in condensed matter and particle physics \cite{Anderson:1972pca,coleman_2015,Witten:2017hdv}. The recent progress of quantum simulation technologies \cite{Hauke2011d,Cirac2012,Alexeev:2020xrq} in controllable platforms such as 
ion traps \cite{Schneider2011,Blatt2012,Monroe2021,Blatt:2021_Demonstrator} has opened the prospect of treating fundamental effects and systems beyond the capability of classical computers. Various trapped-ion experiments have already unveiled static and dynamical properties of quantum matter \cite{Friedenauer2008,Britton_2012,Richerme_2013,Senko_2014,Jurcevic_2014,Jurcevic_2015,Smith2016,Garttner:2016mqj,Jurcevic_2017,Zhang_2017,Hempel_2018,Brydges_2019,Maier2019} as well as lattice gauge theories \cite{Martinez:2016yna,Kokail:2018eiw}. 

In this work, we are interested in using trapped-ion devices to study mesons, which are non-perturbative bound states consisting of two subparticles or charges. They appear prominently in Quantum Chromodynamics (QCD), the theory of strong interactions within the standard model of particle physics, where a quark-antiquark pair is confined by a flux tube. Their properties and phenomenology is of key importance for the understanding of heavy-ion collisions, which provide an experimental way of studying far-from equilibrium dynamics relevant to the physics of the early universe \cite{Friman:2011zz}. Beyond particle physics, mesons exist also in condensed matter systems, in particular Ising spin chain models, where symmetry breaking longitudinal fields \cite{McCoy:1978ta} or long-range interactions \cite{Liu:2018fza,Lerose_2019} can confine domain walls into mesons. 
The existence of mesons in the spectrum has severe consequences for both static and dynamical properties of the QMB system at zero and finite temperature. Some of the diverse implications for entanglement, correlations, and thermalization are theoretically studied in \cite{Greiter:2002,Lake_2009,Kormos2017,James_2019,Mazza_2019,Robinson_2019,Vanderstraeten:2020,Lerose:2019jrs,Banuls:2019qrq,Castro-Alvaredo:2020mzq}. In long-range models, the existence or absence of meson states has also profound implications on the emergence of anomalous cusps in dynamical quantum phase transitions \cite{Halimeh_2020,Defenu:2019dkd,halimeh2021dynamical,hashizume2020dynamical,PhysRevB.102.035115}.

Analog quantum simulations can implement such spin Hamiltonians and therefore provide access to meson features. 
Experimental evidence of dynamically induced magnetic domain wall confinement was first provided in \cite{Tan:2019kya} and \cite{Vovrosh_2021} by analyzing the meson impact on correlation and entanglement spreading after quantum quenches. \change{Recently, the authors of \cite{Schuckert:2020qeo} proposed a protocol to measure fluctuations and dissipations in quantum simulators (see also \cite{Geier2021}), and illustrated a way of obtaining spectral information of the meson system from their relations.} An improved error mitigation technique for the extraction of meson masses on quantum simulators was discussed in \cite{Vovrosh:2021ocf}. 
While these systems are currently most developed in (1+1)-dimensional simulations, their phenomenology can provide important insights that are relevant across dimensions. For example, the recent papers \cite{Verdel:2019chj,Surace:2020ycc,Karpov:2020pqe} explored the capabilities of quantum simulations for real-time string breaking and meson scattering \footnote{See also \cite{Rigobello:2021fxw} for a QED study of meson scattering with tensor network simulations.}. All these studies focused on parameter regimes where either a semiclassical interpretation of mesons in terms of domain walls is possible or a formulation as a simple gauge theory is amenable \footnote{As discussed in detail in \cite{Kormos2017}, the semiclassical regime is valid deep in the ferromagnetic phase, i.e.\ $h<J$ in \eqref{eq:H_NN} and \eqref{eq:H_LR}, when the system is not close to criticality. In the case of nearest-neighbor interactions, isolated spin domains of magnetization opposite to the longitudinal field give rise to a confining potential linear in its spatial extension. Mesonic excitations and their properties can then be estimated from the Bohr--Sommerfeld quantization condition. In a similar way, using a two-kink model, it was shown in \cite{Liu:2018fza} that long-range interactions give rise to an effective confining potential when restricting to the Hilbert space of two domain-wall states. Details on a reformulation of the nearest-neighbor Ising model at very weak transverse field as a nonrelativistic $\mathds{Z}_2$ gauge theory are discussed, e.g., in \cite{Karpov:2020pqe}.}.

Alternatively, \change{meson states can occur also far away from semiclassical regimes}, e.g., close to quantum critical points (QCPs), where an effective (i.e.\ relativistic) quantum field theory (QFT) description is available. Zamolodchikov's E$_8$ model \cite{Zamolodchikov:1989fp} is such an example of an interacting QFT that emerges through longitudinal perturbations at the Ising critical point. The theoretically predicted E$_8$ meson spectrum was first experimentally observed in \cite{Coldea_2010} and found recently renewed interest in \cite{Zou:2020ouw,Zhang:2020enf,Amelin:2020mif}. These experiments were based on neutron scattering measurements and spectroscopic methods in solid-state crystals. 

Here, we instead propose controlled measurements of the E$_8$ meson spectrum on an ion-trap quantum simulator using absorption spectroscopy \cite{Senko_2014,Jurcevic_2015}. For that purpose, we numerically explore the capabilities of experimentally realizable small Ising spin systems to identify the lowest E$_8$ meson states. We show that for sufficiently strong long-range suppression in Ising models, the energy absorption spectrum, which is accessible in the linear response framework, is in close correspondence with the analytical expectation of the E$_8$ QFT. 
We corroborate these findings by a fidelity analysis, which suggest that small systems retain the nature of the first meson across all interaction ranges considered, while the thermodynamic limit may have a transition at a spatial power-law interaction $\sim 1/r$. 
Due to the promising experimental \cite{Martinez:2016yna,Kokail:2018eiw} and theoretical \cite{Hauke2013b,Yang:2016hjn,Muschik_2017,Davoudi:2019bhy,Paulson:2020zjd,Davoudi:2021ney,Zohar:2021nyc} efforts to implement and study gauge theories with ion trap quantum simulations, we see, as an implication of our study, the potential to probe meson physics also in relativistic gauge theories with these technologies.

\section{Ising models and QFTs}

The transverse field Ising model is a famous example of a many-body system exhibiting a quantum phase transition \cite{sachdev_2011}. An additional longitudinal field can break the integrability of the system and introduces interesting new features, in particular mesons, appearing as non-perturbative bound states in the spectrum of the model.
The prototype is the nearest-neighbor (NN) Ising model, defined in terms of Pauli matrices $\sigma^{x,z}_j$ by the Hamiltonian
\begin{equation}  \label{eq:H_NN}
H_{\rm NN} = - J \,\left ( \sum_{j=1}^{N-1} \sigma^z_j \sigma^z_{j+1} + h \sum_{j=1}^N  \sigma^x_j + g \sum_{j=1}^N \sigma^z_j \right) ,
\end{equation}
where the overall energy scale is set by the unit $J$. 
The transverse and longitudinal fields are quantified by the parameters $h$ and $g$, respectively.
The Hamiltonian \eqref{eq:H_NN} is written for $N$ spins at positions $j$ assuming open boundary conditions (obc).
Analogously, one can assume periodic boundary conditions (pbc), defined by $\sigma_{N+1}=\sigma_1$ for a system on a circle, by adding the interaction term $- J \sigma^z_N \sigma^z_{1}$.

In a proper continuum limit, the IR regime of $H_{\rm NN}$ is described by a Majorana fermion QFT with Hamiltonian \cite{Rakovszky:2016ugs}
\small
\begin{equation}
\label{eq:H_IsingQFT}
\hspace{-4 pt} H_\text{IR} = \int_{-\infty}^\infty dx \,  \left\{ \frac{i}{4\pi} \left( \psi\partial_x\psi - \bar\psi\partial_x\bar\psi \right) - \frac{i M_h}{2 \pi}\bar\psi\psi + {\cal C} M_{g}^{15/8} \, \sigma \right\} .
\end{equation}
\normalsize
Here, ${\cal C} \approx 0.062$ is a numerical constant, and $M_{h} \equiv 2 J |1-h|$ and $M_{g} \equiv \, {\cal D} J \, |g|^{8/15}$ with ${\cal D} \approx 5.416$ are mass scales in the transverse and longitudinal direction \cite{Rakovszky:2016ugs,Hodsagi:2018sul}.
The QCP at $\{J=h=1,g=0\}$ translates into $M_h=M_g=0$, in which case the IR is governed by the Ising CFT with central charge $c=1/2$ and scalar primary operators $\epsilon=i\bar\psi\psi$ and $\sigma$ of dimensions $\Delta_\epsilon=1$ and $\Delta_\sigma=1/8$.
For longitudinal relevant perturbations of the Ising CFT, i.e.\ $M_h=0, M_g\ne0$, it is a remarkable prediction of Zamolodchikov that the resulting interacting E$_8$ QFT is also integrable and governed by the exceptional simple Lie algebra of rank 8 \cite{Zamolodchikov:1989fp}.
This QFT contains 8 stable mesons -- fermionic non-perturbative bound states -- whose masses are known as tabulated in table~\ref{tab:e8_masses_ratio} in units of the lightest meson mass $M_1 \equiv M_g$.

\begin{table*}[t]
    \centering
\caption{\label{tab:e8_masses_ratio} Ratios of meson masses for the integrable interacting E$_8$ QFT \cite{Zamolodchikov:1989fp}, which provides an effective description of the nearest-neighbor Ising model along longitudinal perturbations at its critical point.}    
     \begin{tabular}{l*{7}{l}} \hline
 & $M_2/M_1$ & $M_3/M_1$ & $ M_4/M_1$ & $ M_5/M_1$ & $M_6/M_1$ & $M_7/M_1$ & $M_8/M_1$ \\ \hline 
 analytical & 
 $2\cos{\frac{\pi}{5}}$ & 
 $2\cos{\frac{\pi}{30}}$ & 
 $4\cos{\frac{7\pi}{30}}\cos{\frac{\pi}{5}}$ & 
 $4\cos{\frac{2\pi}{15}}\cos{\frac{\pi}{5}}$ & 
 $4\cos{\frac{\pi}{30}}\cos{\frac{\pi}{5}}$ & 
 $8\cos^2{\frac{\pi}{5}}\cos{\frac{7\pi}{30}}$ & 
 $8\cos^2{\frac{\pi}{5}}\cos{\frac{2\pi}{15}}$ \\
 numerical & 1.6180 & 1.9890 & 2.4049 & 2.9563 & 3.2183 & 3.8912 & 4.7834 \\ \hline
\end{tabular}
\end{table*}

On ion-trap quantum simulators, it is experimentally possible to implement a long-range (LR) Ising model, defined by the Hamiltonian \cite{Britton_2012,Richerme_2013,Jurcevic_2014}
\begin{equation}  \label{eq:H_LR}
H_{\mathrm{LR}} = - J \,\left ( \sum_{i<j}^N \frac{1}{\vert i-j\vert^\alpha} \sigma^z_i \sigma^z_{j} + h \sum_{j=1}^N  \sigma^x_j + g \sum_{j=1}^N \sigma^z_j \right),
\end{equation}
where the coefficient $\alpha$ quantifies the LR interaction of two spins at position $i$ and $j$ \footnote{Physical consequences of algebraic long-range interactions in QMB systems for eigenstate thermalization and symmetry properties are studied, e.g., in \cite{russomanno2020longrange}.}.
Similarly to the NN model, one can consider the system for obc and pbc, where in the latter case we assume that two spins at positions $i$ and $j$ interact along their minimal distance on the ring.
For $\alpha\to\infty$, one recovers the NN Hamiltonian \eqref{eq:H_NN}. While experimentally the range $0 \le \alpha \le 3$  is in principle accessible \cite{Porras2004a,Trautmann_2018}, it was observed, e.g., in \cite{Hauke_2013} that already for $\alpha\approx 3$, the physics of the system can resemble closely the NN model.

\section{Energy and absorption spectra}

In what follows, we compare the ideal NN model with the LR model based on numerical diagonalization, to characterize in how far the E$_8$ meson spectrum survives in presence of power-law interactions and for the relatively small systems of few dozens sites to which current experiments on trapped ions are restricted \cite{Monroe2021,Blatt:2021_Demonstrator}.
The basis for the observability in small systems is that the longitudinal field is chosen large enough such that the associated length scale of the first meson $L \sim 1/M_1 \sim \vert g\vert^{-8/15}$ is sufficiently small to be captured by the finite size chain. As already observed in \cite{Kjall_2011} for a realistic model of a solid state crystal, even relatively large longitudinal field values are able to reproduce the E$_8$ spectrum, indicating the strong impact of the QFT regime on the physics of the model.

\subsection{Energy levels}

\begin{figure*}[t]
\centering
 \includegraphics[width=0.49\textwidth]{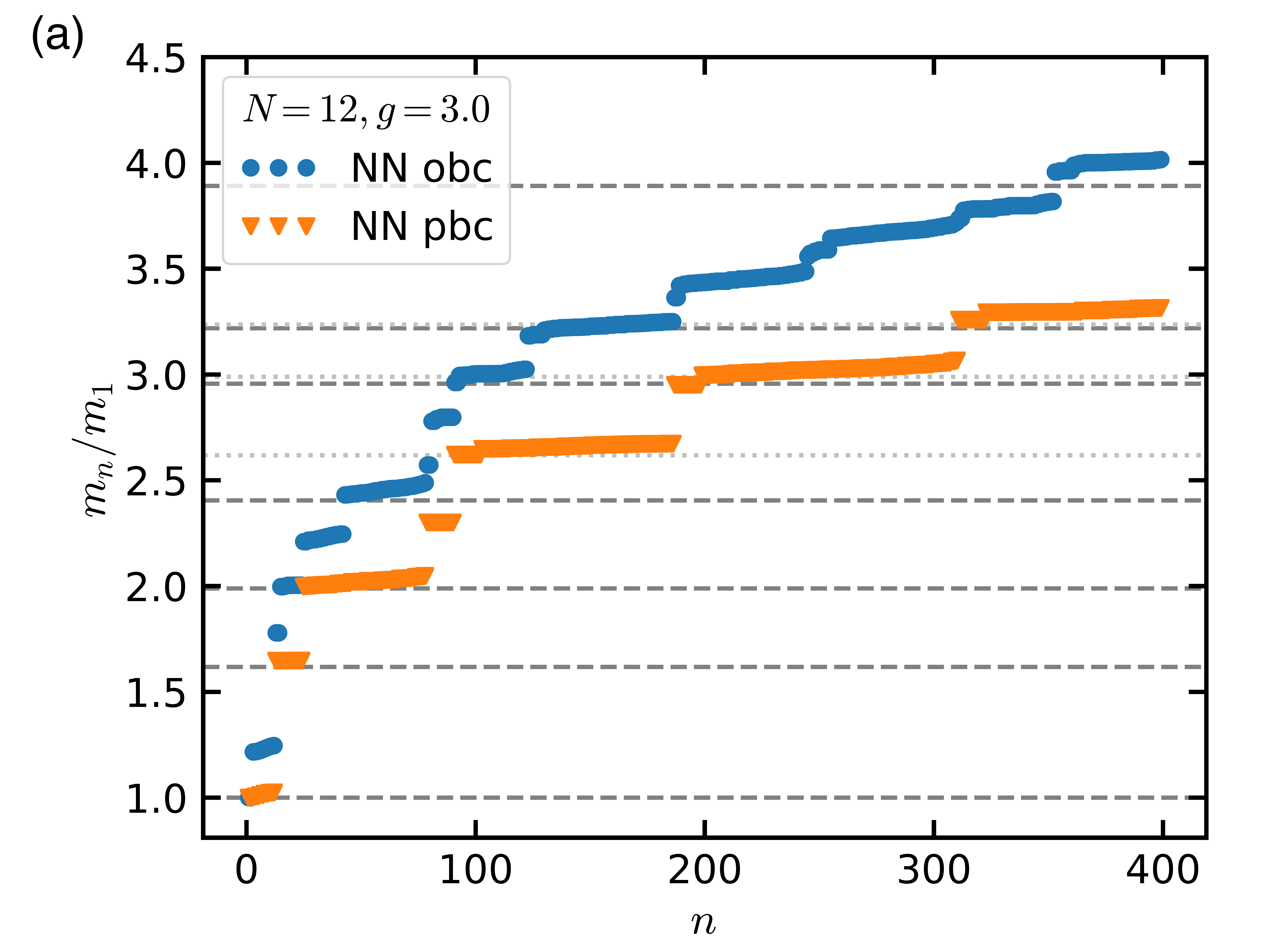} 
 \includegraphics[width=0.49\textwidth]{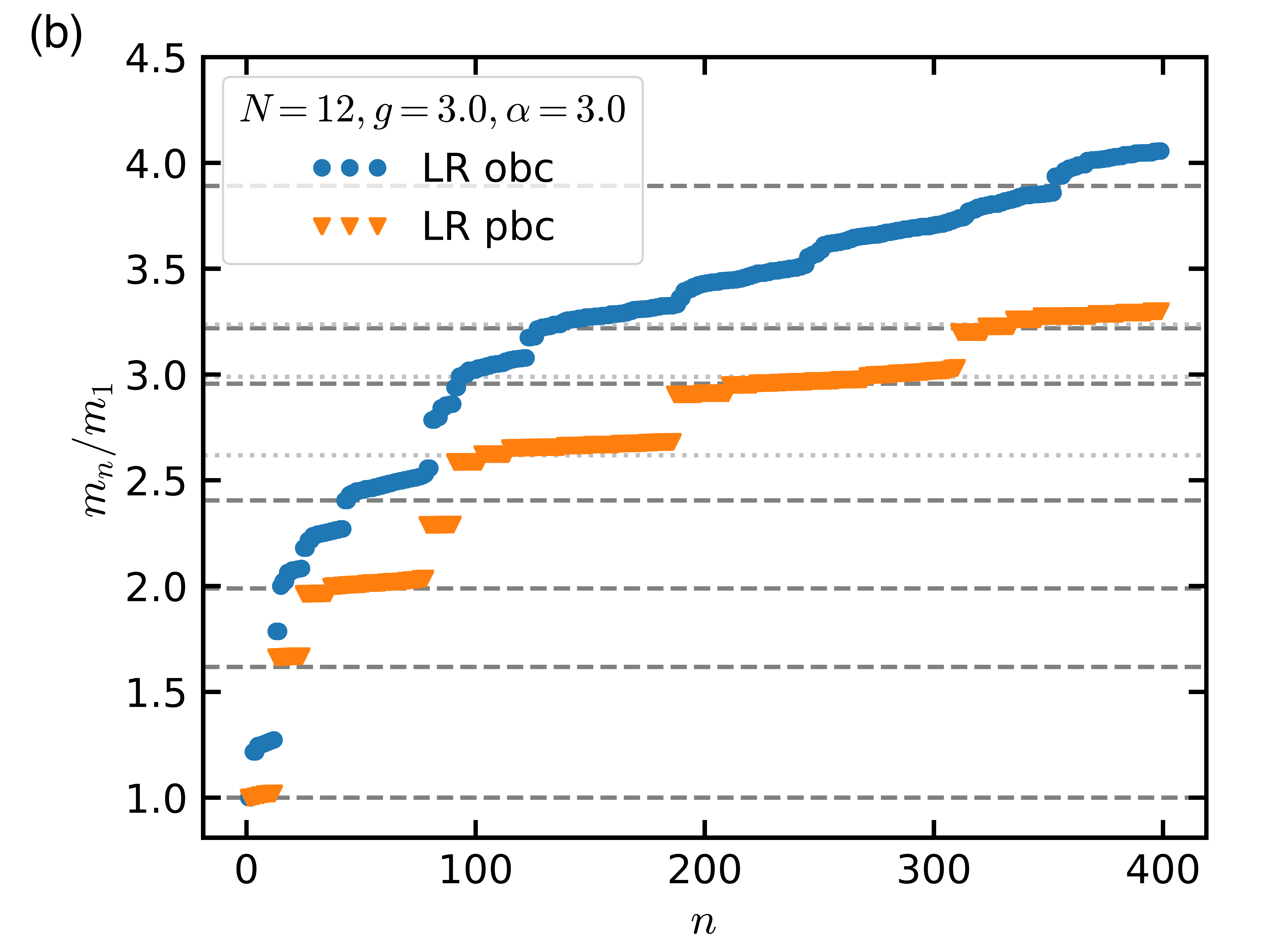} 
 \caption{(Color online) Numerical energy spectra for the NN (a) and LR (b) Ising model with obc (blue circles) and pbc (orange triangles). The normalized mass gaps $m_n/m_1$ of the lowest excited states with level $n \in \mathds{N}$ are shown for the longitudinal field strength $g=3$ in a chain of $N=12$ sites. 
 \change{The transverse field strength is fixed to the critical value $h=1$ throughout this work.} 
 Grey dashed lines represent the analytical E$_8$ meson mass ratios $M_n/M_1$ (cf.\ table~\ref{tab:e8_masses_ratio}). The continuum threshold is at $2M_1$. Grey dotted lines correspond to multiparticle states with masses $M_1+M_2$, $M_1+M_3$, and $2M_2$ (in ascending order).
 While for obc some deviations from the ideal result appear, for pbc even such a small system reproduces well the expected low-lying mass spectrum, for NN as well as for algebraic interactions. 
 }
 \label{fig:energy_spectra_small}
\end{figure*}

In Fig.\,\ref{fig:energy_spectra_small}, the mass gaps $m_n/m_1$ for the lowest $n=1,\ldots,400$ excited eigenstates, normalized to the lowest excited numerical state, are shown  for a chain of $N=12$ spins with an exemplary longitudinal field $g=3$ in the NN (a) and LR (b) Ising model  \footnote{The mass gap $m_n$ of level $n$ is defined as the energy difference to the groundstate, i.e.\ $m_n=E_n-E_0$.}. 
In the finite size system, energy levels appear as bands in the spectrum. In the ideal NN model, pbc (shown as orange triangles) allow for a clean identification of the first 6 meson levels. Apart from an underestimation of the fourth level, the mass ratios agree well with the E$_8$ theory (shown as grey dashed lines). 
The first $n=1,\ldots,N$ eigenvalues can be associated to the the first meson level and follow the momentum dispersion relation in the first Brillouin zone.
In contrast, while obc (blue circles) match particularly some of the higher meson levels, they do not satisfy the ratio of the first E$_8$ masses. We therefore focus in the following on a finite system with pbc on a ring and compare the results to the E$_8$ theory on an infinite line (cf.\ table~\ref{tab:e8_masses_ratio}), i.e.\ we neglect finite volume corrections given by L\"uscher's formula \cite{Luscher:1986}. 

At energies above $2M_1$, multiparticle states exist and form a continuum. Although we do not have a continuum in a finite system, we can nevertheless identify the mass sum $M_1+M_2$ (shown as the lowest grey dotted line). Higher order mass sums are very close to some of the analytical E$_8$ mass ratios.
The LR results for $\alpha=3$ (right panel) resemble the NN profile nearly identically for pbc and slightly smeared-out for obc, and therefore similarly allow one to identify the analytical meson mass ratios.

\subsection{Absorption spectra}

In recent years, methods have been developed to reveal spectra of interacting spin systems in trapped ions akin to neutron-scattering in the solid state \cite{Senko_2014,Jurcevic_2015}.  
Such experimentally measurable absorption spectra can be computed within the framework of linear response theory \footnote{See e.g.\ \cite{jensen1991rare} for an introduction.}. 
Specifically, the mean energy absorption rate $\overline{Q} = \overline{\langle\partial H/\partial t\rangle}$ is proportional to the imaginary (dissipative) part $\chi^{\prime\prime}(\omega) \equiv \chi^{\prime\prime}_{AA}(\omega)$ of the susceptibility,
which is given in general in the Lehmann representation by
\small
\begin{equation} \label{eq:chi}
    \chi^{\prime\prime}_{OA}(\omega) = \pi \sum_{n,m=0}^{2^N-1} \braket{n|A|m}\braket{m|O|n}\left( p_n - p_m\right) \delta[\omega-(E_m-E_n)] .
\end{equation}
\normalsize
Here, the double sum is taken over all eigenstates $\ket{n}$ and $\ket{m}$ of the system, $A$ is an operator that perturbs the Hamiltonian in the time domain, and $O$ is an operator whose response in the system is considered. 
The delta function in eq.\,\eqref{eq:chi} expresses the fact that there is only a contribution to the result when the perturbation frequency $\omega$ equals the energy differences $E_m-E_n$.
For general thermal states, the population factors take the form $p_n = \mathrm e^{-\beta E_n}/Z$, where $Z=\sum_n \mathrm e^{-\beta E_n}$ is the finite temperature partition function. For our studies, we are interested in the zero temperature case where absorption energies are measured with respect to the ground state $\ket{0}$ with energy $E_0$, and $p_0=1$ and $p_n = 0$ for $n>0$. 

In the following, we find that the salient features of the spectrum become accessible with the following straightforwardly measurable operator
\begin{equation}
    \label{eq:operators}
    O = A = \sum_{i=1}^N \sigma_i^z \cos(k r_i),
\end{equation}
where $k \in [-\pi,\pi]$ is  the quasi-momentum and $r_i = a i \equiv i$ the lattice position for unit lattice spacing. For the special case of $k=0$, the imaginary part dynamic susceptibility simplifies (in dimensionless units) to 
\begin{widetext}
\begin{equation} \label{eq:chi_kO}
    \chi^{\prime\prime}(\omega,k=0) = \pi J \sum_{n=0}^{2^N-1} \sum_{i=1}^N \left\vert\braket{0|\sigma_i^z|n}\right\vert^2 \left\{ \delta[\omega-(E_n-E_0)] - \delta[\omega+(E_n-E_0)] \right\}.
\end{equation}
\end{widetext}
In a realistic situation, the energy resolution is restricted by the accessible experimental observation time $t_{\rm obs}$. According to the Wiener--Khintchine theorem \cite{Wiener1930,Khintchine1934}, the delta function is then approximated by a Lorentzian 
\begin{equation} \label{eq:delta}
    \delta[\omega-(E_n-E_0)] \approx \frac{\Gamma}{[\omega-(E_n-E_0)]^2 + \Gamma^2} ,
\end{equation}
with width $\Gamma=1/t_{\rm obs}$.

\begin{figure}[t]
\centering
 \includegraphics[width=\columnwidth]{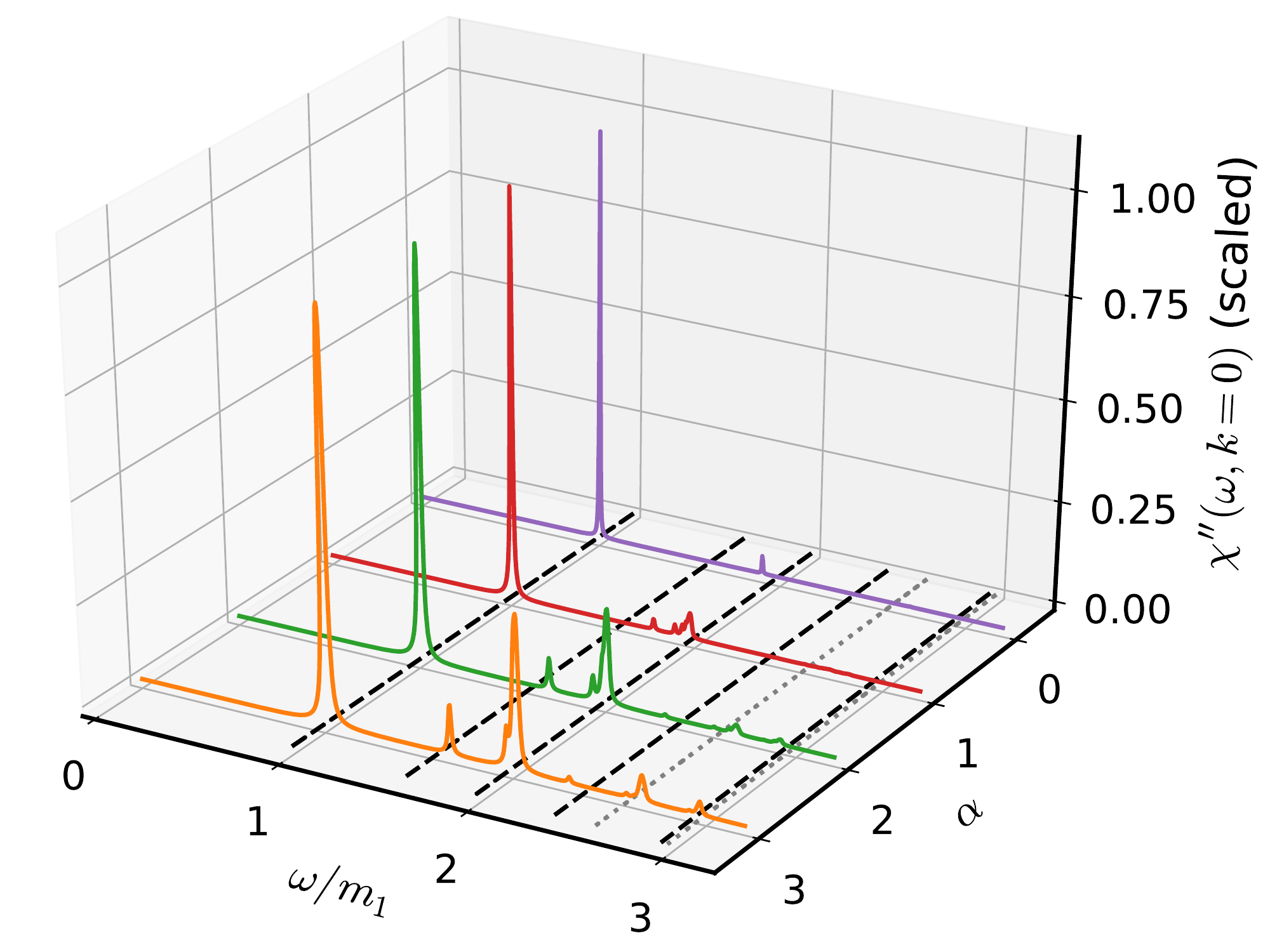} 
 \caption{(Color online) Energy absorption spectrum of the LR model with pbc in dependence on the power-law coefficient $\alpha$. The data are scaled to the maximum of the spectrum. Black dashed lines represent the analytical E$_8$ meson mass ratios (cf.\ table~\ref{tab:e8_masses_ratio}). Grey dotted lines correspond to multiparticle states with masses $M_1+M_2$ and $M_1+M_3$. For the entire range of $\alpha$, a strong peak appears at the lowest meson mass. With increasing $\alpha$, more features become discernible that agree with the analytic E$_8$ meson spectrum of the QFT at the critical point of the NN Ising model. 
 Numerical parameters: $N=18$ (pbc), $\Gamma/J=0.1$, $g=3$.  
 }
 \label{fig:alpha3D}
\end{figure}

We numerically calculate the energy absorption spectrum according to Eqs.\,\eqref{eq:chi_kO} and \eqref{eq:delta} for the realistic value $\Gamma/J=0.1$ (see Sec.~\ref{sec:trappedions}) on a chain of $N=18$ sites, which is the largest system size that we can achieve by iterative eigensolvers for sparse matrices, while keeping a large portion of the spectrum \footnote{We refer to the appendices for further details on the finite size, long-range and longitudinal field dependence of the energy and absorption spectra discussed in this section.}. 
Figure\,\ref{fig:alpha3D} shows the energy absorption spectrum in the LR model as a function of the frequency in dependence of the coefficient $\alpha$. 
For low $\alpha$, only the first meson mass can unequivocally be discerned. 
As $\alpha$ is increased, peaks at the analytical E$_8$ meson mass ratios are formed, whereby the first meson retains the largest spectral density. The continuum threshold at $2M_1$ overlaps with the third meson peak. Above, also the mass sum $M_1+M_2$ is identifiable while the fifth meson peak overlaps with the mass sum $M_1+M_3$.

\begin{figure}[t]
\centering
 \includegraphics[width=\columnwidth]{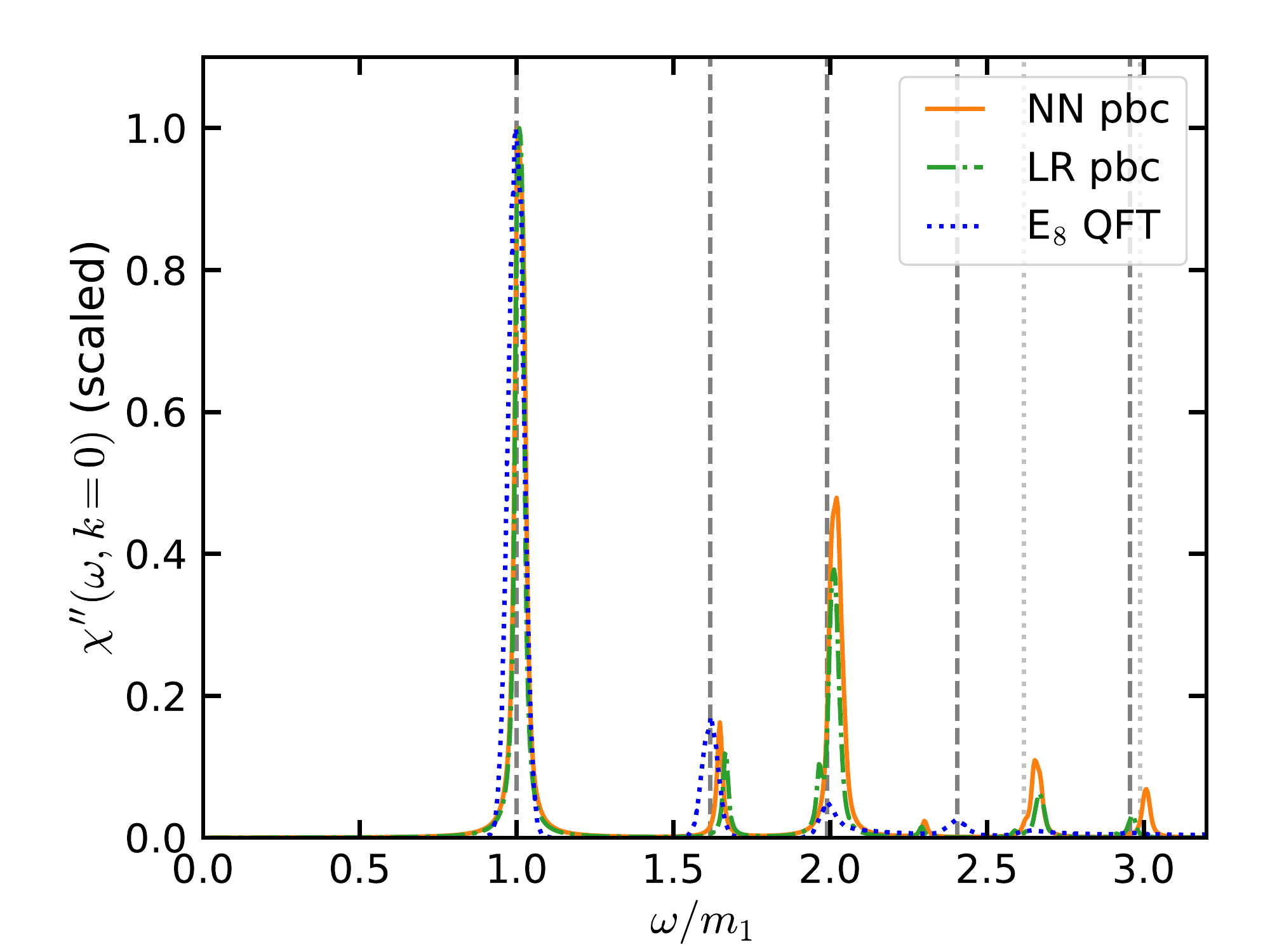} 
 \caption{(Color online) Comparison of the energy absorption spectrum in the NN and LR model with the analytical E$_8$ dynamical structure function from \cite{wang2021spin}. The data are scaled to the maximum of the spectrum. Grey dashed lines represent the analytical E$_8$ meson mass ratios (cf.\ table~\ref{tab:e8_masses_ratio}). Grey dotted lines correspond to multiparticle states with masses $M_1+M_2$ and $M_1+M_3$.
 From the numerical absorption spectra, meson peaks can be identified very close to their expected analytical ratios. The LR model allows one to resolve the quantitative ratio of the first to the second meson peak height of the QFT prediction with nearly the same precision as the NN model.
 Numerical parameters: $N=18$ (pbc), $\Gamma/J=0.1$, $g=3$, $\alpha=3$.
 }
 \label{fig:E8comparison}
\end{figure}

In Fig.\,\ref{fig:E8comparison}, the energy absorption spectrum in the LR model with $\alpha=3$ (green dash-dotted curve) is compared to the NN model (orange solid curve) for one selected value of the longitudinal field. The analogon of this spectral density in the E$_8$ QFT is the dynamical structure function, which has been calculated recently in \cite{wang2021spin}. The corresponding spectrum is shown as the blue dotted curve for a similar frequency broadening. In the Ising model data, the first 5 meson states and the mass sum $M_1+M_2$ are visible as peaks with (apart from the 4th level) good quantitative agreement to the analytical mass ratios. While in the exact E$_8$ spectrum the meson peak heights are continuously decreasing, the finite size data are not able to reproduce this feature above the continuum threshold. However, the ratio of the first to the second meson peak height is even in good quantitative agreement with the analytical prediction.

\change{\section{Fidelity analysis}}

The existence of a clear band structure in the energy spectrum of the LR Ising model (cf.\ Fig.\,\ref{fig:energy_spectra_small}) as well as peaks in the absorption spectrum (cf.\ Fig.\,\ref{fig:alpha3D}) even for small values of $\alpha$ raises the question whether the underlying quantum states still can be interpreted as mesons and whether they resemble their counterparts in the E$_8$ regime of the NN Ising model (corresponding to $\alpha=\infty$) for finite values of $\alpha$ even on the deeper level of quantum information measures. Regarding the first point, it is well known \cite{Liu:2018fza,Lerose_2019} that LR interactions confine domain walls in the Ising model. It is therefore in principle justified to interpret the existence of discrete band structures as meson states. In the present case, we have additionally also the effect of the longitudinal field. We address the resemblance with the E$_8$ regime of the NN model for this case using the \textit{fidelity} $F(\alpha)$ and \textit{fidelity susceptibility} $\chi_F(\alpha)$. These quantities have been used previously for ground \cite{Gu_2008,GU_2010} and excited states \cite{khaloufrivera2021quantum,russomanno2020longrange} as a theoretical framework to identify and characterize quantum phase transitions. Here, we use them to detect if there is a fundamental change in the meson structure as the LR coefficient $\alpha$ is varied.

We consider the system with pbc, in which the first meson band consists of the first $n=1,\ldots,N$ excited eigenstates of the Hamiltonian. Since they have different degeneracies in the LR and NN model, the overlap of some of these states is not well-defined and hence numerically not unique. In fact, only the first excited state ($n=1$) is nondegenerate in all cases and allows us to define the fidelity as
\begin{equation} \label{eq:F}
    F(\alpha) =  \vert\! \braket{\phi_1(\alpha)|\phi_1(\alpha=\infty)} \!\vert , 
\end{equation}
where $\phi_1$ denotes the first excited state in the LR and NN model, respectively. Furthermore, following \cite{Gu_2008,GU_2010} the fidelity susceptibility is defined as
\begin{equation} \label{eq:chi_F}
    \chi_F(\alpha) = -\frac{\partial^2 F(\alpha,\delta\alpha)}{\partial(\delta\alpha)^2}\bigg\vert_{\delta\alpha=0} = \lim_{\delta\alpha\to0} \frac{-2 \ln F(\alpha,\delta\alpha)}{(\delta\alpha)^2} ,
\end{equation}
where $F(\alpha,\delta\alpha) = \vert\! \braket{\phi_1(\alpha)|\phi_1(\alpha+\delta\alpha)}\!\vert$. In our numerics, we use the second relation with the numerical value $\delta\alpha=0.01$ and probe the range $0 \le \alpha \le 3$. 

\begin{figure*}[t]
\centering
 \includegraphics[width=0.49\textwidth]{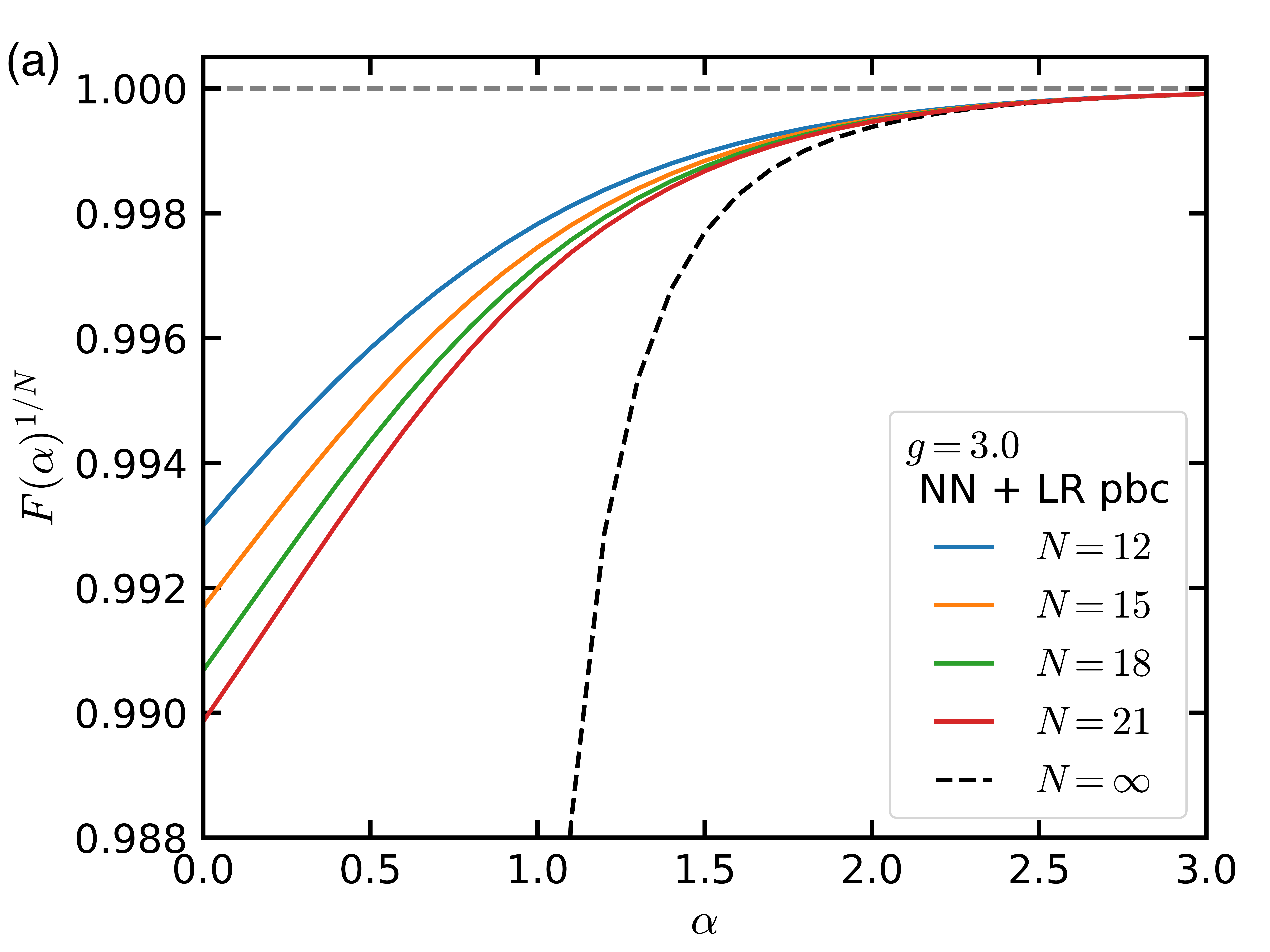} 
 \includegraphics[width=0.49\textwidth]{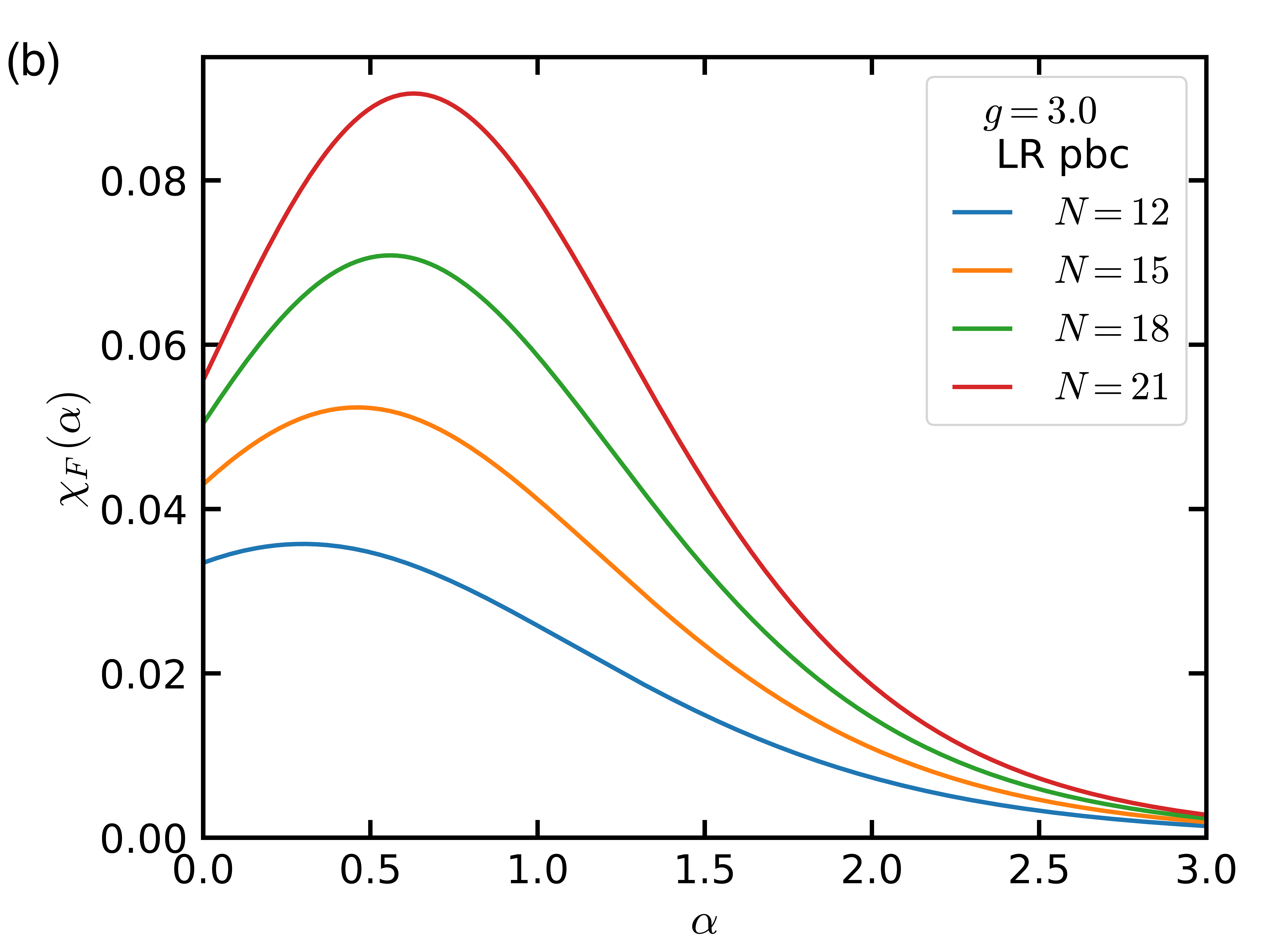} 
 \caption{(Color online) Dependence of the fidelity (a), defined in \eqref{eq:F}, and fidelity susceptibility (b), defined in \eqref{eq:chi_F}, on the LR coefficient $\alpha$. 
 (a): In finite-size systems (colored solid lines, increasing $N$ from top to bottom), the first meson state retains a large overlap to the one of the NN model ($\alpha=\infty$). An extrapolation to $N\to\infty$ (black dashed line) indicates a transition in the nature of the meson state in the thermodynamic limit, occurring at some value of $\alpha\lesssim 1.5$. 
 (b): In agreement with this finding, the fidelity susceptibility shows a peak that becomes sharper with system size (colored solid lines, increasing $N$ from bottom to top). Assuming a scaling with $N^{-1}$, we obtain a peak position of $\alpha_{\rm max} \approx 1.07 \pm 0.02$ in the thermodynamic limit.
 }
 \label{fig:fidelity}
\end{figure*}

The results for the fidelity per site $f(\alpha) \equiv F(\alpha)^{1/N}$ and the fidelity susceptibility $\chi_F(\alpha)$ are shown in Fig.\,\ref{fig:fidelity} for several chain lengths at the longitudinal field value $g=3$. For all finite system sizes under consideration (colored solid curves), which are within experimental scope, for $\alpha \gtrsim 2$, $f(\alpha)$ lies close to the maximal value of 1, and even at all-to-all LR interactions ($\alpha=0$, lowest solid curve) the fidelity per site decreases at most by 1\%. These findings indicate that the quantum nature of the first excited state in the LR model resembles very closely its counterpart in the NN case, at least for finite system sizes. 

In addition to the finite-size results, we extrapolate the data to the thermodynamic limit $N \to \infty$ by making the scaling ansatz $f(\alpha) = f_\infty(\alpha) + c(\alpha){N^{-b(\alpha)}}$. We leave the exponent $b(\alpha)$ together with $c(\alpha)$ and $f_\infty(\alpha)$ as free fit parameter. The result for $f_\infty(\alpha)$, representing the prediction for the thermodynamic limit, is shown as the black dashed curve in the left panel of Fig.\,\ref{fig:fidelity}. While it shows fast convergence for large $\alpha$, it decreases rapidly for $\alpha \lesssim 1.5$, indicating a transition in the nature of the meson state for large system sizes. 
Similarly, the fidelity susceptibility, shown in the right panel, exhibits a peak at small values of $\alpha$ and then decreases towards 0 for strong LR suppression. Such a peak is suggestive of a transition in the first excited meson state occurring at some intermediate value of $\alpha$. As the system size increases, the peak position $\alpha_{\rm max}$ moves towards larger values of $\alpha$. Assuming a scaling with $N^{-1}$, we can extract the value $\alpha_{\rm max} \approx 1.07 \pm 0.02$ for $N\to\infty$ in the thermodynamic limit. This range seems to agree with the rapid decrease of the fidelity in the left panel. 

Thus, while finite size systems retain the same physics across all considered values of $\alpha$, the scaling analysis suggest the appearance of interesting new physics for the first meson state in the LR versus NN model, which would be worth to study on its own. This finding agrees with the fact that the Ising model with variable-range interactions in a transverse field shows a transition in its quench dynamics at $\alpha= 1$ \cite{Hauke_2013,Cevolani2015}. Besides hinting at interesting physics in the excited states, this result suggests that the mesons for at least $\alpha\gtrsim 1$ remain smoothly connected to the NN meson even at large system sizes.

\vspace{1cm}
\section{Meson mass identifications}

\begin{figure}[t]
\centering
 \includegraphics[width=0.49\textwidth]{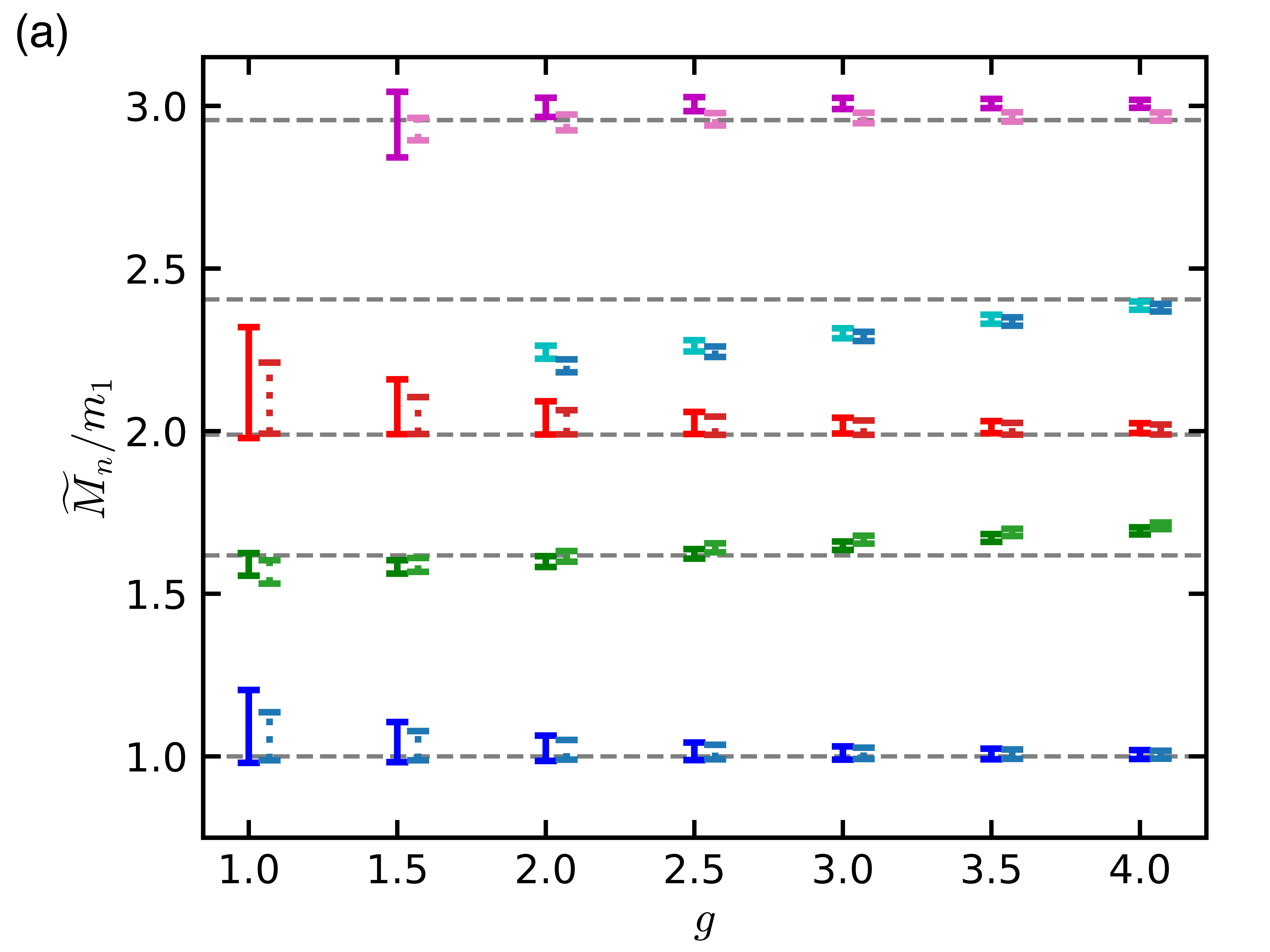} 
 \includegraphics[width=0.49\textwidth]{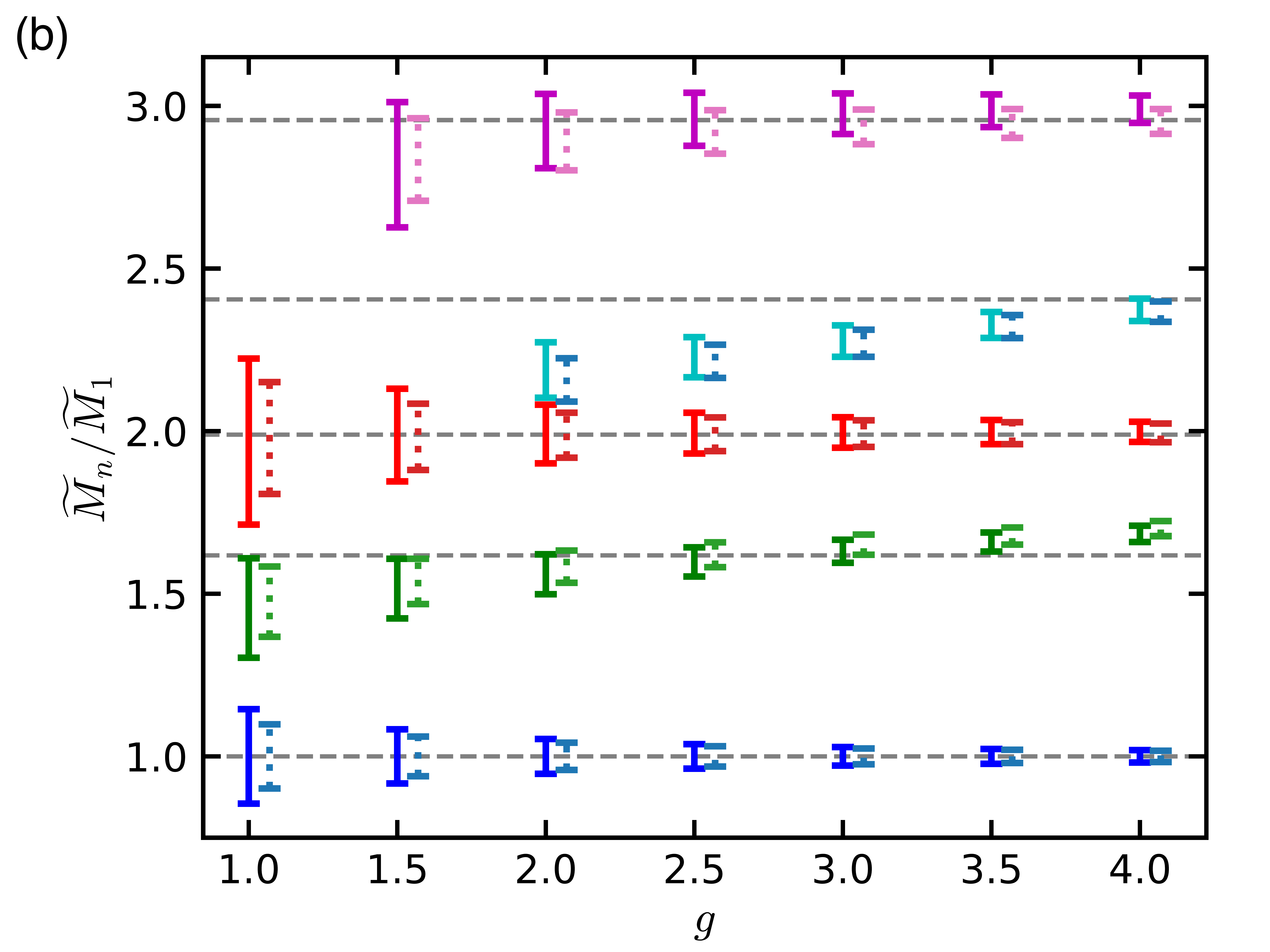} 
 \caption{(Color online) Extracted meson mass ratios $\widetilde{M}_n/m_1$ and $\widetilde{M}_n/\widetilde{M}_1$ from the energy absorption spectra in dependence of the longitudinal field g. The results are expressed in units of the mass gap $m_1$ (a) and the first extracted meson mass $\widetilde{M}_1$ (b). The level $n=1,2,\ldots,5$ increases from bottom to top. Solid errorbars are for the NN model, dotted ones for the LR model (shown slightly displaced for graphical purposes). Grey dashed lines represent the the analytical E$_8$ meson mass ratios $M_n/M_1$ (cf.\ table~\ref{tab:e8_masses_ratio}). Numerical parameters: $N=18$ (pbc), $\Gamma/J=0.1$, $\alpha=3$.
 Once $g$ is sufficiently large, the meson mass ratios can be reliably extracted even in small systems. 
 }
 \label{fig:masses_field_dependence}
\end{figure}

For the previous discussions, one specific longitudinal field value was chosen. In this section, we extract the meson masses from the energy absorption spectrum in dependence of $g$ \footnote{Further details on the underlying spectra are discussed in appendix~\ref{app:field_dependence}.}. 
The results are presented in Fig.\,\ref{fig:masses_field_dependence}. Individual meson masses $\widetilde{M}_n$ are obtained from a Gaussian fit to each peak in units of the mass gap $m_1$ of the first excited state with an uncertainty corresponding to its full width at half maximum (shown in panel (a)). Since the individual energy of an eigenstate is experimentally not accessible, we additionally express the results with respect to the extracted mass $\widetilde{M}_1$ of the first meson by propagating its uncertainty (shown in panel (b)).
With increasing value of $g$, the uncertainty of the meson mass decreases, allowing for a more precise identification of the analytical E$_8$ mass ratios $M_n/M_1$ up to the fifth level for both the NN (solid errorbars) and the LR model (dotted errorbars). The fourth meson is constantly underestimated except for the largest considered longitudinal field strength. Overall, the numerical data of the finite size system are closest to the E$_8$ QFT in the range $3 \le g \le 4$, with an even smaller uncertainty for the LR model. \change{Moreover, one can observe that the measurable ratio in panel (b) even allows for a slightly better consistency with the analytical E$_8$ mass ratios. The resulting uncertainties (peak widths) follow as a property of the spectrum in combination with the chosen inverse observation time $\Gamma/J=0.1$.}

\section{Quantum simulation in trapped ions}
\label{sec:trappedions}

\begin{figure}[t]
\centering
 \includegraphics[width=0.49\textwidth]{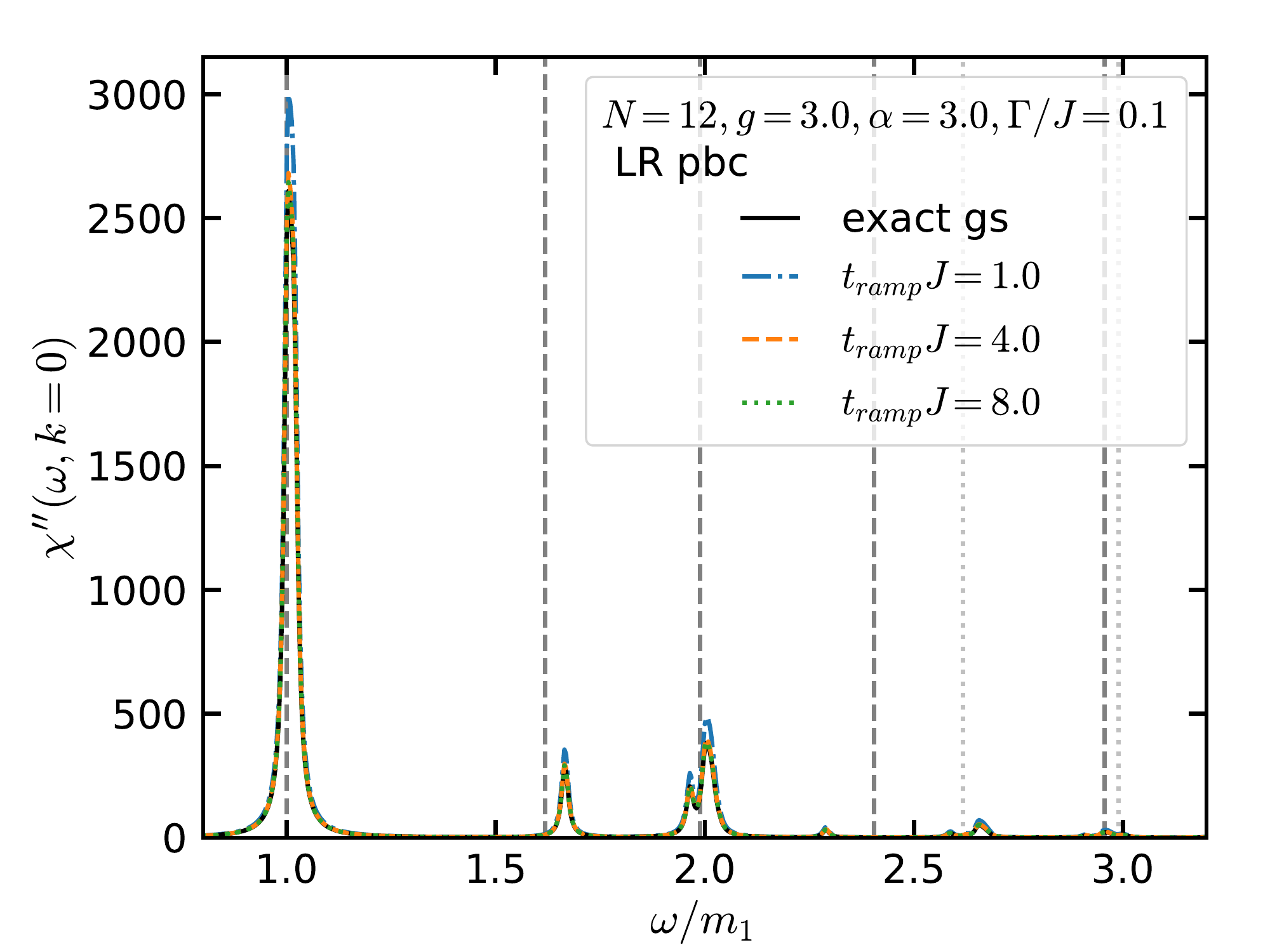} 
 \caption{(Color online) Comparison of the energy absorption spectra for adiabatically prepared groundstates with different time durations $t_{ramp}$ of a linear ramp (colored dashed curves) to the exact ground state case in the LR model (black solid curve). Background lines are as in Fig.~\ref{fig:E8comparison}. 
 Even for a relatively fast ramp of $t_{ramp}J = 1$, the qualitative agreement of the spectrum with the exact case is very good and allows for the identification of E$_8$ meson mass ratios.
 }
 \label{fig:adiabatic_comparison}
\end{figure}

In trapped-ion quantum simulators, effective magnetic models are routinely realized by encoding the basis states $\ket{\uparrow}$ and $\ket{\downarrow}$ of spins 1/2 in two long-lived hyperfine states and inducing effective spin--spin interactions $\sim J$ through a phonon bus, e.g., using a Moelmer--Soerensen-type laser or microwave beam \cite{MoelmerSoerensen2000}. Effective magnetic fields $\sim h,g$ can be realized by a detuning of the Moelmer--Soerensen beams \cite{Richerme_2013,Jurcevic_2014,Jurcevic_2015,Wang2012,Hauke_2015} or by additional lasers that are tuned off-resonantly to the carrier transition \cite{Smith2016,Maier2019}.  

An experimental protocol to measure the E$_8$ spectrum in such a system is as follows. 
First, the effective spins are prepared in the electronic ground state, corresponding to the fully polarized state $\ket{\uparrow,..,\uparrow}$, the ground state at $g=\infty$. By slowly decreasing $g$ and turning on $J$ and $h$, the system is adiabatically transferred to the ground state at the desired parameter values. Such a procedure can produce considerable excitations when crossing a quantum phase transition \cite{Richerme_2013}. In the present scenario, instead, \change{even though we are interested in the QFT regime emanating from the critical point, we find} the final value of the transverse field $g$ \change{can be} large, ensuring a large many-body gap on the order of $M_{g} \equiv \, {\cal D} J \, |g|^{8/15}$. For $g=3$, we have $M_{g}=9.7 J$ and thus the initial state preparation can occur adiabatically in times much shorter than $\hbar/J$, which in turn are much shorter than typical coherence times. 
Alternatively, ground states in trapped-ion quantum computers can be prepared to good precision using variational algorithms \cite{Kokail:2018eiw}. 

After initialization, the system is perturbed with a time-dependent magnetic field, which again can be realized by periodically modulating the detuning of the Moelmer--Soerensen beams or by a time-modulated AC-Stark shift. Using single-site addressing, site dependent AC-Stark shifts with switching times much faster than the timescales of the internal dynamics (on the order of $\hbar/J$) have already been demonstrated experimentally \cite{Smith2016,Maier2019}. It is thus possible to perturb the effective spin system with an operator of the type defined in Eq.~\eqref{eq:operators}. 

Two spectroscopy protocols are thinkable. Either, the perturbation is turned on abruptly, e.g., as a step function and the subsequent time evolution of the same observable is tracked, which for local observables of the type $O$ can be done by standard fluorescence measurements~\cite{Schindler_2013,Blatt:2021_Demonstrator,Monroe:2019asq}. A Fourier transform then yields the desired imaginary part of the dynamic susceptibility, $\chi''(\omega)$, defined in Eq.~\eqref{eq:chi_kO} \cite{jensen1991rare}. 
Alternatively, the perturbation can be modulated temporally with a $\cos(\omega t)$. By tracking the absorbed energy per unit time, which amounts to the measurement of few-body correlators and which has already been demonstrated experimentally \cite{Kokail:2018eiw}, again $\chi''(\omega)$ is obtained. 

Typical trapped-ion experiments on many-body spin systems generate long-range interactions \cite{Britton_2012,Richerme_2013,Senko_2014,Jurcevic_2014,Jurcevic_2015}. 
In linear chains with open boundary conditions, for not too large systems these approximate a spatial power-law decay to good precision \cite{Nevado_2016,Trautmann_2018}, and any deviations from the desired power-law interactions can be mitigated by shaping of the interactions, e.g., by additional laser beams \cite{Korenblit2012,Hauke_2015,Davoudi:2019bhy}, periodic driving \cite{Nevado_2017,Manovitz2020}, or trap-shaping techniques~\cite{Bermudez2012b,Zippilli2014,Yang:2016hjn,Gutierrez2019}. 
It is nowadays also possible to prepare ions in ring conformations, thus enabling the realization of periodic boundary conditions \cite{Horstmann2010,Horstmann2011,Li_2017}. 
While the theoretical range of power-law decay exponents is $0\leq \alpha \leq 3$ \cite{Porras2004a}, the experimentally most favorable power-law decays are at $\alpha=0$ when working with the axial center-of-mass mode or in an intermediate range when interactions are transmitted by the radial phonon modes. For example, in \cite{Jurcevic_2014} the range of $0.75\leq \alpha \leq 1.75$ has been accessed, which -- as the spectra and fidelity analysis reported in the previous sections show -- enables access to meson spectra that closely approach the physics of the ideal NN model.  

\change{The robustness of the adiabatic ground state preparation and subsequent spectroscopic analysis is explicitly demonstrated in Fig.~\ref{fig:adiabatic_comparison} for the LR model. Here, we consider the previously discussed parameters $\alpha=g=3$ for a small chain of $N=12$ sites. Starting from the fully polarized state, the ground state in the E$_8$ regime is adiabatically prepared by increasing the transverse field from $h=0$ up to $h=1$ using a linear ramp profile of duration $t_{ramp}$. We assume $J\mathrm d t=0.01$, from which $\mathrm d h$ follows through the relation $\mathrm dh/\mathrm dt = 1/t_{ramp}$. The figure shows the resulting absorption spectra for several values of $t_{ramp} \lesssim M_g$ (colored dashed curves) in comparison to the case when the exact ground state is calculated and employed in eq.~\eqref{eq:chi_kO} (black solid curve). The curves for $t_{ramp}J = \{4,8\}$ are nearly coinciding with the exact result. Only for a fast ramp with $t_{ramp}J = 1$ the peak heights are marginally overestimated but still allow for a precise meson identification.}

\section{Summary and outlook}

In this letter, we have demonstrated that the relativistic E$_8$ QFT can be identified experimentally on ion-trap quantum simulators. Surprisingly small systems of only 12 to 18 sites with pbc, which implement the LR quantum Ising model at the experimentally largest possible LR suppression, resemble the NN Ising model closely and allow for the identification of E$_8$ meson states. For longer-ranged interactions, while most meson states disappear from the spectrum, the lowest meson remains a strong feature. 
We have calculated the energy absorption spectrum based on linear response theory and showed that it shares qualitative and quantitative features with its QFT counterpart. Single and multiparticle meson states appear as peaks in the energy absorption spectrum, which allow for a precise extraction of analytically predicted E$_8$ meson mass ratios even for large longitudinal field values. 
As a fidelity analysis shows, for small systems the nature of the first meson changes only insignificantly across all values of $\alpha$ considered, while we find indications for a transition in the meson state in the thermodynamic limit at a critical value around $\alpha_c\approx 1$.  
We have also discussed an protocol adapted to existing trapped-ion technologies to experimentally access the meson spectra. While we have focused on the zero momentum case, this procedure can be extended to derive also relativistic dispersion relations at finite momenta\change{, which would enable, e.g., to extract the meson dispersion relation}.

We have focused in our study on the E$_8$ regime, which appears in a parameter region of the simple Ising model (longitudinal perturbations at the QCP) and has been experimentally verified previously in solid state crystals. Using ion-trap based quantum simulation technologies opens, however, a new avenue to address relativistic meson physics also in more complicated gauge theories, see, e.g., \cite{Banuls:2019bmf} for an overview of recent progress in the field. Furthermore, ion-trap quantum simulations allow for studies of finite-temperature systems \cite{Zhu:2019bri,Mildenberger_MA}, which offers, for example, the possibility to study the rich physics of meson melting \cite{Rothkopf:2019ipj}, a process for which currently no complete microscopic understanding is available \,\footnote{See \cite{meson:project} for a related study of this phenomenon from a tensor network perspective in quantum spin chains.}.

\begin{acknowledgments}
The Gravity, Quantum Fields and Information group at AEI is generously supported by the Alexander von Humboldt Foundation and the Federal Ministry for Education and Research through the Sofja Kovalevskaja Award. JK is partially supported by the International Max Planck Research School for Mathematical and Physical Aspects of Gravitation, Cosmology and Quantum Field Theory. The work of JK is supported in part by a fellowship from the Studienstiftung des deutschen Volkes (German Academic Scholarship Foundation). 
PH is supported by Provincia Autonoma di Trento, the ERC Starting Grant StrEnQTh (project ID 804305), the Google Research Scholar Award ProGauge, and Q@TN — Quantum Science and Technology in Trento.
\end{acknowledgments}

\appendix
\section{Finite size effects}
\label{app:finite_size}

In this appendix we elaborate on some further details about the finite size dependence of the results presented in the main text.

\begin{figure}[h]
\centering
 \includegraphics[width=0.49\columnwidth]{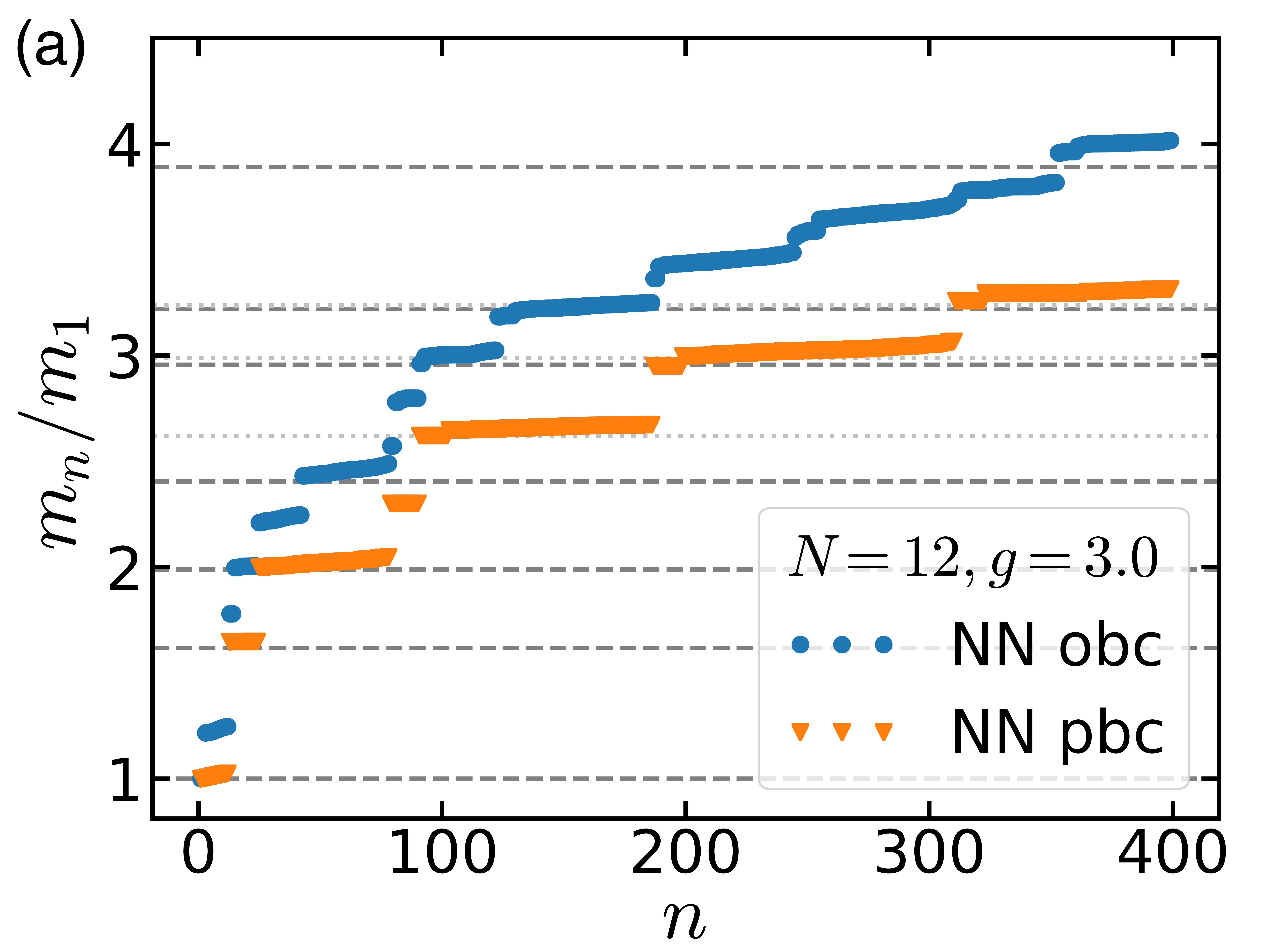} 
 \includegraphics[width=0.49\columnwidth]{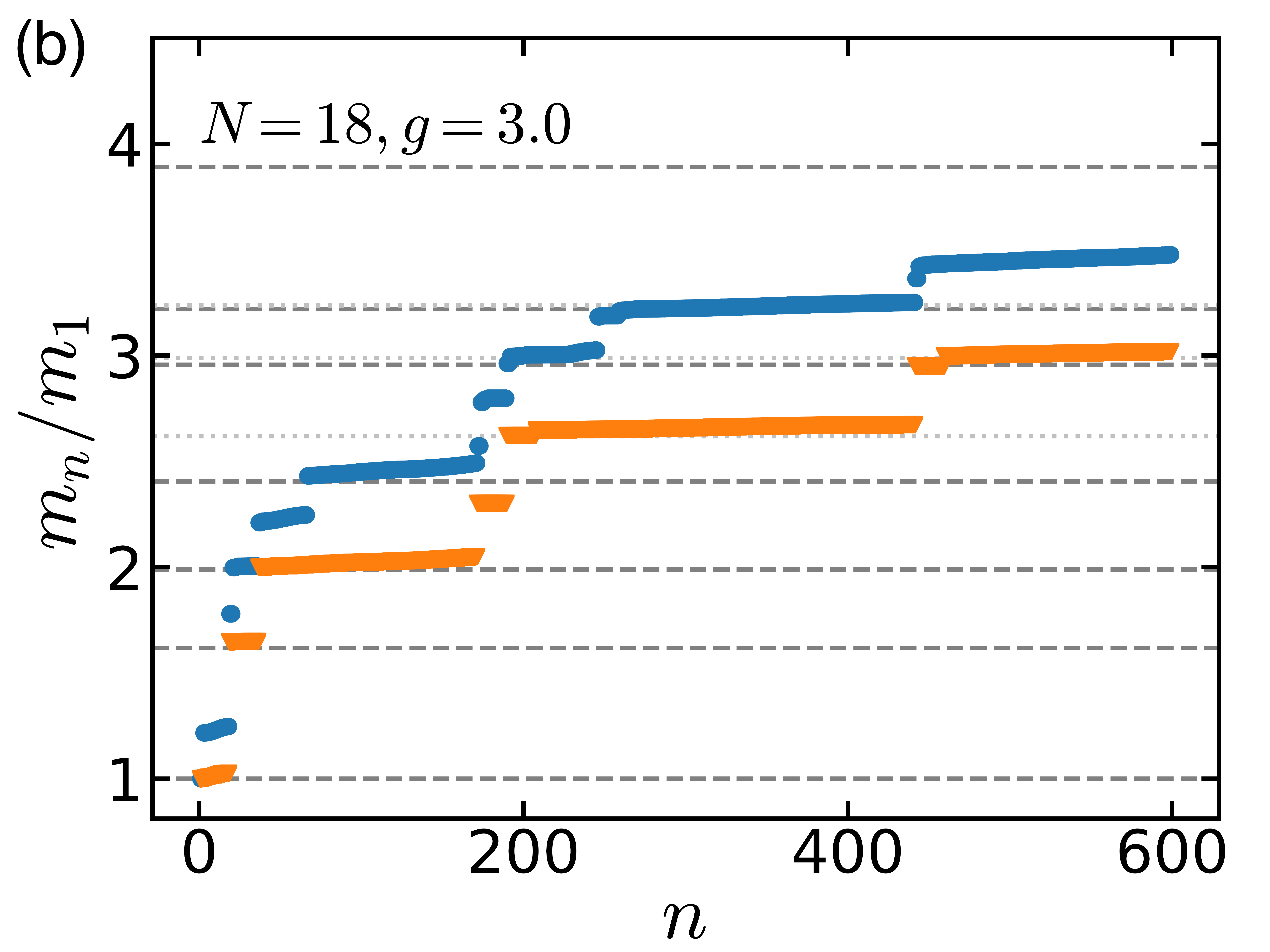} 
 \includegraphics[width=0.49\columnwidth]{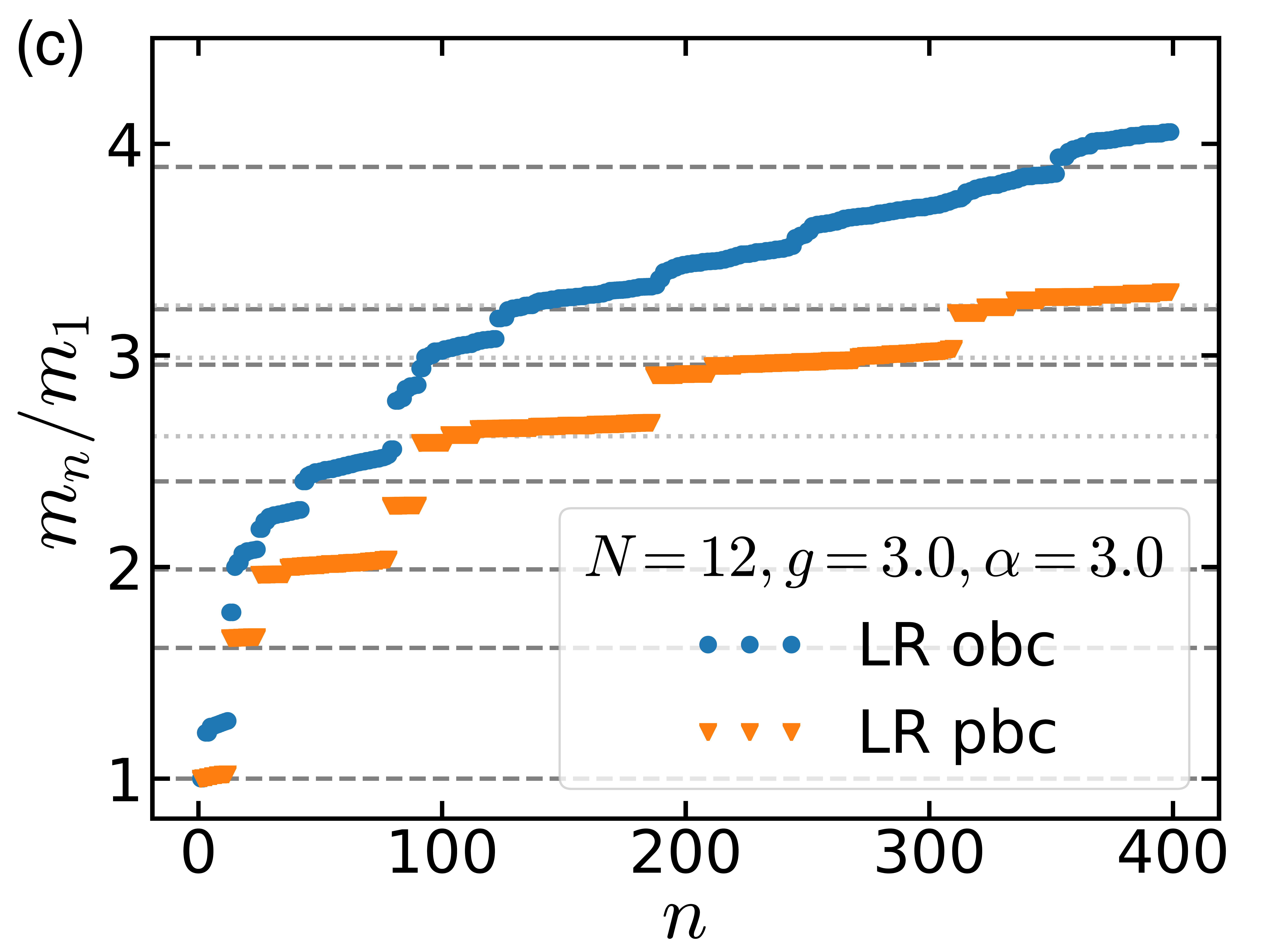} 
 \includegraphics[width=0.49\columnwidth]{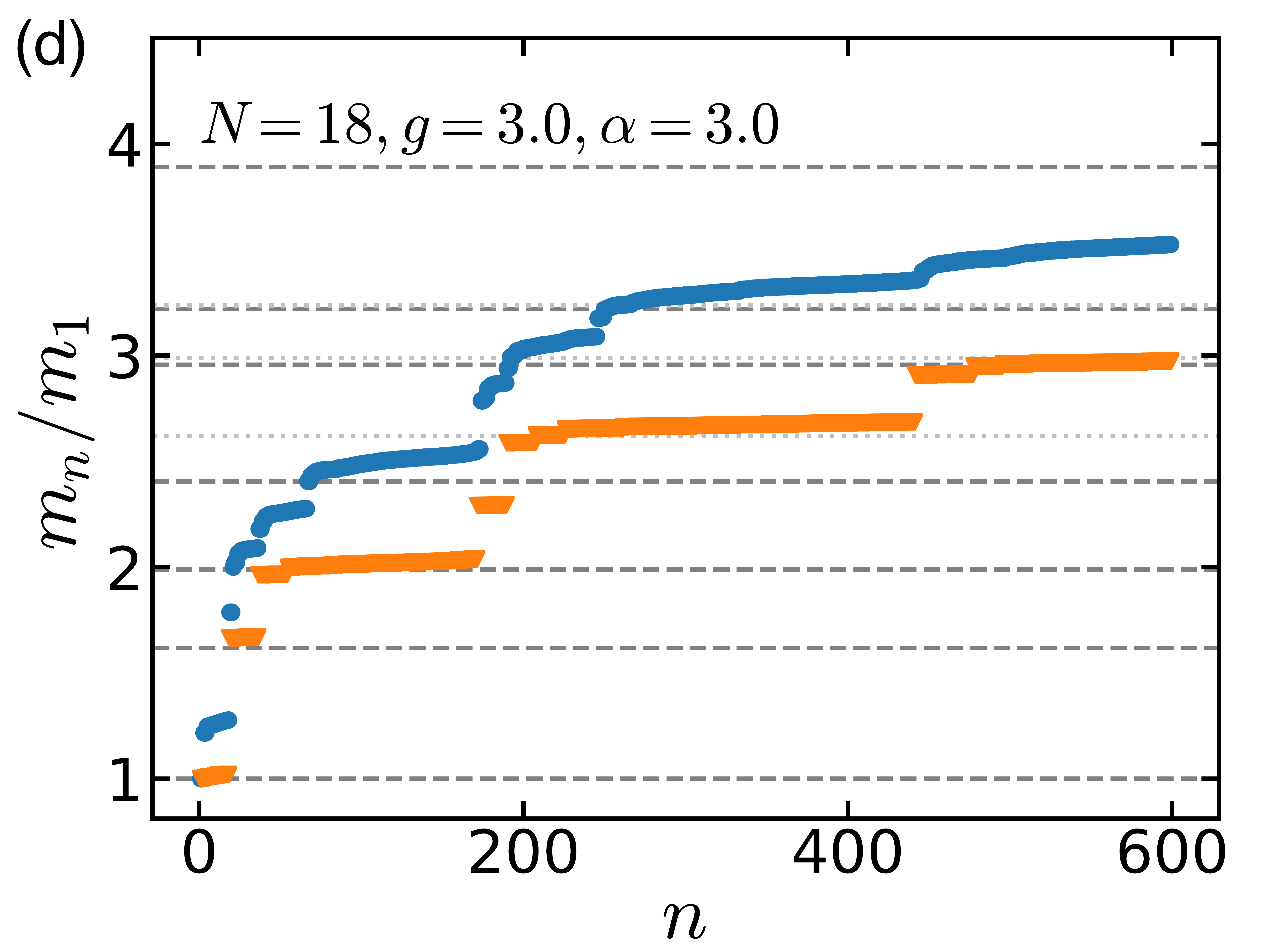} 
 \caption{(Color online) Numerical energy spectra (normalized mass gaps of excited states) for the NN (a,b) and LR (c,d) Ising model with obc (blue circles) and pbc (orange triangles). The left column is for a chain of size $N=12$, the right for $N=18$. Grey dashed lines represent the the analytical E$_8$ meson mass ratios (cf.\ table~\ref{tab:e8_masses_ratio}). Grey dotted lines correspond to multiparticle states with masses $M_1+M_2$, $M_1+M_3$ and $2M_2$ (in ascending order).}
 \label{fig:spectra_finitesize}
\end{figure}

In Fig.~\ref{fig:spectra_finitesize}, the normalized mass gaps of a chain of size $N=12$ (left column  (a,c) as in Fig.\,\ref{fig:energy_spectra_small} of the main text) are compared to a chain with $N=18$ sites (right column (b,d)). The longitudinal field $g=3$ and the largest experimentally accessible decay parameter $\alpha=3$ are considered.
Due to the exponential difference in the total number of eigenstates, different portions of the spectrum are available for a comparable number of excited states. 
For both obc and pbc, the effect of the finite size difference seems to be very mild in the energy spectrum.
Observe also that for obc, higher bands in the LR model seem to resemble a continuous branch. There are only mild differences between the NN model (top row (a,b)) and the LR model at $\alpha=3$ (bottom row (c,d)).

\begin{figure}[h]
\centering
 \includegraphics[width=0.49\textwidth]{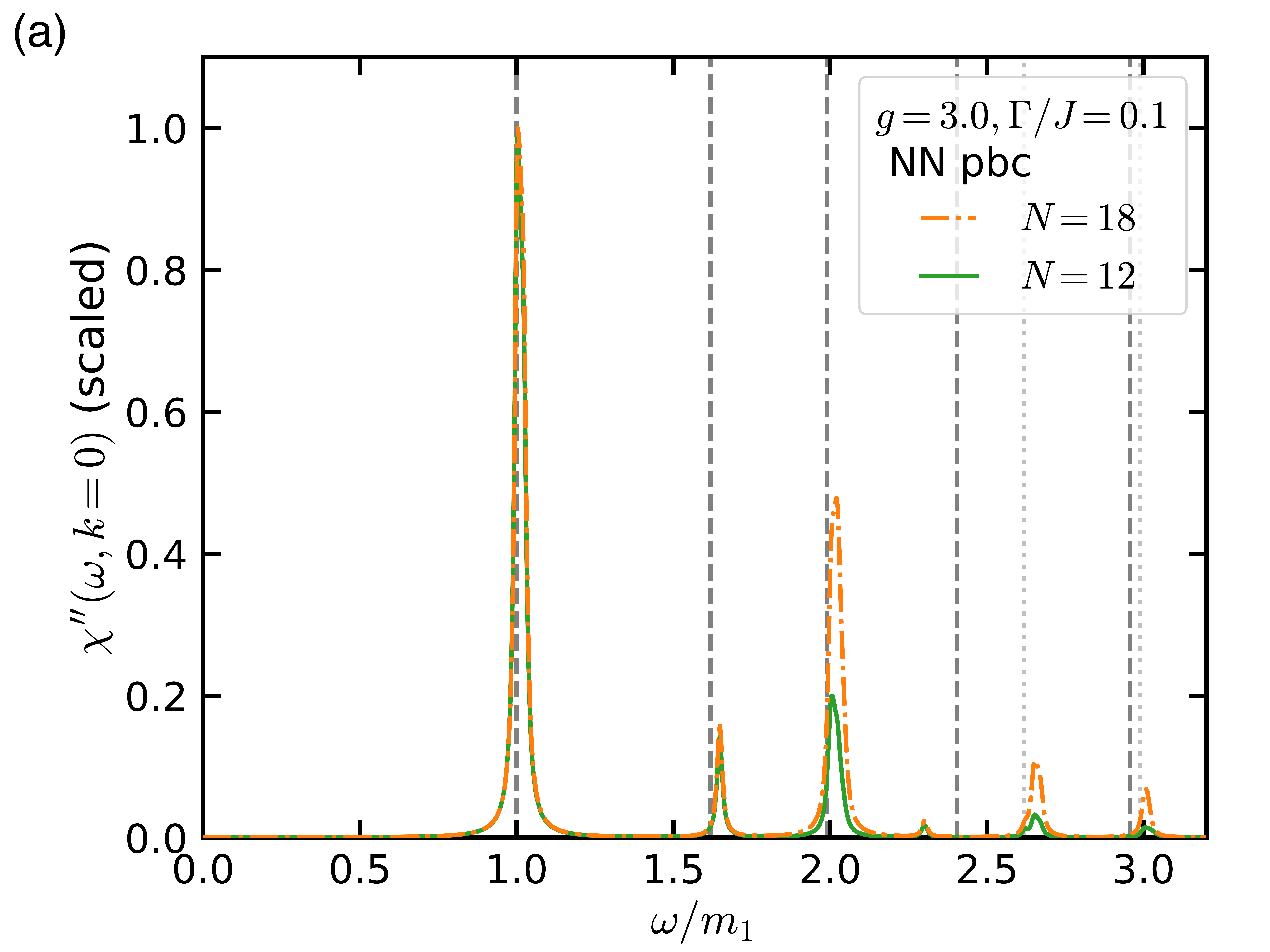} 
 \includegraphics[width=0.49\textwidth]{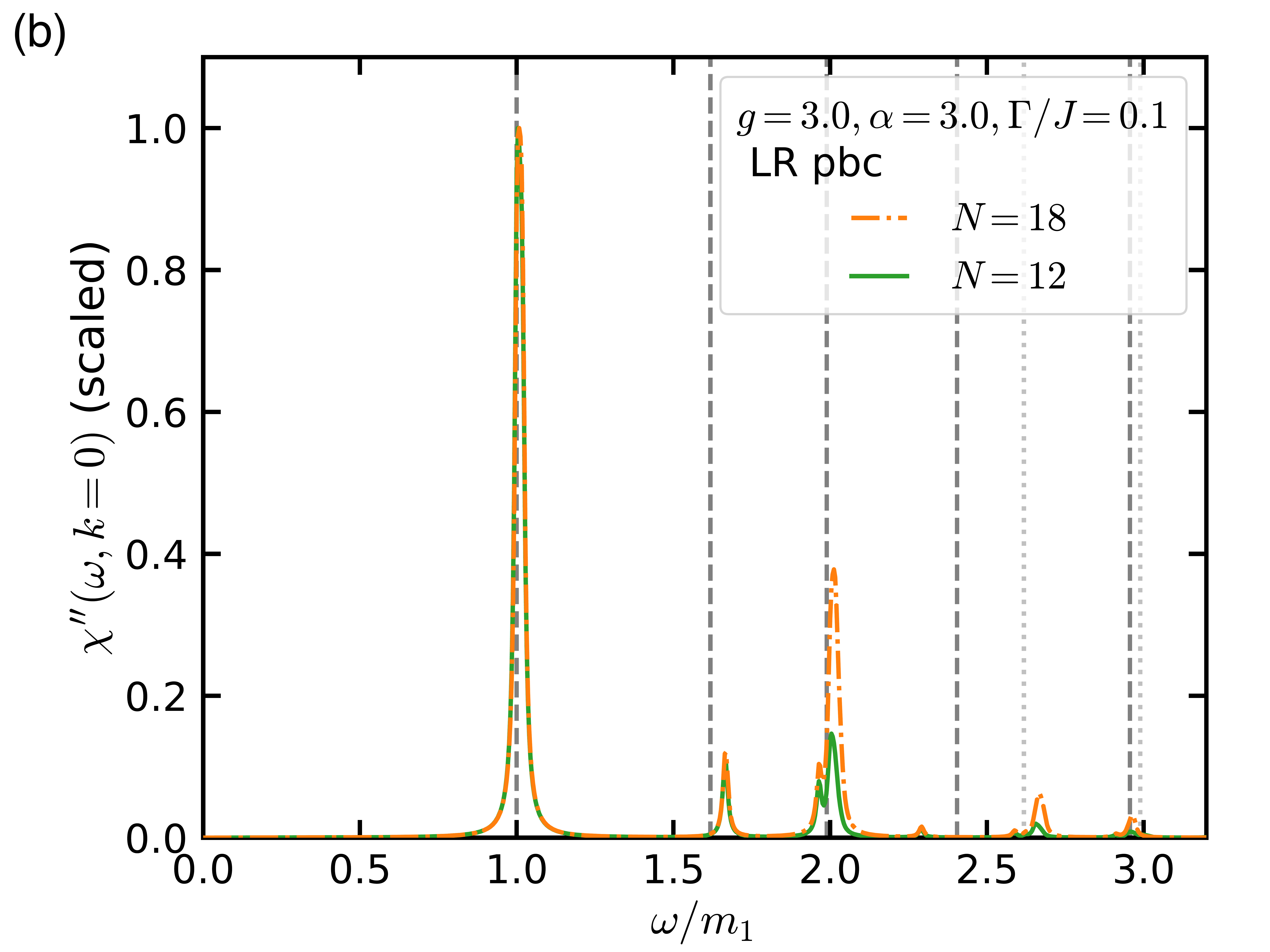} 
 \caption{(Color online) Comparison of the energy absorption spectrum in the NN (a) and LR Ising model (b) for different chain sizes $N$. The data are scaled to the maximum of the spectrum. Grey dashed lines represent the the analytical E$_8$ meson mass ratios (cf.\ table~\ref{tab:e8_masses_ratio}). Grey dotted lines correspond to multiparticle states with masses $M_1+M_2$ and $M_1+M_3$.}
 \label{fig:absorption_finitesize}
\end{figure}

The underlying eigenstates give rise to the energy absorption spectra shown in Fig.\,\ref{fig:absorption_finitesize}. There are nearly no visible differences for the first two meson peaks. Only above the continuum threshold, differences in multiparticle states occur. We therefore conclude that the quantitative agreement with the analytical E$_8$ result for the dynamical structure function, which is described in the main text, is a stable feature for both the NN and LR model. 

\section{Long-range dependence}
\label{app:alpha_dependence}

In this appendix, we analyze the effect of the LR coefficient $\alpha$ on the physics discussed in the main text.

\begin{figure}[h]
\centering
 \includegraphics[width=0.49\columnwidth]{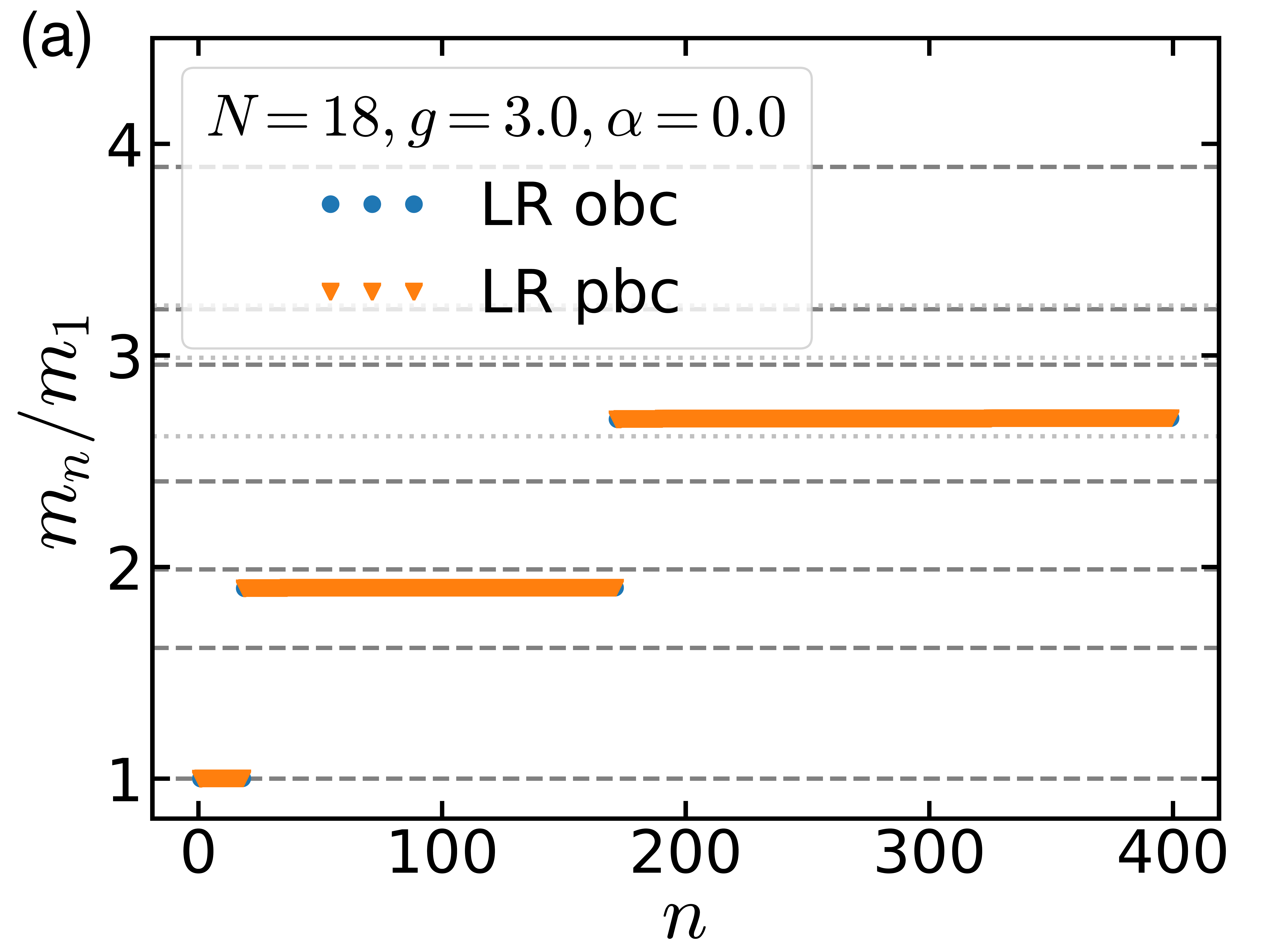} 
 \includegraphics[width=0.49\columnwidth]{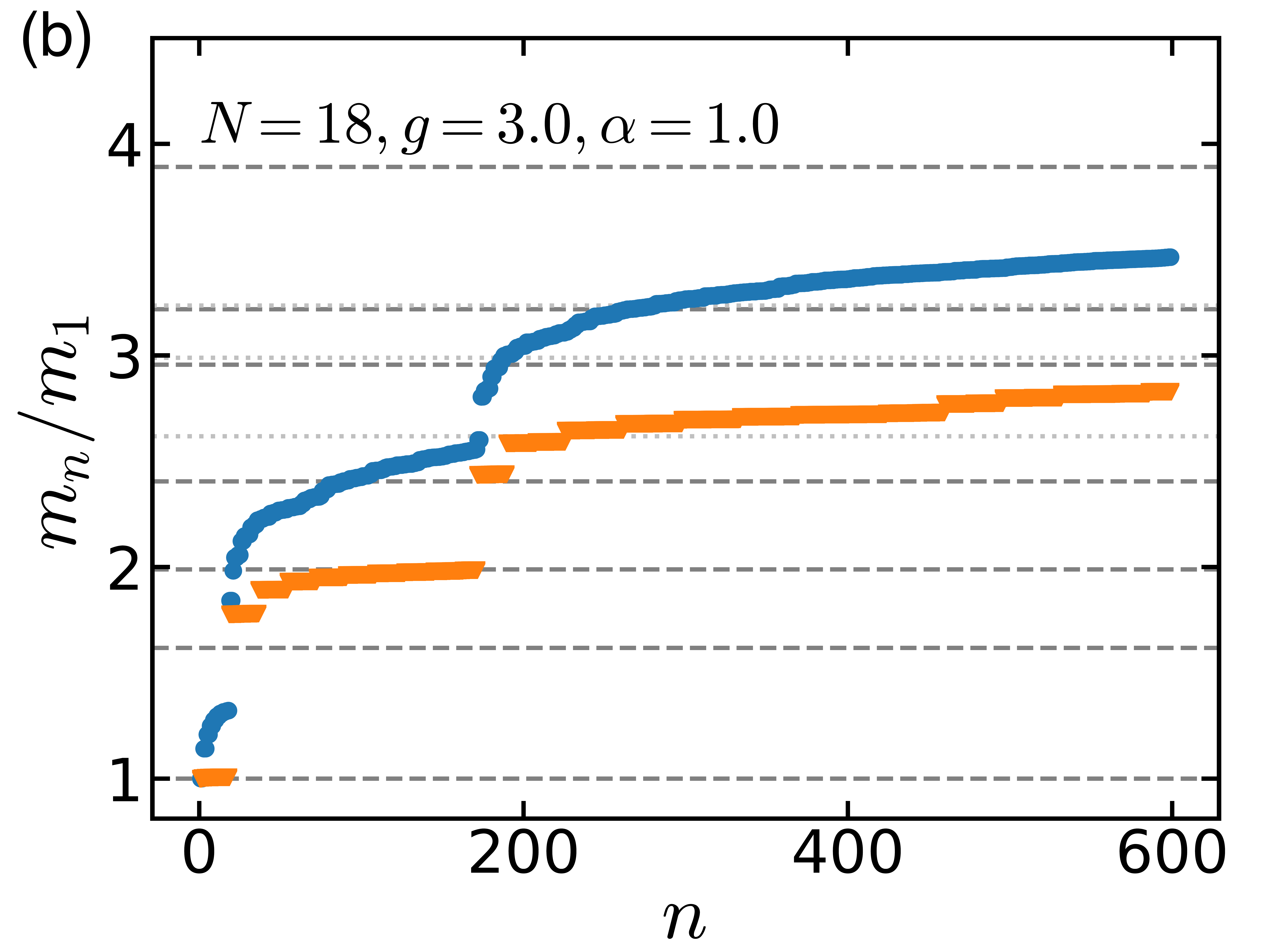} 
 \includegraphics[width=0.49\columnwidth]{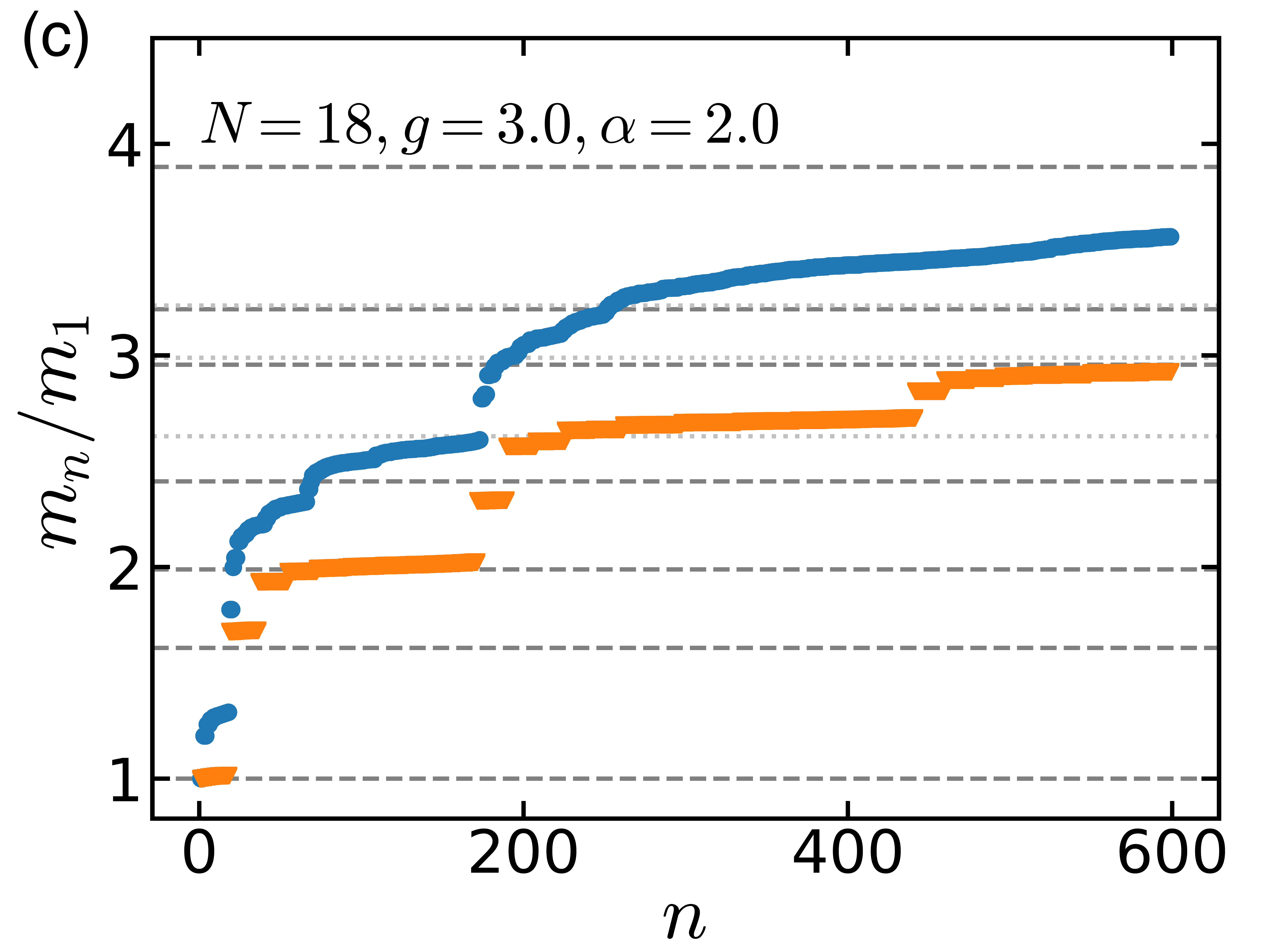} 
 \includegraphics[width=0.49\columnwidth]{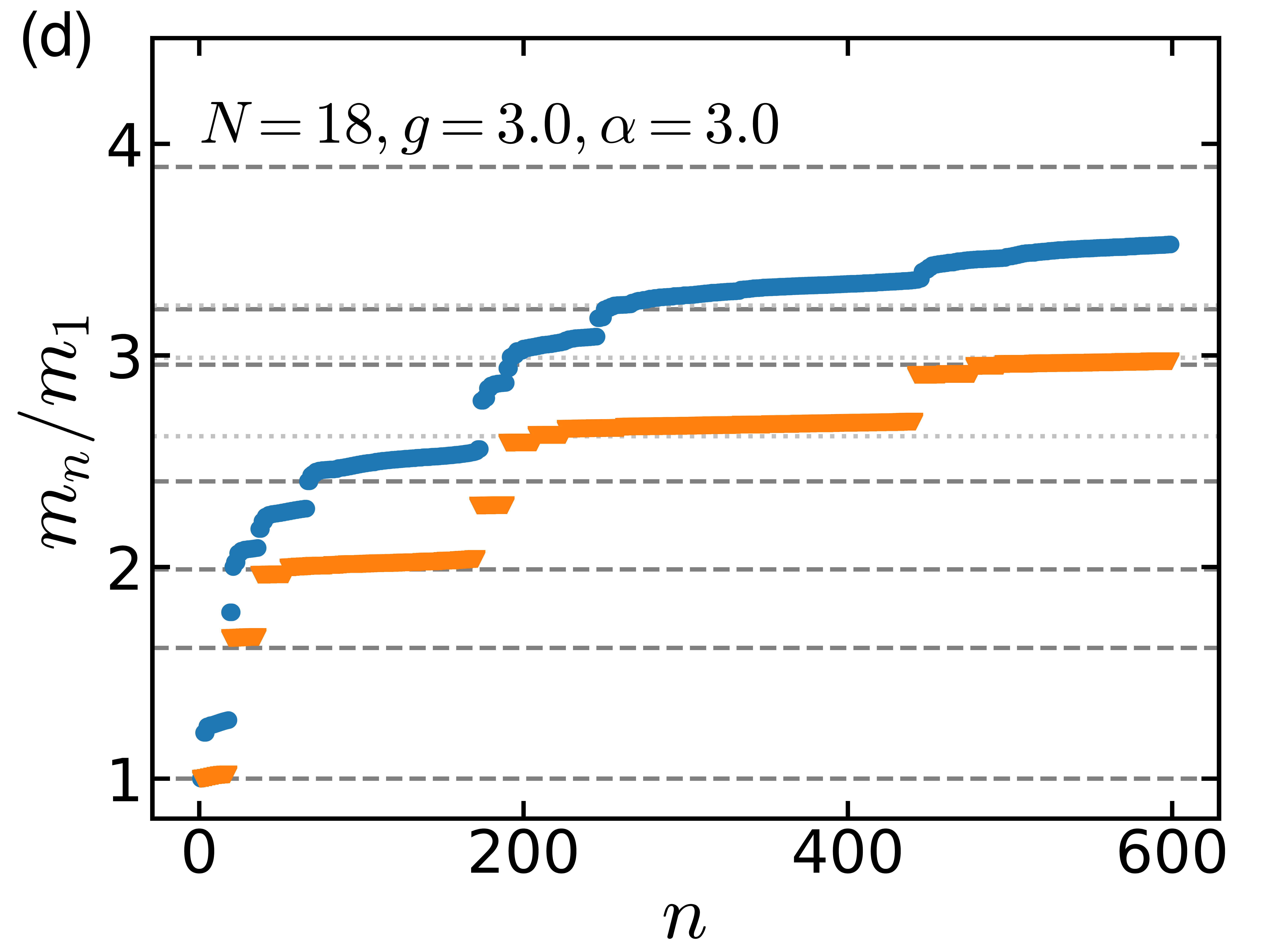} 
 \caption{(Color online) Effect of the LR coefficient $\alpha$ on the numerical energy spectrum (normalized mass gaps of excited states) with obc (blue circles) and pbc (orange triangles). Background lines are as in Fig.\,\ref{fig:spectra_finitesize}.}
 \label{fig:alpha_effect}
\end{figure}

Figure \ref{fig:alpha_effect} displays the energy spectrum as a function of $\alpha$.
We vary the parameter in the range $\alpha=0$ (all-to-all LR interactions) up to the previously used $\alpha=3$ (strong LR suppression). At $\alpha=0$, three identical degenerate branches are visible within the considered portion of the spectrum, for obc and pbc.
For increasing values of $\alpha$, semi-continuous branches in the obc spectrum are more and more split into discrete bands and new bands appear in the case of pbc. Already for $\alpha \ge 2$, the bands in the pbc spectrum resemble the E$_8$ lines. 

\begin{figure}[t]
\centering
 \includegraphics[width=0.49\textwidth]{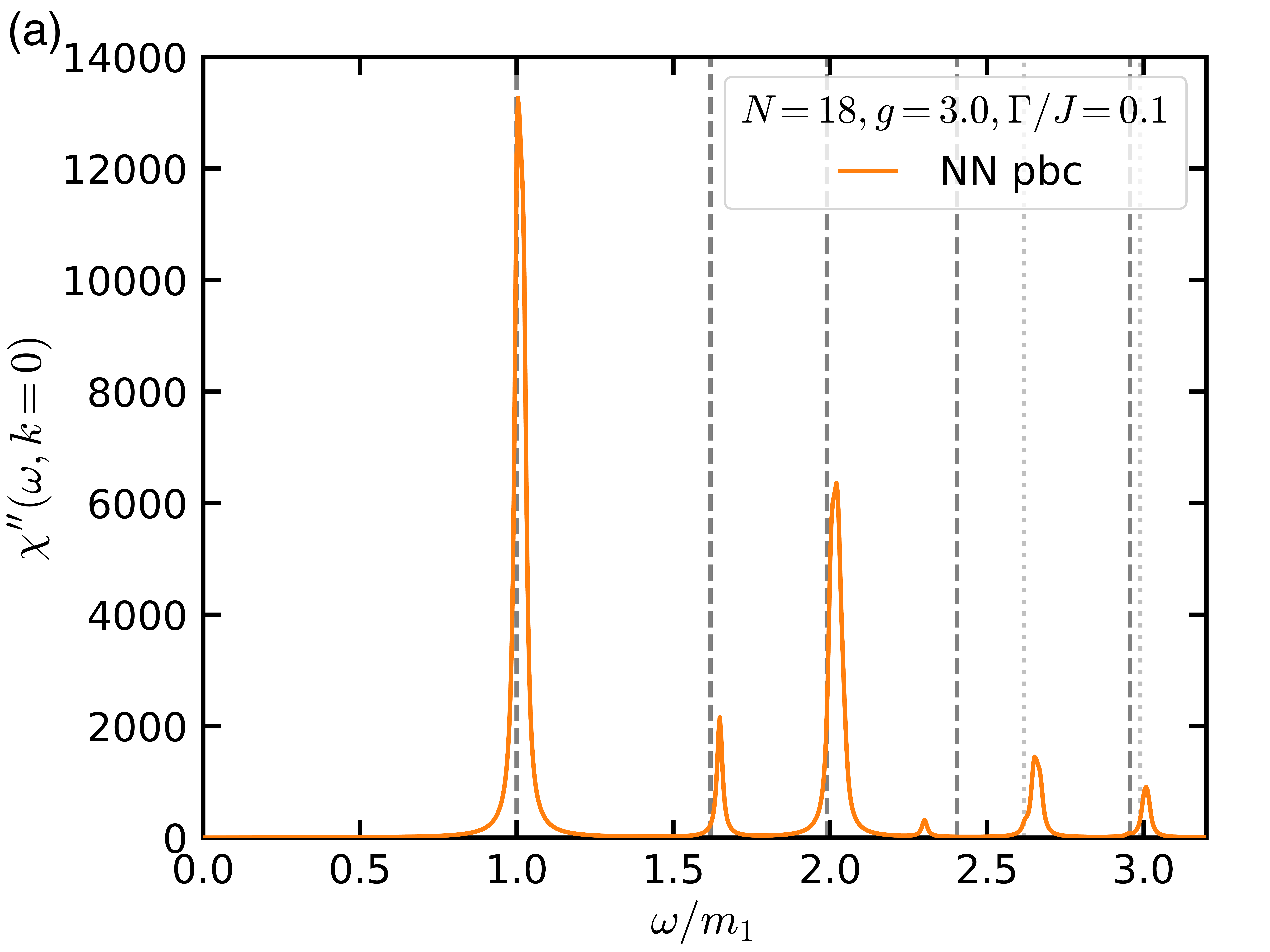} 
 \includegraphics[width=0.49\textwidth]{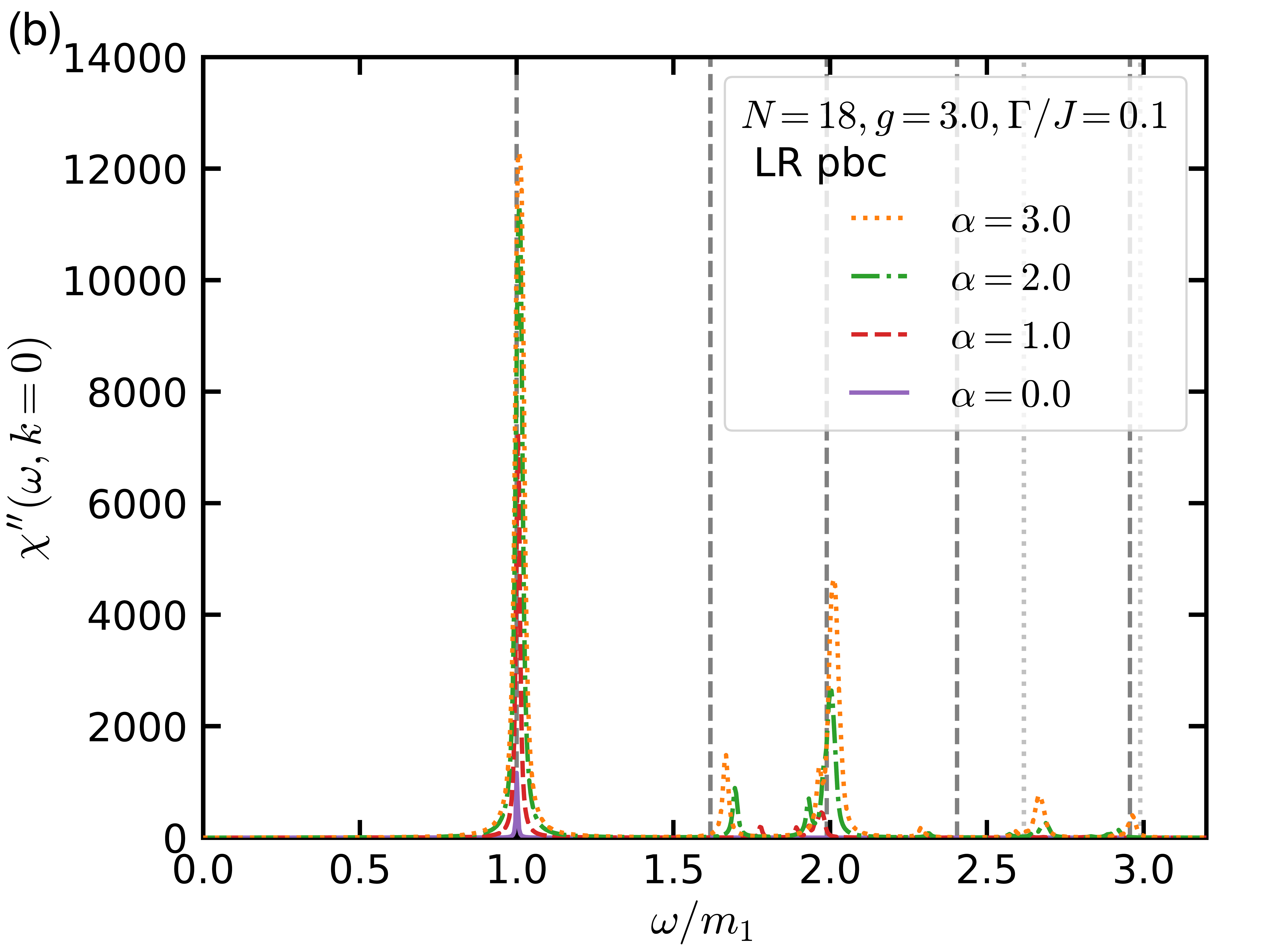} 
 \caption{(Color online) Comparison of the absorption spectrum in the NN (a) and LR Ising model (b). In the LR model, the coefficient $\alpha$ is varied over the whole experimentally accessible range of values. Background lines are as in Fig.\,\ref{fig:absorption_finitesize}.}
 \label{fig:absorption_alpha}
\end{figure}

The resulting energy absorption spectra are shown in Fig.\,\ref{fig:absorption_alpha} (b) in absolute units for a quantitative comparison to the NN model (a). As $\alpha$ is increased, the intensities of the individual peaks are increased and the peak positions resemble the E$_8$ mass ratios. Observe that there is only a very small difference in the absolute height of the first peak for the NN versus the LR model at $\alpha=3$.

\section{Longitudinal field dependence}
\label{app:field_dependence}

In this appendix we provide further details about the longitudinal field dependence of the absorption spectra. 

\begin{figure}[h]
\centering
 \includegraphics[width=0.49\columnwidth]{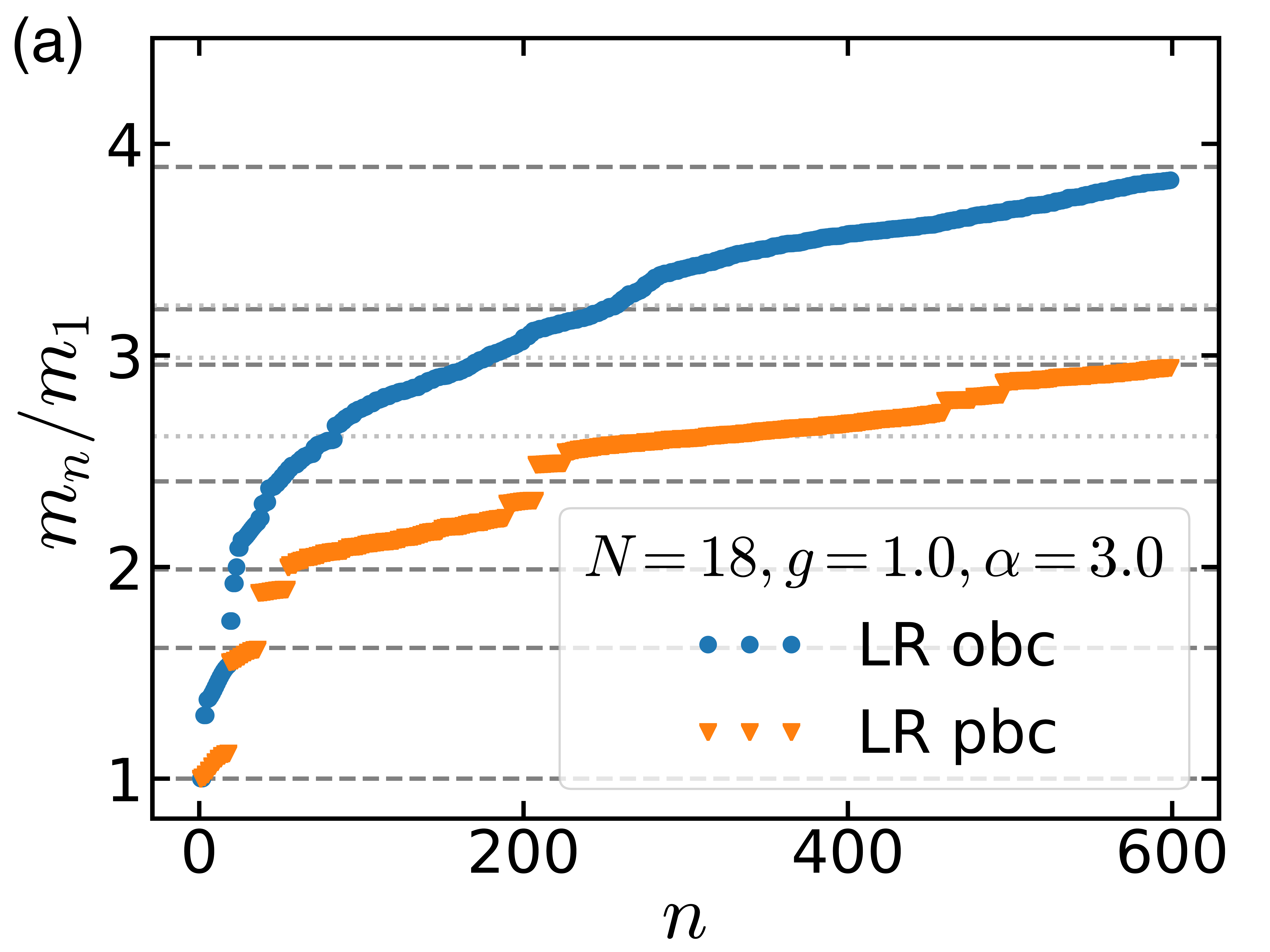} 
 \includegraphics[width=0.49\columnwidth]{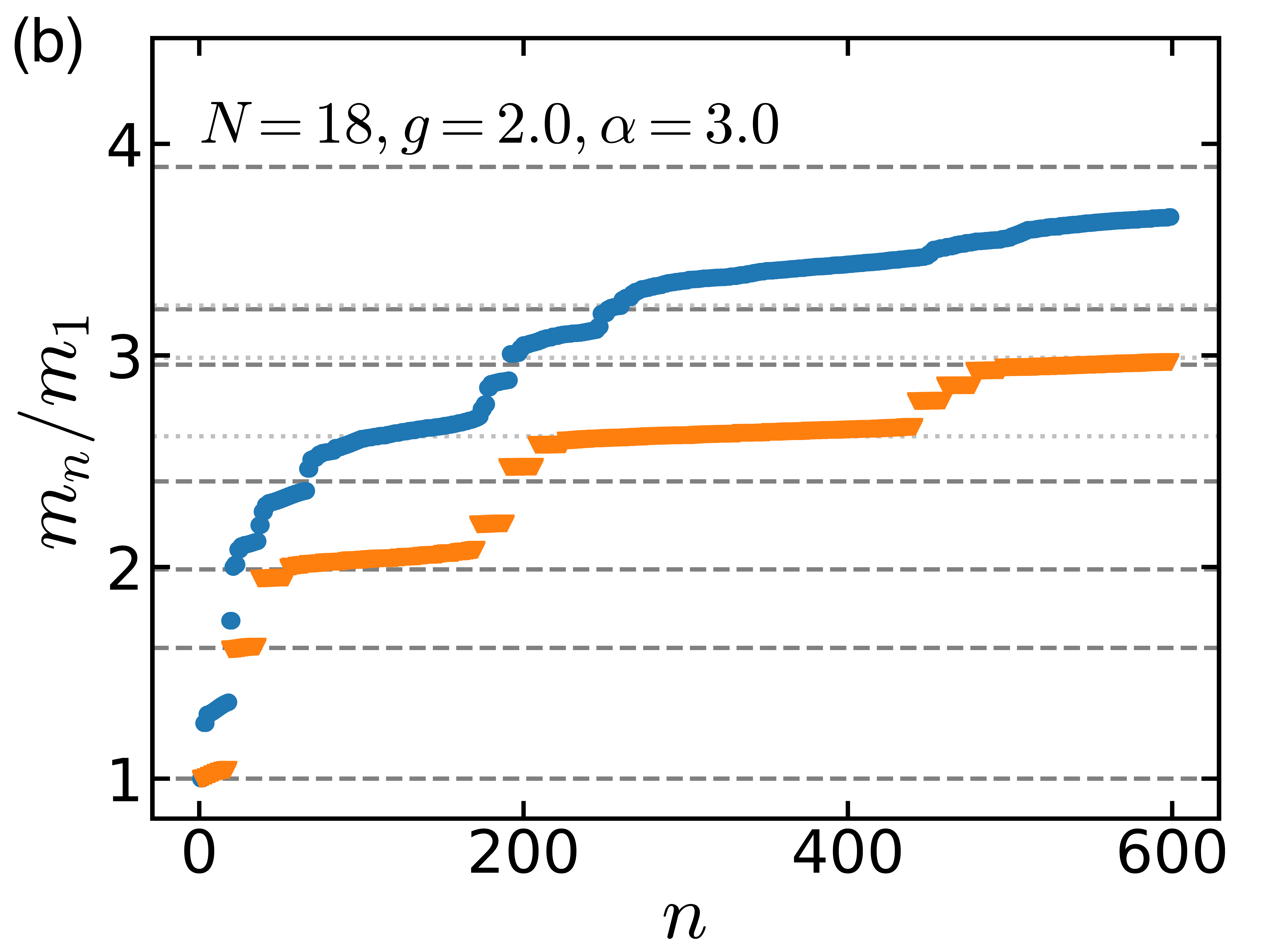} 
 \includegraphics[width=0.49\columnwidth]{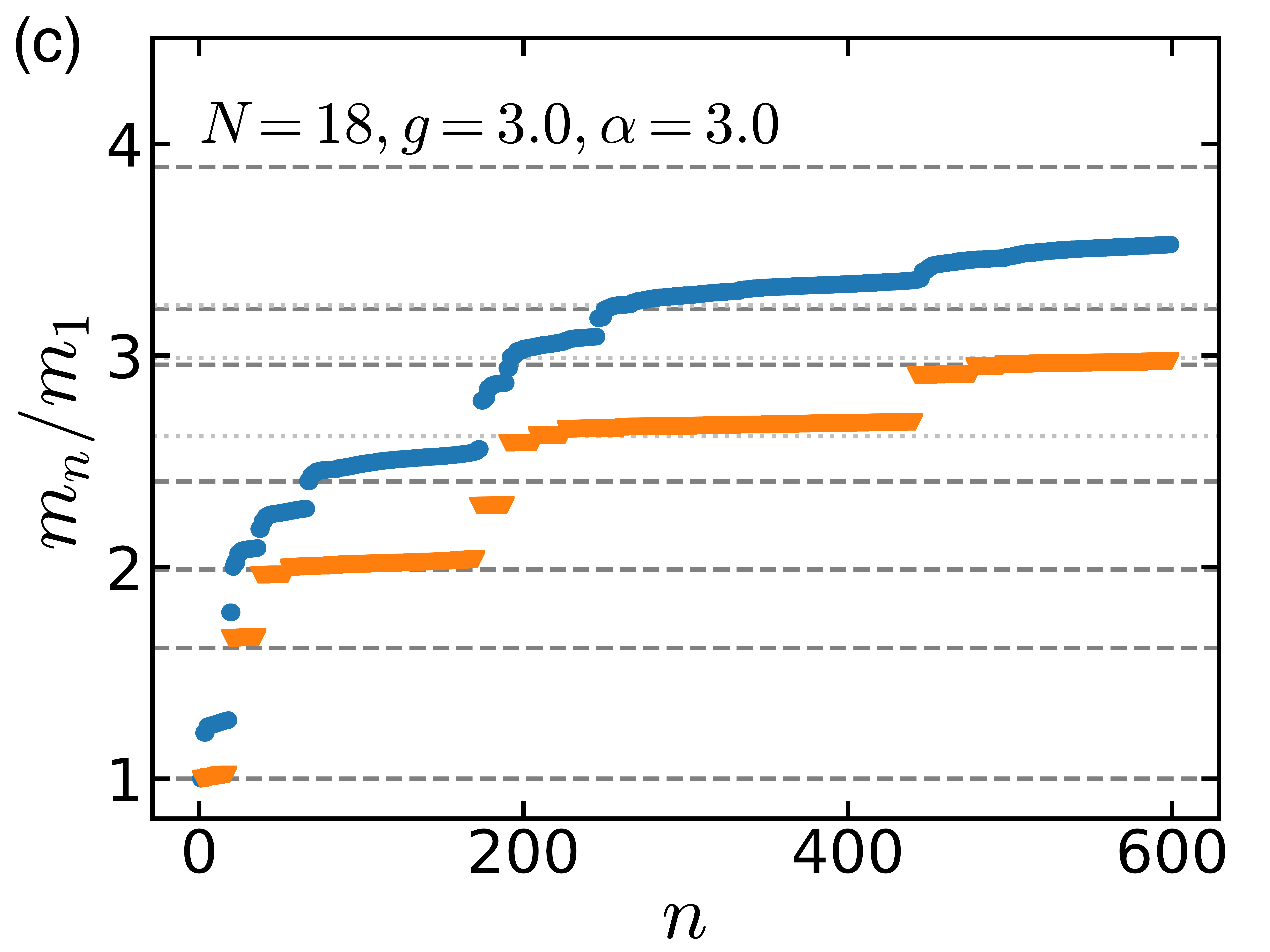} 
 \includegraphics[width=0.49\columnwidth]{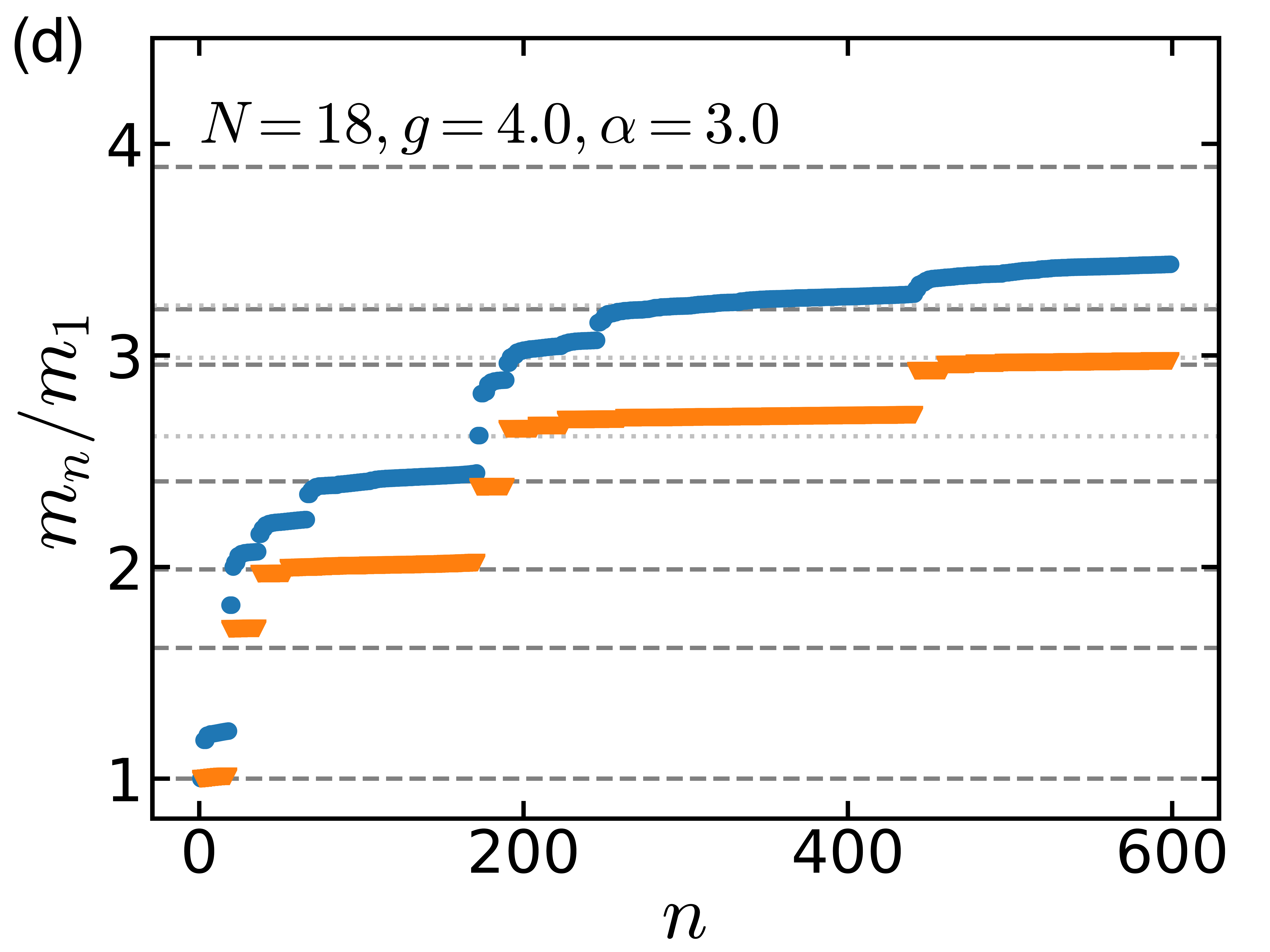} 
 \caption{(Color online) Effect of the longitudinal field value $g$ on the numerical energy spectrum (normalized mass gaps of excited states) in the LR model with obc (blue circles) and pbc (orange triangles). Background lines are as in Fig.\,\ref{fig:spectra_finitesize}.}
 \label{fig:LR_spectra_field_dependence}
\end{figure}

Figure \ref{fig:LR_spectra_field_dependence} shows the energy spectra of the LR model at $\alpha=3$ for several longitudinal field values. With increasing field strength, it becomes visible that a large semi-continuous band breaks apart into several discrete bands, which flatten out at the expected analytical E$_8$ mass ratios for pbc.

\begin{figure*}[h]
\centering
 \includegraphics[width=0.49\textwidth]{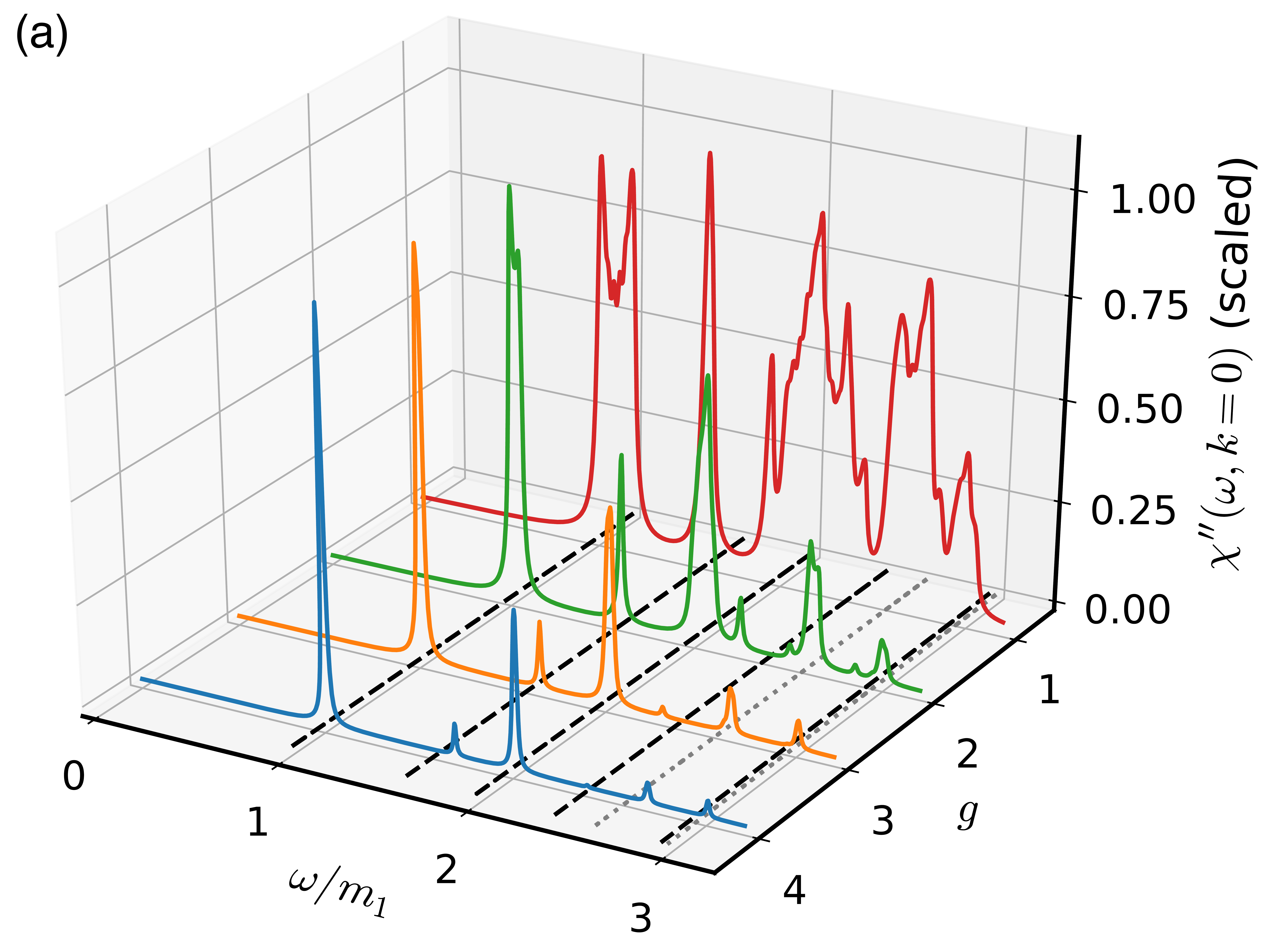} 
 \includegraphics[width=0.49\textwidth]{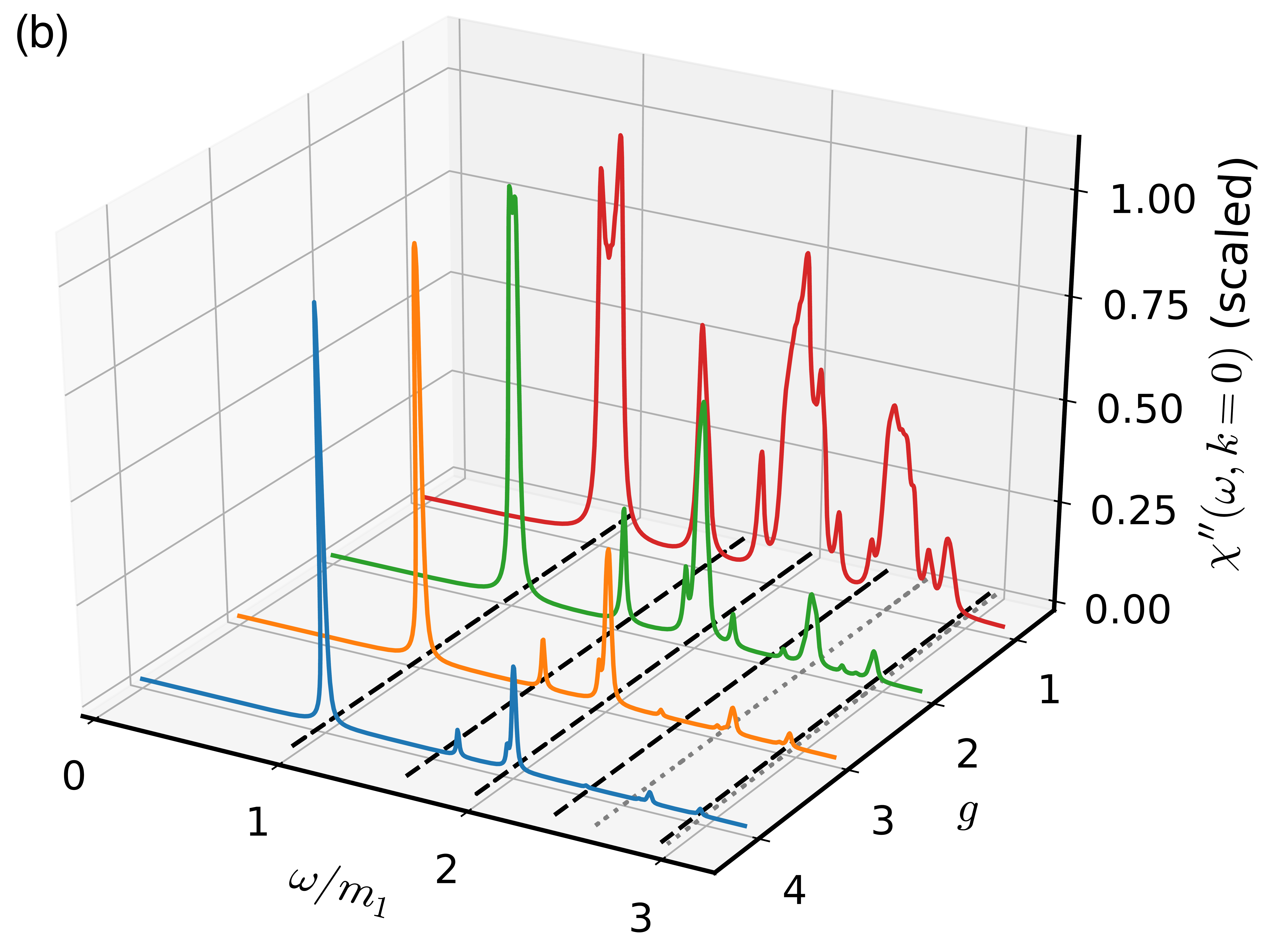} 
 \includegraphics[width=0.49\textwidth]{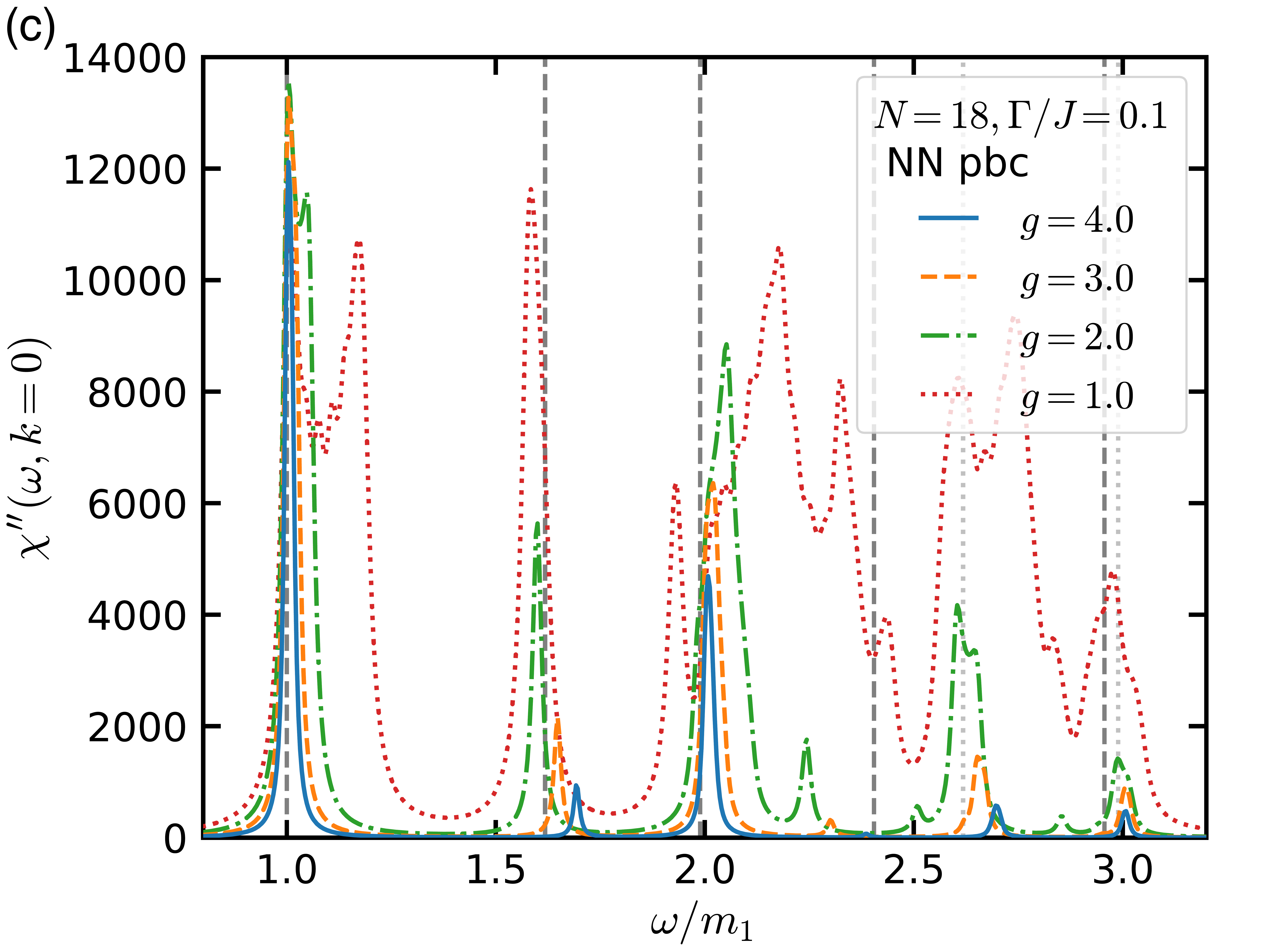} 
 \includegraphics[width=0.49\textwidth]{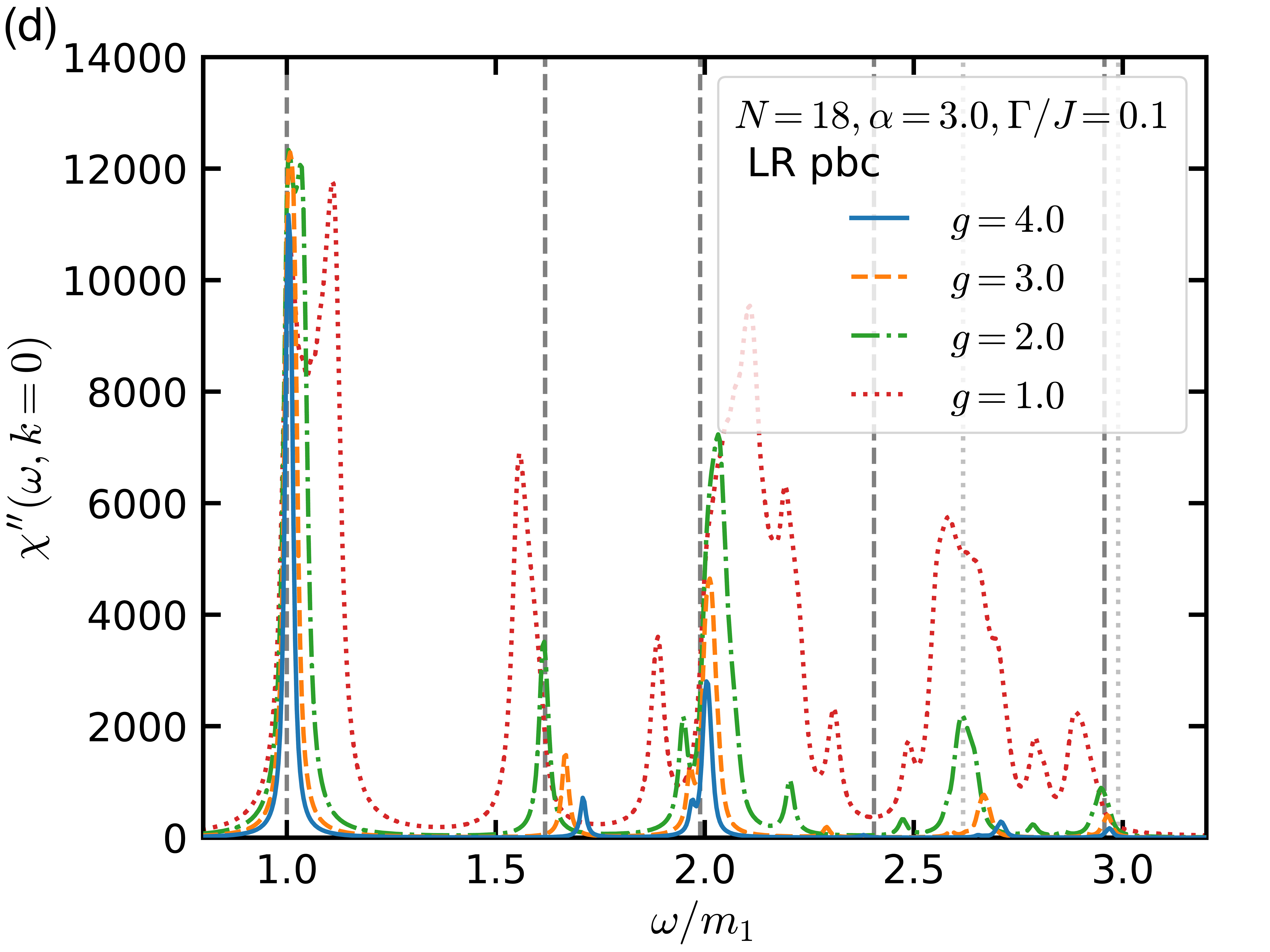} 
 \caption{(Color online) Energy absorption spectrum of the NN (a,c) and LR model (b,d) with pbc in dependence on the longitudinal field value $g$. In (a,b) the data are scaled to the maximum of the spectrum, in (c,d) in absolute units. Dashed lines represent the analytical E$_8$ meson mass ratios (cf.\ table~\ref{tab:e8_masses_ratio}), dotted lines correspond to multiparticle states with masses $M_1+M_2$ and $M_1+M_3$. Numerical parameters: $N=18$ (pbc), $\Gamma/J=0.1$, $\alpha=3$.}
 \label{fig:absorption_field3D}
\end{figure*}

The resulting energy absorption spectra are compared to the NN model in Fig.\,\ref{fig:absorption_field3D}. For a better visual presentation, we show both the scaled spectra in the top panels (a,b) as well in absolute units in the bottom panels (c,d) for a quantitative comparison.
With increasing longitudinal field, the peaks get narrower and allow for the identification of the proper meson mass ratios and sums. For the system size under consideration, $N=18$, one can infer that a longitudinal field $g \ge 2$ is necessary to capture all associated of the low-lying meson length scales in the system. The NN spectrum differs qualitatively from the LR model only at the smallest depicted field value $g=1$ (red curves). At larger values, the quantitative differences in the absolute scale are very small.

\clearpage
\bibliographystyle{jk_ref_layout_wTitle} 
\bibliography{literature}

\begin{thebibliography}{110}%
\makeatletter
\providecommand \@ifxundefined [1]{%
 \@ifx{#1\undefined}
}%
\providecommand \@ifnum [1]{%
 \ifnum #1\expandafter \@firstoftwo
 \else \expandafter \@secondoftwo
 \fi
}%
\providecommand \@ifx [1]{%
 \ifx #1\expandafter \@firstoftwo
 \else \expandafter \@secondoftwo
 \fi
}%
\providecommand \natexlab [1]{#1}%
\providecommand \emph  [1]{``#1''}%
\providecommand \bibnamefont  [1]{#1}%
\providecommand \bibfnamefont [1]{#1}%
\providecommand \citenamefont [1]{#1}%
\providecommand \href@noop [0]{\@secondoftwo}%
\providecommand \href [0]{\begingroup \@sanitize@url \@href}%
\providecommand \@href[1]{\@@startlink{#1}\@@href}%
\providecommand \@@href[1]{\endgroup#1\@@endlink}%
\providecommand \@sanitize@url [0]{\catcode `\\12\catcode `\$12\catcode
  `\&12\catcode `\#12\catcode `\^12\catcode `\_12\catcode `\%12\relax}%
\providecommand \@@startlink[1]{}%
\providecommand \@@endlink[0]{}%
\providecommand \url  [0]{\begingroup\@sanitize@url \@url }%
\providecommand \@url [1]{\endgroup\@href {#1}{\urlprefix }}%
\providecommand \urlprefix  [0]{URL }%
\providecommand \Eprint [0]{\href }%
\providecommand \doibase [0]{http://dx.doi.org/}%
\providecommand \selectlanguage [0]{\@gobble}%
\providecommand \bibinfo  [0]{\@secondoftwo}%
\providecommand \bibfield  [0]{\@secondoftwo}%
\providecommand \translation [1]{[#1]}%
\providecommand \BibitemOpen [0]{}%
\providecommand \bibitemStop [0]{}%
\providecommand \bibitemNoStop [0]{.\EOS\space}%
\providecommand \EOS [0]{\spacefactor3000\relax}%
\providecommand \BibitemShut  [1]{\csname bibitem#1\endcsname}%
\let\auto@bib@innerbib\@empty
\bibitem [{\citenamefont {Anderson}(1972)}]{Anderson:1972pca}%
  \BibitemOpen
  \bibfield  {author} {\bibinfo {author} {\bibfnamefont {P.~W.}\ \bibnamefont
  {Anderson}},\ }\bibfield  {title} {\emph {\bibinfo {title} {{More Is
  Different}},}\ }\href {\doibase 10.1126/science.177.4047.393} {\bibfield
  {journal} {\bibinfo  {journal} {Science}\ }\textbf {\bibinfo {volume}
  {177}},\ \bibinfo {pages} {393} (\bibinfo {year} {1972})}
\bibitem [{\citenamefont {Coleman}(2015)}]{coleman_2015}%
  \BibitemOpen
  \bibfield  {author} {\bibinfo {author} {\bibfnamefont {P.}~\bibnamefont
  {Coleman}},\ }\href {\doibase 10.1017/CBO9781139020916} {\emph {\bibinfo
  {title} {Introduction to Many-Body Physics}}}\ (\bibinfo  {publisher}
  {Cambridge University Press},\ \bibinfo {year} {2015})
\bibitem [{\citenamefont {Witten}(2018)}]{Witten:2017hdv}%
  \BibitemOpen
  \bibfield  {author} {\bibinfo {author} {\bibfnamefont {E.}~\bibnamefont
  {Witten}},\ }\bibfield  {title} {\emph {\bibinfo {title} {{Symmetry and
  Emergence}},}\ }\href {\doibase 10.1038/nphys4348} {\bibfield  {journal}
  {\bibinfo  {journal} {Nature Phys.}\ }\textbf {\bibinfo {volume} {14}},\
  \bibinfo {pages} {116} (\bibinfo {year} {2018})},\ \Eprint
  {http://arxiv.org/abs/1710.01791} {arXiv:1710.01791 [hep-th]}
\bibitem [{\citenamefont {Hauke}\ \emph {et~al.}(2012)\citenamefont {Hauke},
  \citenamefont {Cucchietti}, \citenamefont {Tagliacozzo}, \citenamefont
  {Deutsch},\ and\ \citenamefont {Lewenstein}}]{Hauke2011d}%
  \BibitemOpen
  \bibfield  {author} {\bibinfo {author} {\bibfnamefont {P.}~\bibnamefont
  {Hauke}}, \bibinfo {author} {\bibfnamefont {F.~M.}\ \bibnamefont
  {Cucchietti}}, \bibinfo {author} {\bibfnamefont {L.}~\bibnamefont
  {Tagliacozzo}}, \bibinfo {author} {\bibfnamefont {I.}~\bibnamefont
  {Deutsch}}, \ and\ \bibinfo {author} {\bibfnamefont {M.}~\bibnamefont
  {Lewenstein}},\ }\bibfield  {title} {\emph {\bibinfo {title} {Can one trust
  quantum simulators?}}\ }\href@noop {} {\bibfield  {journal} {\bibinfo
  {journal} {Rep. Prog. Phys.}\ }\textbf {\bibinfo {volume} {75}},\ \bibinfo
  {pages} {082401} (\bibinfo {year} {2012})}
\bibitem [{\citenamefont {Cirac}\ and\ \citenamefont
  {Zoller}(2012)}]{Cirac2012}%
  \BibitemOpen
  \bibfield  {author} {\bibinfo {author} {\bibfnamefont {J.~I.}\ \bibnamefont
  {Cirac}}\ and\ \bibinfo {author} {\bibfnamefont {P.}~\bibnamefont {Zoller}},\
  }\bibfield  {title} {\emph {\bibinfo {title} {Goals and opportunities in
  quantum simulation},}\ }\href {\doibase doi:10.1038/nphys2275} {\bibfield
  {journal} {\bibinfo  {journal} {Nat. Phys.}\ }\textbf {\bibinfo {volume}
  {8}},\ \bibinfo {pages} {264} (\bibinfo {year} {2012})}
\bibitem [{\citenamefont {Alexeev}\ \emph {et~al.}(2021)\citenamefont {Alexeev}
  \emph {et~al.}}]{Alexeev:2020xrq}%
  \BibitemOpen
  \bibfield  {author} {\bibinfo {author} {\bibfnamefont {Y.}~\bibnamefont
  {Alexeev}} \emph {et~al.},\ }\bibfield  {title} {\emph {\bibinfo {title}
  {{Quantum Computer Systems for Scientific Discovery}},}\ }\href {\doibase
  10.1103/PRXQuantum.2.017001} {\bibfield  {journal} {\bibinfo  {journal} {P.
  R. X. Quantum.}\ }\textbf {\bibinfo {volume} {2}},\ \bibinfo {pages} {017001}
  (\bibinfo {year} {2021})},\ \Eprint {http://arxiv.org/abs/1912.07577}
  {arXiv:1912.07577 [quant-ph]}
\bibitem [{\citenamefont {Schneider}, \citenamefont {Porras},\ and\
  \citenamefont {Schaetz}(2011)}]{Schneider2011}%
  \BibitemOpen
  \bibfield  {author} {\bibinfo {author} {\bibfnamefont {C.}~\bibnamefont
  {Schneider}}, \bibinfo {author} {\bibfnamefont {D.}~\bibnamefont {Porras}}, \
  and\ \bibinfo {author} {\bibfnamefont {T.}~\bibnamefont {Schaetz}},\
  }\bibfield  {title} {\emph {\bibinfo {title} {{Many-Body Physics with Trapped
  Ions}},}\ }\href@noop {} {\  (\bibinfo {year} {2011})},\ \Eprint
  {http://arxiv.org/abs/1106.2597} {arXiv:1106.2597 [quant-ph]}
\bibitem [{\citenamefont {Blatt}\ and\ \citenamefont {Roos}(2012)}]{Blatt2012}%
  \BibitemOpen
  \bibfield  {author} {\bibinfo {author} {\bibfnamefont {R.}~\bibnamefont
  {Blatt}}\ and\ \bibinfo {author} {\bibfnamefont {C.~F.}\ \bibnamefont
  {Roos}},\ }\bibfield  {title} {\emph {\bibinfo {title} {Quantum simulations
  with trapped ions},}\ }\href {\doibase doi:10.1038/nphys2252} {\bibfield
  {journal} {\bibinfo  {journal} {Nat. Phys.}\ }\textbf {\bibinfo {volume}
  {8}},\ \bibinfo {pages} {277} (\bibinfo {year} {2012})}
\bibitem [{\citenamefont {Monroe}\ \emph
  {et~al.}(2021{\natexlab{a}})\citenamefont {Monroe}, \citenamefont {Campbell},
  \citenamefont {Duan}, \citenamefont {Gong}, \citenamefont {Gorshkov},
  \citenamefont {Hess}, \citenamefont {Islam}, \citenamefont {Kim},
  \citenamefont {Linke}, \citenamefont {Pagano}, \citenamefont {Richerme},
  \citenamefont {Senko},\ and\ \citenamefont {Yao}}]{Monroe2021}%
  \BibitemOpen
  \bibfield  {author} {\bibinfo {author} {\bibfnamefont {C.}~\bibnamefont
  {Monroe}}, \bibinfo {author} {\bibfnamefont {W.~C.}\ \bibnamefont
  {Campbell}}, \bibinfo {author} {\bibfnamefont {L.-M.}\ \bibnamefont {Duan}},
  \bibinfo {author} {\bibfnamefont {Z.-X.}\ \bibnamefont {Gong}}, \bibinfo
  {author} {\bibfnamefont {A.~V.}\ \bibnamefont {Gorshkov}}, \bibinfo {author}
  {\bibfnamefont {P.~W.}\ \bibnamefont {Hess}}, \bibinfo {author}
  {\bibfnamefont {R.}~\bibnamefont {Islam}}, \bibinfo {author} {\bibfnamefont
  {K.}~\bibnamefont {Kim}}, \bibinfo {author} {\bibfnamefont {N.~M.}\
  \bibnamefont {Linke}}, \bibinfo {author} {\bibfnamefont {G.}~\bibnamefont
  {Pagano}}, \bibinfo {author} {\bibfnamefont {P.}~\bibnamefont {Richerme}},
  \bibinfo {author} {\bibfnamefont {C.}~\bibnamefont {Senko}}, \ and\ \bibinfo
  {author} {\bibfnamefont {N.~Y.}\ \bibnamefont {Yao}},\ }\bibfield  {title}
  {\emph {\bibinfo {title} {Programmable quantum simulations of spin systems
  with trapped ions},}\ }\href {\doibase 10.1103/RevModPhys.93.025001}
  {\bibfield  {journal} {\bibinfo  {journal} {Rev. Mod. Phys.}\ }\textbf
  {\bibinfo {volume} {93}},\ \bibinfo {pages} {025001} (\bibinfo {year}
  {2021}{\natexlab{a}})}
\bibitem [{\citenamefont {Pogorelov}\ \emph {et~al.}(2021)\citenamefont
  {Pogorelov}, \citenamefont {Feldker}, \citenamefont {Marciniak},
  \citenamefont {Postler}, \citenamefont {Jacob}, \citenamefont
  {Krieglsteiner}, \citenamefont {Podlesnic}, \citenamefont {Meth},
  \citenamefont {Negnevitsky}, \citenamefont {Stadler}, \citenamefont
  {H\"ofer}, \citenamefont {W\"achter}, \citenamefont {Lakhmanskiy},
  \citenamefont {Blatt}, \citenamefont {Schindler},\ and\ \citenamefont
  {Monz}}]{Blatt:2021_Demonstrator}%
  \BibitemOpen
  \bibfield  {author} {\bibinfo {author} {\bibfnamefont {I.}~\bibnamefont
  {Pogorelov}}, \bibinfo {author} {\bibfnamefont {T.}~\bibnamefont {Feldker}},
  \bibinfo {author} {\bibfnamefont {C.~D.}\ \bibnamefont {Marciniak}}, \bibinfo
  {author} {\bibfnamefont {L.}~\bibnamefont {Postler}}, \bibinfo {author}
  {\bibfnamefont {G.}~\bibnamefont {Jacob}}, \bibinfo {author} {\bibfnamefont
  {O.}~\bibnamefont {Krieglsteiner}}, \bibinfo {author} {\bibfnamefont
  {V.}~\bibnamefont {Podlesnic}}, \bibinfo {author} {\bibfnamefont
  {M.}~\bibnamefont {Meth}}, \bibinfo {author} {\bibfnamefont {V.}~\bibnamefont
  {Negnevitsky}}, \bibinfo {author} {\bibfnamefont {M.}~\bibnamefont
  {Stadler}}, \bibinfo {author} {\bibfnamefont {B.}~\bibnamefont {H\"ofer}},
  \bibinfo {author} {\bibfnamefont {C.}~\bibnamefont {W\"achter}}, \bibinfo
  {author} {\bibfnamefont {K.}~\bibnamefont {Lakhmanskiy}}, \bibinfo {author}
  {\bibfnamefont {R.}~\bibnamefont {Blatt}}, \bibinfo {author} {\bibfnamefont
  {P.}~\bibnamefont {Schindler}}, \ and\ \bibinfo {author} {\bibfnamefont
  {T.}~\bibnamefont {Monz}},\ }\bibfield  {title} {\emph {\bibinfo {title}
  {Compact ion-trap quantum computing demonstrator},}\ }\href {\doibase
  10.1103/PRXQuantum.2.020343} {\bibfield  {journal} {\bibinfo  {journal} {PRX
  Quantum}\ }\textbf {\bibinfo {volume} {2}},\ \bibinfo {pages} {020343}
  (\bibinfo {year} {2021})}
\bibitem [{\citenamefont {Friedenauer}\ \emph {et~al.}(2008)\citenamefont
  {Friedenauer}, \citenamefont {Schmitz}, \citenamefont {Glueckert},
  \citenamefont {Porras},\ and\ \citenamefont {Schaetz}}]{Friedenauer2008}%
  \BibitemOpen
  \bibfield  {author} {\bibinfo {author} {\bibfnamefont {A.}~\bibnamefont
  {Friedenauer}}, \bibinfo {author} {\bibfnamefont {H.}~\bibnamefont
  {Schmitz}}, \bibinfo {author} {\bibfnamefont {J.~T.}\ \bibnamefont
  {Glueckert}}, \bibinfo {author} {\bibfnamefont {D.}~\bibnamefont {Porras}}, \
  and\ \bibinfo {author} {\bibfnamefont {T.}~\bibnamefont {Schaetz}},\
  }\bibfield  {title} {\emph {\bibinfo {title} {Simulating a quantum magnet
  with trapped ions.}}\ }\href@noop {} {\bibfield  {journal} {\bibinfo
  {journal} {Nat. Phys.}\ }\textbf {\bibinfo {volume} {4}},\ \bibinfo {pages}
  {757} (\bibinfo {year} {2008})}
\bibitem [{\citenamefont {Britton}\ \emph {et~al.}(2012)\citenamefont
  {Britton}, \citenamefont {Sawyer}, \citenamefont {Keith}, \citenamefont
  {Wang}, \citenamefont {Freericks}, \citenamefont {Uys}, \citenamefont
  {Biercuk},\ and\ \citenamefont {Bollinger}}]{Britton_2012}%
  \BibitemOpen
  \bibfield  {author} {\bibinfo {author} {\bibfnamefont {J.~W.}\ \bibnamefont
  {Britton}}, \bibinfo {author} {\bibfnamefont {B.~C.}\ \bibnamefont {Sawyer}},
  \bibinfo {author} {\bibfnamefont {A.~C.}\ \bibnamefont {Keith}}, \bibinfo
  {author} {\bibfnamefont {C.-C.~J.}\ \bibnamefont {Wang}}, \bibinfo {author}
  {\bibfnamefont {J.~K.}\ \bibnamefont {Freericks}}, \bibinfo {author}
  {\bibfnamefont {H.}~\bibnamefont {Uys}}, \bibinfo {author} {\bibfnamefont
  {M.~J.}\ \bibnamefont {Biercuk}}, \ and\ \bibinfo {author} {\bibfnamefont
  {J.~J.}\ \bibnamefont {Bollinger}},\ }\bibfield  {title} {\emph {\bibinfo
  {title} {Engineered two-dimensional ising interactions in a trapped-ion
  quantum simulator with hundreds of spins},}\ }\href {\doibase
  10.1038/nature10981} {\bibfield  {journal} {\bibinfo  {journal} {Nature}\
  }\textbf {\bibinfo {volume} {484}},\ \bibinfo {pages} {489–492} (\bibinfo
  {year} {2012})}
\bibitem [{\citenamefont {Richerme}\ \emph {et~al.}(2013)\citenamefont
  {Richerme}, \citenamefont {Senko}, \citenamefont {Korenblit}, \citenamefont
  {Smith}, \citenamefont {Lee}, \citenamefont {Islam}, \citenamefont
  {Campbell},\ and\ \citenamefont {Monroe}}]{Richerme_2013}%
  \BibitemOpen
  \bibfield  {author} {\bibinfo {author} {\bibfnamefont {P.}~\bibnamefont
  {Richerme}}, \bibinfo {author} {\bibfnamefont {C.}~\bibnamefont {Senko}},
  \bibinfo {author} {\bibfnamefont {S.}~\bibnamefont {Korenblit}}, \bibinfo
  {author} {\bibfnamefont {J.}~\bibnamefont {Smith}}, \bibinfo {author}
  {\bibfnamefont {A.}~\bibnamefont {Lee}}, \bibinfo {author} {\bibfnamefont
  {R.}~\bibnamefont {Islam}}, \bibinfo {author} {\bibfnamefont {W.~C.}\
  \bibnamefont {Campbell}}, \ and\ \bibinfo {author} {\bibfnamefont
  {C.}~\bibnamefont {Monroe}},\ }\bibfield  {title} {\emph {\bibinfo {title}
  {Quantum catalysis of magnetic phase transitions in a quantum simulator},}\
  }\href {\doibase 10.1103/physrevlett.111.100506} {\bibfield  {journal}
  {\bibinfo  {journal} {Physical Review Letters}\ }\textbf {\bibinfo {volume}
  {111}} (\bibinfo {year} {2013}),\ 10.1103/physrevlett.111.100506},\ \Eprint
  {http://arxiv.org/abs/1303.6983} {arXiv:1303.6983 [quant-ph]}
\bibitem [{\citenamefont {Senko}\ \emph {et~al.}(2014)\citenamefont {Senko},
  \citenamefont {Smith}, \citenamefont {Richerme}, \citenamefont {Lee},
  \citenamefont {Campbell},\ and\ \citenamefont {Monroe}}]{Senko_2014}%
  \BibitemOpen
  \bibfield  {author} {\bibinfo {author} {\bibfnamefont {C.}~\bibnamefont
  {Senko}}, \bibinfo {author} {\bibfnamefont {J.}~\bibnamefont {Smith}},
  \bibinfo {author} {\bibfnamefont {P.}~\bibnamefont {Richerme}}, \bibinfo
  {author} {\bibfnamefont {A.}~\bibnamefont {Lee}}, \bibinfo {author}
  {\bibfnamefont {W.~C.}\ \bibnamefont {Campbell}}, \ and\ \bibinfo {author}
  {\bibfnamefont {C.}~\bibnamefont {Monroe}},\ }\bibfield  {title} {\emph
  {\bibinfo {title} {Coherent imaging spectroscopy of a quantum many-body spin
  system},}\ }\href {\doibase 10.1126/science.1251422} {\bibfield  {journal}
  {\bibinfo  {journal} {Science}\ }\textbf {\bibinfo {volume} {345}},\ \bibinfo
  {pages} {430–433} (\bibinfo {year} {2014})},\ \Eprint
  {http://arxiv.org/abs/1401.5751} {arXiv:1401.5751 [quant-ph]}
\bibitem [{\citenamefont {Jurcevic}\ \emph {et~al.}(2014)\citenamefont
  {Jurcevic}, \citenamefont {Lanyon}, \citenamefont {Hauke}, \citenamefont
  {Hempel}, \citenamefont {Zoller}, \citenamefont {Blatt},\ and\ \citenamefont
  {Roos}}]{Jurcevic_2014}%
  \BibitemOpen
  \bibfield  {author} {\bibinfo {author} {\bibfnamefont {P.}~\bibnamefont
  {Jurcevic}}, \bibinfo {author} {\bibfnamefont {B.~P.}\ \bibnamefont
  {Lanyon}}, \bibinfo {author} {\bibfnamefont {P.}~\bibnamefont {Hauke}},
  \bibinfo {author} {\bibfnamefont {C.}~\bibnamefont {Hempel}}, \bibinfo
  {author} {\bibfnamefont {P.}~\bibnamefont {Zoller}}, \bibinfo {author}
  {\bibfnamefont {R.}~\bibnamefont {Blatt}}, \ and\ \bibinfo {author}
  {\bibfnamefont {C.~F.}\ \bibnamefont {Roos}},\ }\bibfield  {title} {\emph
  {\bibinfo {title} {Quasiparticle engineering and entanglement propagation in
  a quantum many-body system},}\ }\href {\doibase 10.1038/nature13461}
  {\bibfield  {journal} {\bibinfo  {journal} {Nature}\ }\textbf {\bibinfo
  {volume} {511}},\ \bibinfo {pages} {202–205} (\bibinfo {year} {2014})}
\bibitem [{\citenamefont {Jurcevic}\ \emph {et~al.}(2015)\citenamefont
  {Jurcevic}, \citenamefont {Hauke}, \citenamefont {Maier}, \citenamefont
  {Hempel}, \citenamefont {Lanyon}, \citenamefont {Blatt},\ and\ \citenamefont
  {Roos}}]{Jurcevic_2015}%
  \BibitemOpen
  \bibfield  {author} {\bibinfo {author} {\bibfnamefont {P.}~\bibnamefont
  {Jurcevic}}, \bibinfo {author} {\bibfnamefont {P.}~\bibnamefont {Hauke}},
  \bibinfo {author} {\bibfnamefont {C.}~\bibnamefont {Maier}}, \bibinfo
  {author} {\bibfnamefont {C.}~\bibnamefont {Hempel}}, \bibinfo {author}
  {\bibfnamefont {B.}~\bibnamefont {Lanyon}}, \bibinfo {author} {\bibfnamefont
  {R.}~\bibnamefont {Blatt}}, \ and\ \bibinfo {author} {\bibfnamefont
  {C.}~\bibnamefont {Roos}},\ }\bibfield  {title} {\emph {\bibinfo {title}
  {Spectroscopy of interacting quasiparticles in trapped ions},}\ }\href
  {\doibase 10.1103/physrevlett.115.100501} {\bibfield  {journal} {\bibinfo
  {journal} {Physical Review Letters}\ }\textbf {\bibinfo {volume} {115}}
  (\bibinfo {year} {2015}),\ 10.1103/physrevlett.115.100501},\ \Eprint
  {http://arxiv.org/abs/1505.02066} {arXiv:1505.02066 [quant-ph]}
\bibitem [{\citenamefont {Smith}\ \emph {et~al.}(2016)\citenamefont {Smith},
  \citenamefont {Lee}, \citenamefont {Richerme}, \citenamefont {Neyenhuis},
  \citenamefont {Hess}, \citenamefont {Hauke}, \citenamefont {Heyl},
  \citenamefont {Huse},\ and\ \citenamefont {Monroe}}]{Smith2016}%
  \BibitemOpen
  \bibfield  {author} {\bibinfo {author} {\bibfnamefont {J.}~\bibnamefont
  {Smith}}, \bibinfo {author} {\bibfnamefont {A.}~\bibnamefont {Lee}}, \bibinfo
  {author} {\bibfnamefont {P.}~\bibnamefont {Richerme}}, \bibinfo {author}
  {\bibfnamefont {B.}~\bibnamefont {Neyenhuis}}, \bibinfo {author}
  {\bibfnamefont {P.~W.}\ \bibnamefont {Hess}}, \bibinfo {author}
  {\bibfnamefont {P.}~\bibnamefont {Hauke}}, \bibinfo {author} {\bibfnamefont
  {M.}~\bibnamefont {Heyl}}, \bibinfo {author} {\bibfnamefont {D.~A.}\
  \bibnamefont {Huse}}, \ and\ \bibinfo {author} {\bibfnamefont
  {C.}~\bibnamefont {Monroe}},\ }\bibfield  {title} {\emph {\bibinfo {title}
  {Many-body localization in a quantum simulator with programmable random
  disorder},}\ }\href {\doibase 10.1038/nphys3783} {\bibfield  {journal}
  {\bibinfo  {journal} {Nature Physics}\ }\textbf {\bibinfo {volume} {12}},\
  \bibinfo {pages} {907–911} (\bibinfo {year} {2016})}
\bibitem [{\citenamefont {G\"arttner}\ \emph {et~al.}(2017)\citenamefont
  {G\"arttner}, \citenamefont {Bohnet}, \citenamefont {Safavi-Naini},
  \citenamefont {Wall}, \citenamefont {Bollinger},\ and\ \citenamefont
  {Rey}}]{Garttner:2016mqj}%
  \BibitemOpen
  \bibfield  {author} {\bibinfo {author} {\bibfnamefont {M.}~\bibnamefont
  {G\"arttner}}, \bibinfo {author} {\bibfnamefont {J.~G.}\ \bibnamefont
  {Bohnet}}, \bibinfo {author} {\bibfnamefont {A.}~\bibnamefont
  {Safavi-Naini}}, \bibinfo {author} {\bibfnamefont {M.~L.}\ \bibnamefont
  {Wall}}, \bibinfo {author} {\bibfnamefont {J.~J.}\ \bibnamefont {Bollinger}},
  \ and\ \bibinfo {author} {\bibfnamefont {A.~M.}\ \bibnamefont {Rey}},\
  }\bibfield  {title} {\emph {\bibinfo {title} {{Measuring out-of-time-order
  correlations and multiple quantum spectra in a trapped ion quantum
  magnet}},}\ }\href {\doibase 10.1038/nphys4119} {\bibfield  {journal}
  {\bibinfo  {journal} {Nature Phys.}\ }\textbf {\bibinfo {volume} {13}},\
  \bibinfo {pages} {781} (\bibinfo {year} {2017})},\ \Eprint
  {http://arxiv.org/abs/1608.08938} {arXiv:1608.08938 [quant-ph]}
\bibitem [{\citenamefont {Jurcevic}\ \emph {et~al.}(2017)\citenamefont
  {Jurcevic}, \citenamefont {Shen}, \citenamefont {Hauke}, \citenamefont
  {Maier}, \citenamefont {Brydges}, \citenamefont {Hempel}, \citenamefont
  {Lanyon}, \citenamefont {Heyl}, \citenamefont {Blatt},\ and\ \citenamefont
  {Roos}}]{Jurcevic_2017}%
  \BibitemOpen
  \bibfield  {author} {\bibinfo {author} {\bibfnamefont {P.}~\bibnamefont
  {Jurcevic}}, \bibinfo {author} {\bibfnamefont {H.}~\bibnamefont {Shen}},
  \bibinfo {author} {\bibfnamefont {P.}~\bibnamefont {Hauke}}, \bibinfo
  {author} {\bibfnamefont {C.}~\bibnamefont {Maier}}, \bibinfo {author}
  {\bibfnamefont {T.}~\bibnamefont {Brydges}}, \bibinfo {author} {\bibfnamefont
  {C.}~\bibnamefont {Hempel}}, \bibinfo {author} {\bibfnamefont
  {B.}~\bibnamefont {Lanyon}}, \bibinfo {author} {\bibfnamefont
  {M.}~\bibnamefont {Heyl}}, \bibinfo {author} {\bibfnamefont {R.}~\bibnamefont
  {Blatt}}, \ and\ \bibinfo {author} {\bibfnamefont {C.}~\bibnamefont {Roos}},\
  }\bibfield  {title} {\emph {\bibinfo {title} {Direct observation of dynamical
  quantum phase transitions in an interacting many-body system},}\ }\href
  {\doibase 10.1103/physrevlett.119.080501} {\bibfield  {journal} {\bibinfo
  {journal} {Physical Review Letters}\ }\textbf {\bibinfo {volume} {119}}
  (\bibinfo {year} {2017}),\ 10.1103/physrevlett.119.080501}
\bibitem [{\citenamefont {Zhang}\ \emph {et~al.}(2017)\citenamefont {Zhang},
  \citenamefont {Pagano}, \citenamefont {Hess}, \citenamefont {Kyprianidis},
  \citenamefont {Becker}, \citenamefont {Kaplan}, \citenamefont {Gorshkov},
  \citenamefont {Gong},\ and\ \citenamefont {Monroe}}]{Zhang_2017}%
  \BibitemOpen
  \bibfield  {author} {\bibinfo {author} {\bibfnamefont {J.}~\bibnamefont
  {Zhang}}, \bibinfo {author} {\bibfnamefont {G.}~\bibnamefont {Pagano}},
  \bibinfo {author} {\bibfnamefont {P.~W.}\ \bibnamefont {Hess}}, \bibinfo
  {author} {\bibfnamefont {A.}~\bibnamefont {Kyprianidis}}, \bibinfo {author}
  {\bibfnamefont {P.}~\bibnamefont {Becker}}, \bibinfo {author} {\bibfnamefont
  {H.}~\bibnamefont {Kaplan}}, \bibinfo {author} {\bibfnamefont {A.~V.}\
  \bibnamefont {Gorshkov}}, \bibinfo {author} {\bibfnamefont {Z.-X.}\
  \bibnamefont {Gong}}, \ and\ \bibinfo {author} {\bibfnamefont
  {C.}~\bibnamefont {Monroe}},\ }\bibfield  {title} {\emph {\bibinfo {title}
  {Observation of a many-body dynamical phase transition with a 53-qubit
  quantum simulator},}\ }\href {\doibase 10.1038/nature24654} {\bibfield
  {journal} {\bibinfo  {journal} {Nature}\ }\textbf {\bibinfo {volume} {551}},\
  \bibinfo {pages} {601–604} (\bibinfo {year} {2017})}
\bibitem [{\citenamefont {Hempel}\ \emph {et~al.}(2018)\citenamefont {Hempel},
  \citenamefont {Maier}, \citenamefont {Romero}, \citenamefont {McClean},
  \citenamefont {Monz}, \citenamefont {Shen}, \citenamefont {Jurcevic},
  \citenamefont {Lanyon}, \citenamefont {Love}, \citenamefont {Babbush},\ and\
  \citenamefont {et~al.}}]{Hempel_2018}%
  \BibitemOpen
  \bibfield  {author} {\bibinfo {author} {\bibfnamefont {C.}~\bibnamefont
  {Hempel}}, \bibinfo {author} {\bibfnamefont {C.}~\bibnamefont {Maier}},
  \bibinfo {author} {\bibfnamefont {J.}~\bibnamefont {Romero}}, \bibinfo
  {author} {\bibfnamefont {J.}~\bibnamefont {McClean}}, \bibinfo {author}
  {\bibfnamefont {T.}~\bibnamefont {Monz}}, \bibinfo {author} {\bibfnamefont
  {H.}~\bibnamefont {Shen}}, \bibinfo {author} {\bibfnamefont {P.}~\bibnamefont
  {Jurcevic}}, \bibinfo {author} {\bibfnamefont {B.~P.}\ \bibnamefont
  {Lanyon}}, \bibinfo {author} {\bibfnamefont {P.}~\bibnamefont {Love}},
  \bibinfo {author} {\bibfnamefont {R.}~\bibnamefont {Babbush}}, \ and\
  \bibinfo {author} {\bibnamefont {et~al.}},\ }\bibfield  {title} {\emph
  {\bibinfo {title} {Quantum chemistry calculations on a trapped-ion quantum
  simulator},}\ }\href {\doibase 10.1103/physrevx.8.031022} {\bibfield
  {journal} {\bibinfo  {journal} {Physical Review X}\ }\textbf {\bibinfo
  {volume} {8}} (\bibinfo {year} {2018}),\ 10.1103/physrevx.8.031022}
\bibitem [{\citenamefont {Brydges}\ \emph {et~al.}(2019)\citenamefont
  {Brydges}, \citenamefont {Elben}, \citenamefont {Jurcevic}, \citenamefont
  {Vermersch}, \citenamefont {Maier}, \citenamefont {Lanyon}, \citenamefont
  {Zoller}, \citenamefont {Blatt},\ and\ \citenamefont {Roos}}]{Brydges_2019}%
  \BibitemOpen
  \bibfield  {author} {\bibinfo {author} {\bibfnamefont {T.}~\bibnamefont
  {Brydges}}, \bibinfo {author} {\bibfnamefont {A.}~\bibnamefont {Elben}},
  \bibinfo {author} {\bibfnamefont {P.}~\bibnamefont {Jurcevic}}, \bibinfo
  {author} {\bibfnamefont {B.}~\bibnamefont {Vermersch}}, \bibinfo {author}
  {\bibfnamefont {C.}~\bibnamefont {Maier}}, \bibinfo {author} {\bibfnamefont
  {B.~P.}\ \bibnamefont {Lanyon}}, \bibinfo {author} {\bibfnamefont
  {P.}~\bibnamefont {Zoller}}, \bibinfo {author} {\bibfnamefont
  {R.}~\bibnamefont {Blatt}}, \ and\ \bibinfo {author} {\bibfnamefont {C.~F.}\
  \bibnamefont {Roos}},\ }\bibfield  {title} {\emph {\bibinfo {title} {{Probing
  R\'enyi entanglement entropy via randomized measurements}},}\ }\href
  {\doibase 10.1126/science.aau4963} {\bibfield  {journal} {\bibinfo  {journal}
  {Science}\ }\textbf {\bibinfo {volume} {364}},\ \bibinfo {pages} {260–263}
  (\bibinfo {year} {2019})}
\bibitem [{\citenamefont {Maier}\ \emph {et~al.}(2019)\citenamefont {Maier},
  \citenamefont {Brydges}, \citenamefont {Jurcevic}, \citenamefont {Trautmann},
  \citenamefont {Hempel}, \citenamefont {Lanyon}, \citenamefont {Hauke},
  \citenamefont {Blatt},\ and\ \citenamefont {Roos}}]{Maier2019}%
  \BibitemOpen
  \bibfield  {author} {\bibinfo {author} {\bibfnamefont {C.}~\bibnamefont
  {Maier}}, \bibinfo {author} {\bibfnamefont {T.}~\bibnamefont {Brydges}},
  \bibinfo {author} {\bibfnamefont {P.}~\bibnamefont {Jurcevic}}, \bibinfo
  {author} {\bibfnamefont {N.}~\bibnamefont {Trautmann}}, \bibinfo {author}
  {\bibfnamefont {C.}~\bibnamefont {Hempel}}, \bibinfo {author} {\bibfnamefont
  {B.~P.}\ \bibnamefont {Lanyon}}, \bibinfo {author} {\bibfnamefont
  {P.}~\bibnamefont {Hauke}}, \bibinfo {author} {\bibfnamefont
  {R.}~\bibnamefont {Blatt}}, \ and\ \bibinfo {author} {\bibfnamefont {C.~F.}\
  \bibnamefont {Roos}},\ }\bibfield  {title} {\emph {\bibinfo {title}
  {Environment-assisted quantum transport in a 10-qubit network},}\ }\href
  {\doibase 10.1103/physrevlett.122.050501} {\bibfield  {journal} {\bibinfo
  {journal} {Physical Review Letters}\ }\textbf {\bibinfo {volume} {122}}
  (\bibinfo {year} {2019}),\ 10.1103/physrevlett.122.050501}
\bibitem [{\citenamefont {Martinez}\ \emph {et~al.}(2016)\citenamefont
  {Martinez} \emph {et~al.}}]{Martinez:2016yna}%
  \BibitemOpen
  \bibfield  {author} {\bibinfo {author} {\bibfnamefont {E.~A.}\ \bibnamefont
  {Martinez}} \emph {et~al.},\ }\bibfield  {title} {\emph {\bibinfo {title}
  {{Real-time dynamics of lattice gauge theories with a few-qubit quantum
  computer}},}\ }\href {\doibase 10.1038/nature18318} {\bibfield  {journal}
  {\bibinfo  {journal} {Nature}\ }\textbf {\bibinfo {volume} {534}},\ \bibinfo
  {pages} {516} (\bibinfo {year} {2016})},\ \Eprint
  {http://arxiv.org/abs/1605.04570} {arXiv:1605.04570 [quant-ph]}
\bibitem [{\citenamefont {Kokail}\ \emph {et~al.}(2019)\citenamefont {Kokail}
  \emph {et~al.}}]{Kokail:2018eiw}%
  \BibitemOpen
  \bibfield  {author} {\bibinfo {author} {\bibfnamefont {C.}~\bibnamefont
  {Kokail}} \emph {et~al.},\ }\bibfield  {title} {\emph {\bibinfo {title}
  {{Self-verifying variational quantum simulation of lattice models}},}\ }\href
  {\doibase 10.1038/s41586-019-1177-4} {\bibfield  {journal} {\bibinfo
  {journal} {Nature}\ }\textbf {\bibinfo {volume} {569}},\ \bibinfo {pages}
  {355} (\bibinfo {year} {2019})},\ \Eprint {http://arxiv.org/abs/1810.03421}
  {arXiv:1810.03421 [quant-ph]}
\bibitem [{\citenamefont {Friman}\ \emph {et~al.}(2011)\citenamefont {Friman},
  \citenamefont {Hohne}, \citenamefont {Knoll}, \citenamefont {Leupold},
  \citenamefont {Randrup}, \citenamefont {Rapp},\ and\ \citenamefont
  {Senger}}]{Friman:2011zz}%
  \BibitemOpen
  \bibfield  {author} {\bibinfo {author} {\bibfnamefont {B.}~\bibnamefont
  {Friman}}, \bibinfo {author} {\bibfnamefont {C.}~\bibnamefont {Hohne}},
  \bibinfo {author} {\bibfnamefont {J.}~\bibnamefont {Knoll}}, \bibinfo
  {author} {\bibfnamefont {S.}~\bibnamefont {Leupold}}, \bibinfo {author}
  {\bibfnamefont {J.}~\bibnamefont {Randrup}}, \bibinfo {author} {\bibfnamefont
  {R.}~\bibnamefont {Rapp}}, \ and\ \bibinfo {author} {\bibfnamefont
  {P.}~\bibnamefont {Senger}},\ }\bibfield  {title} {\emph {\bibinfo {title}
  {{The CBM Physics Book: Compressed Baryonic Matter in Laboratory
  Experiments}},}\ }\href {\doibase 10.1007/978-3-642-13293-3} {\bibfield
  {journal} {\bibinfo  {journal} {Lect. Notes Phys.}\ }\textbf {\bibinfo
  {volume} {814}},\ \bibinfo {pages} {1} (\bibinfo {year} {2011})}
\bibitem [{\citenamefont {McCoy}\ and\ \citenamefont
  {Wu}(1978)}]{McCoy:1978ta}%
  \BibitemOpen
  \bibfield  {author} {\bibinfo {author} {\bibfnamefont {B.~M.}\ \bibnamefont
  {McCoy}}\ and\ \bibinfo {author} {\bibfnamefont {T.~T.}\ \bibnamefont {Wu}},\
  }\bibfield  {title} {\emph {\bibinfo {title} {{Two-dimensional Ising Field
  Theory in a Magnetic Field: Breakup of the Cut in the Two Point Function}},}\
  }\href {\doibase 10.1103/PhysRevD.18.1259} {\bibfield  {journal} {\bibinfo
  {journal} {Phys. Rev. D}\ }\textbf {\bibinfo {volume} {18}},\ \bibinfo
  {pages} {1259} (\bibinfo {year} {1978})}
\bibitem [{\citenamefont {Liu}\ \emph {et~al.}(2019)\citenamefont {Liu},
  \citenamefont {Lundgren}, \citenamefont {Titum}, \citenamefont {Pagano},
  \citenamefont {Zhang}, \citenamefont {Monroe},\ and\ \citenamefont
  {Gorshkov}}]{Liu:2018fza}%
  \BibitemOpen
  \bibfield  {author} {\bibinfo {author} {\bibfnamefont {F.}~\bibnamefont
  {Liu}}, \bibinfo {author} {\bibfnamefont {R.}~\bibnamefont {Lundgren}},
  \bibinfo {author} {\bibfnamefont {P.}~\bibnamefont {Titum}}, \bibinfo
  {author} {\bibfnamefont {G.}~\bibnamefont {Pagano}}, \bibinfo {author}
  {\bibfnamefont {J.}~\bibnamefont {Zhang}}, \bibinfo {author} {\bibfnamefont
  {C.}~\bibnamefont {Monroe}}, \ and\ \bibinfo {author} {\bibfnamefont {A.~V.}\
  \bibnamefont {Gorshkov}},\ }\bibfield  {title} {\emph {\bibinfo {title}
  {{Confined Quasiparticle Dynamics in Long-Range Interacting Quantum Spin
  Chains}},}\ }\href {\doibase 10.1103/PhysRevLett.122.150601} {\bibfield
  {journal} {\bibinfo  {journal} {Phys. Rev. Lett.}\ }\textbf {\bibinfo
  {volume} {122}},\ \bibinfo {pages} {150601} (\bibinfo {year} {2019})},\
  \Eprint {http://arxiv.org/abs/1810.02365} {arXiv:1810.02365
  [cond-mat.quant-gas]}
\bibitem [{\citenamefont {Lerose}\ \emph
  {et~al.}(2019{\natexlab{a}})\citenamefont {Lerose}, \citenamefont {Zunkovic},
  \citenamefont {Silva},\ and\ \citenamefont {Gambassi}}]{Lerose_2019}%
  \BibitemOpen
  \bibfield  {author} {\bibinfo {author} {\bibfnamefont {A.}~\bibnamefont
  {Lerose}}, \bibinfo {author} {\bibfnamefont {B.}~\bibnamefont {Zunkovic}},
  \bibinfo {author} {\bibfnamefont {A.}~\bibnamefont {Silva}}, \ and\ \bibinfo
  {author} {\bibfnamefont {A.}~\bibnamefont {Gambassi}},\ }\bibfield  {title}
  {\emph {\bibinfo {title} {Quasilocalized excitations induced by long-range
  interactions in translationally invariant quantum spin chains},}\ }\href
  {\doibase 10.1103/physrevb.99.121112} {\bibfield  {journal} {\bibinfo
  {journal} {Physical Review B}\ }\textbf {\bibinfo {volume} {99}} (\bibinfo
  {year} {2019}{\natexlab{a}}),\ 10.1103/physrevb.99.121112},\ \Eprint
  {http://arxiv.org/abs/1811.05513} {arXiv:1811.05513 [cond-mat.quant-gas]}
\bibitem [{\citenamefont {Greiter}(2002)}]{Greiter:2002}%
  \BibitemOpen
  \bibfield  {author} {\bibinfo {author} {\bibfnamefont {M.}~\bibnamefont
  {Greiter}},\ }\bibfield  {title} {\emph {\bibinfo {title} {Fictitious flux
  confinement: Magnetic pairing in coupled spin chains or planes},}\ }\href
  {\doibase 10.1103/PhysRevB.66.054505} {\bibfield  {journal} {\bibinfo
  {journal} {Phys. Rev. B}\ }\textbf {\bibinfo {volume} {66}},\ \bibinfo
  {pages} {054505} (\bibinfo {year} {2002})}
\bibitem [{\citenamefont {Lake}\ \emph {et~al.}(2009)\citenamefont {Lake},
  \citenamefont {Tsvelik}, \citenamefont {Notbohm}, \citenamefont
  {Alan~Tennant}, \citenamefont {Perring}, \citenamefont {Reehuis},
  \citenamefont {Sekar}, \citenamefont {Krabbes},\ and\ \citenamefont
  {B\"uchner}}]{Lake_2009}%
  \BibitemOpen
  \bibfield  {author} {\bibinfo {author} {\bibfnamefont {B.}~\bibnamefont
  {Lake}}, \bibinfo {author} {\bibfnamefont {A.~M.}\ \bibnamefont {Tsvelik}},
  \bibinfo {author} {\bibfnamefont {S.}~\bibnamefont {Notbohm}}, \bibinfo
  {author} {\bibfnamefont {D.}~\bibnamefont {Alan~Tennant}}, \bibinfo {author}
  {\bibfnamefont {T.~G.}\ \bibnamefont {Perring}}, \bibinfo {author}
  {\bibfnamefont {M.}~\bibnamefont {Reehuis}}, \bibinfo {author} {\bibfnamefont
  {C.}~\bibnamefont {Sekar}}, \bibinfo {author} {\bibfnamefont
  {G.}~\bibnamefont {Krabbes}}, \ and\ \bibinfo {author} {\bibfnamefont
  {B.}~\bibnamefont {B\"uchner}},\ }\bibfield  {title} {\emph {\bibinfo {title}
  {Confinement of fractional quantum number particles in a condensed-matter
  system},}\ }\href {\doibase 10.1038/nphys1462} {\bibfield  {journal}
  {\bibinfo  {journal} {Nature Physics}\ }\textbf {\bibinfo {volume} {6}},\
  \bibinfo {pages} {50–55} (\bibinfo {year} {2009})}
\bibitem [{\citenamefont {Kormos}\ \emph {et~al.}(2017)\citenamefont {Kormos},
  \citenamefont {Collura}, \citenamefont {Takacs},\ and\ \citenamefont
  {Calabrese}}]{Kormos2017}%
  \BibitemOpen
  \bibfield  {author} {\bibinfo {author} {\bibfnamefont {M.}~\bibnamefont
  {Kormos}}, \bibinfo {author} {\bibfnamefont {M.}~\bibnamefont {Collura}},
  \bibinfo {author} {\bibfnamefont {G.}~\bibnamefont {Takacs}}, \ and\ \bibinfo
  {author} {\bibfnamefont {P.}~\bibnamefont {Calabrese}},\ }\bibfield  {title}
  {\emph {\bibinfo {title} {Real-time confinement following a quantum quench to
  a non-integrable model},}\ }\href {\doibase
  https://doi.org/10.1038/nphys3934} {\bibfield  {journal} {\bibinfo  {journal}
  {Nature Phys.}\ }\textbf {\bibinfo {volume} {13}},\ \bibinfo {pages}
  {246–249} (\bibinfo {year} {2017})},\ \Eprint
  {http://arxiv.org/abs/1604.03571} {arXiv:1604.03571 [cond-mat.stat-mech]}
\bibitem [{\citenamefont {James}, \citenamefont {Konik},\ and\ \citenamefont
  {Robinson}(2019)}]{James_2019}%
  \BibitemOpen
  \bibfield  {author} {\bibinfo {author} {\bibfnamefont {A.~J.~A.}\
  \bibnamefont {James}}, \bibinfo {author} {\bibfnamefont {R.~M.}\ \bibnamefont
  {Konik}}, \ and\ \bibinfo {author} {\bibfnamefont {N.~J.}\ \bibnamefont
  {Robinson}},\ }\bibfield  {title} {\emph {\bibinfo {title} {Nonthermal states
  arising from confinement in one and two dimensions},}\ }\href {\doibase
  10.1103/PhysRevLett.122.130603} {\bibfield  {journal} {\bibinfo  {journal}
  {Phys. Rev. Lett.}\ }\textbf {\bibinfo {volume} {122}},\ \bibinfo {pages}
  {130603} (\bibinfo {year} {2019})},\ \Eprint
  {http://arxiv.org/abs/1804.09990} {arXiv:1804.09990 [cond-mat.stat-mech]}
\bibitem [{\citenamefont {Mazza}\ \emph {et~al.}(2019)\citenamefont {Mazza},
  \citenamefont {Perfetto}, \citenamefont {Lerose}, \citenamefont {Collura},\
  and\ \citenamefont {Gambassi}}]{Mazza_2019}%
  \BibitemOpen
  \bibfield  {author} {\bibinfo {author} {\bibfnamefont {P.~P.}\ \bibnamefont
  {Mazza}}, \bibinfo {author} {\bibfnamefont {G.}~\bibnamefont {Perfetto}},
  \bibinfo {author} {\bibfnamefont {A.}~\bibnamefont {Lerose}}, \bibinfo
  {author} {\bibfnamefont {M.}~\bibnamefont {Collura}}, \ and\ \bibinfo
  {author} {\bibfnamefont {A.}~\bibnamefont {Gambassi}},\ }\bibfield  {title}
  {\emph {\bibinfo {title} {Suppression of transport in nondisordered quantum
  spin chains due to confined excitations},}\ }\href {\doibase
  10.1103/physrevb.99.180302} {\bibfield  {journal} {\bibinfo  {journal}
  {Physical Review B}\ }\textbf {\bibinfo {volume} {99}} (\bibinfo {year}
  {2019}),\ 10.1103/physrevb.99.180302}
\bibitem [{\citenamefont {Robinson}, \citenamefont {James},\ and\ \citenamefont
  {Konik}(2019)}]{Robinson_2019}%
  \BibitemOpen
  \bibfield  {author} {\bibinfo {author} {\bibfnamefont {N.~J.}\ \bibnamefont
  {Robinson}}, \bibinfo {author} {\bibfnamefont {A.~J.~A.}\ \bibnamefont
  {James}}, \ and\ \bibinfo {author} {\bibfnamefont {R.~M.}\ \bibnamefont
  {Konik}},\ }\bibfield  {title} {\emph {\bibinfo {title} {Signatures of rare
  states and thermalization in a theory with confinement},}\ }\href {\doibase
  10.1103/PhysRevB.99.195108} {\bibfield  {journal} {\bibinfo  {journal} {Phys.
  Rev. B}\ }\textbf {\bibinfo {volume} {99}},\ \bibinfo {pages} {195108}
  (\bibinfo {year} {2019})},\ \Eprint {http://arxiv.org/abs/1808.10782}
  {arXiv:1808.10782 [cond-mat.stat-mech]}
\bibitem [{\citenamefont {Vanderstraeten}\ \emph {et~al.}(2020)\citenamefont
  {Vanderstraeten}, \citenamefont {Wybo}, \citenamefont {Chepiga},
  \citenamefont {Verstraete},\ and\ \citenamefont
  {Mila}}]{Vanderstraeten:2020}%
  \BibitemOpen
  \bibfield  {author} {\bibinfo {author} {\bibfnamefont {L.}~\bibnamefont
  {Vanderstraeten}}, \bibinfo {author} {\bibfnamefont {E.}~\bibnamefont
  {Wybo}}, \bibinfo {author} {\bibfnamefont {N.}~\bibnamefont {Chepiga}},
  \bibinfo {author} {\bibfnamefont {F.}~\bibnamefont {Verstraete}}, \ and\
  \bibinfo {author} {\bibfnamefont {F.}~\bibnamefont {Mila}},\ }\bibfield
  {title} {\emph {\bibinfo {title} {Spinon confinement and deconfinement in
  spin-1 chains},}\ }\href {\doibase 10.1103/PhysRevB.101.115138} {\bibfield
  {journal} {\bibinfo  {journal} {Phys. Rev. B}\ }\textbf {\bibinfo {volume}
  {101}},\ \bibinfo {pages} {115138} (\bibinfo {year} {2020})}
\bibitem [{\citenamefont {Lerose}\ \emph
  {et~al.}(2019{\natexlab{b}})\citenamefont {Lerose}, \citenamefont {Surace},
  \citenamefont {Mazza}, \citenamefont {Perfetto}, \citenamefont {Collura},\
  and\ \citenamefont {Gambassi}}]{Lerose:2019jrs}%
  \BibitemOpen
  \bibfield  {author} {\bibinfo {author} {\bibfnamefont {A.}~\bibnamefont
  {Lerose}}, \bibinfo {author} {\bibfnamefont {F.~M.}\ \bibnamefont {Surace}},
  \bibinfo {author} {\bibfnamefont {P.~P.}\ \bibnamefont {Mazza}}, \bibinfo
  {author} {\bibfnamefont {G.}~\bibnamefont {Perfetto}}, \bibinfo {author}
  {\bibfnamefont {M.}~\bibnamefont {Collura}}, \ and\ \bibinfo {author}
  {\bibfnamefont {A.}~\bibnamefont {Gambassi}},\ }\bibfield  {title} {\emph
  {\bibinfo {title} {{Quasilocalized dynamics from confinement of quantum
  excitations}},}\ }\href {\doibase 10.1103/PhysRevB.102.041118} {\bibfield
  {journal} {\bibinfo  {journal} {Phys. Rev. B}\ }\textbf {\bibinfo {volume}
  {102}},\ \bibinfo {pages} {041118} (\bibinfo {year} {2019}{\natexlab{b}})},\
  \Eprint {http://arxiv.org/abs/1911.07877} {arXiv:1911.07877
  [cond-mat.stat-mech]}
\bibitem [{\citenamefont {Banuls}\ \emph {et~al.}(2020)\citenamefont {Banuls},
  \citenamefont {Heller}, \citenamefont {Jansen}, \citenamefont {Knaute},\ and\
  \citenamefont {Svensson}}]{Banuls:2019qrq}%
  \BibitemOpen
  \bibfield  {author} {\bibinfo {author} {\bibfnamefont {M.~C.}\ \bibnamefont
  {Banuls}}, \bibinfo {author} {\bibfnamefont {M.~P.}\ \bibnamefont {Heller}},
  \bibinfo {author} {\bibfnamefont {K.}~\bibnamefont {Jansen}}, \bibinfo
  {author} {\bibfnamefont {J.}~\bibnamefont {Knaute}}, \ and\ \bibinfo {author}
  {\bibfnamefont {V.}~\bibnamefont {Svensson}},\ }\bibfield  {title} {\emph
  {\bibinfo {title} {{From spin chains to real-time thermal field theory using
  tensor networks}},}\ }\href {\doibase 10.1103/PhysRevResearch.2.033301}
  {\bibfield  {journal} {\bibinfo  {journal} {Phys. Rev. Res.}\ }\textbf
  {\bibinfo {volume} {2}},\ \bibinfo {pages} {033301} (\bibinfo {year}
  {2020})},\ \Eprint {http://arxiv.org/abs/1912.08836} {arXiv:1912.08836
  [hep-th]}
\bibitem [{\citenamefont {Castro-Alvaredo}\ \emph {et~al.}(2020)\citenamefont
  {Castro-Alvaredo}, \citenamefont {Lencses}, \citenamefont {Szecsenyi},\ and\
  \citenamefont {Viti}}]{Castro-Alvaredo:2020mzq}%
  \BibitemOpen
  \bibfield  {author} {\bibinfo {author} {\bibfnamefont {O.~A.}\ \bibnamefont
  {Castro-Alvaredo}}, \bibinfo {author} {\bibfnamefont {M.}~\bibnamefont
  {Lencses}}, \bibinfo {author} {\bibfnamefont {I.~M.}\ \bibnamefont
  {Szecsenyi}}, \ and\ \bibinfo {author} {\bibfnamefont {J.}~\bibnamefont
  {Viti}},\ }\bibfield  {title} {\emph {\bibinfo {title} {{Entanglement
  Oscillations near a Quantum Critical Point}},}\ }\href {\doibase
  10.1103/PhysRevLett.124.230601} {\bibfield  {journal} {\bibinfo  {journal}
  {Phys. Rev. Lett.}\ }\textbf {\bibinfo {volume} {124}},\ \bibinfo {pages}
  {230601} (\bibinfo {year} {2020})},\ \Eprint
  {http://arxiv.org/abs/2001.10007} {arXiv:2001.10007 [cond-mat.stat-mech]}
\bibitem [{\citenamefont {Halimeh}\ \emph {et~al.}(2020)\citenamefont
  {Halimeh}, \citenamefont {Van~Damme}, \citenamefont {Zauner-Stauber},\ and\
  \citenamefont {Vanderstraeten}}]{Halimeh_2020}%
  \BibitemOpen
  \bibfield  {author} {\bibinfo {author} {\bibfnamefont {J.~C.}\ \bibnamefont
  {Halimeh}}, \bibinfo {author} {\bibfnamefont {M.}~\bibnamefont {Van~Damme}},
  \bibinfo {author} {\bibfnamefont {V.}~\bibnamefont {Zauner-Stauber}}, \ and\
  \bibinfo {author} {\bibfnamefont {L.}~\bibnamefont {Vanderstraeten}},\
  }\bibfield  {title} {\emph {\bibinfo {title} {Quasiparticle origin of
  dynamical quantum phase transitions},}\ }\href {\doibase
  10.1103/physrevresearch.2.033111} {\bibfield  {journal} {\bibinfo  {journal}
  {Physical Review Research}\ }\textbf {\bibinfo {volume} {2}},\ \bibinfo
  {pages} {033111} (\bibinfo {year} {2020})},\ \Eprint
  {http://arxiv.org/abs/1810.07187} {arXiv:1810.07187 [cond-mat.str-el]}
\bibitem [{\citenamefont {Defenu}, \citenamefont {Enss},\ and\ \citenamefont
  {Halimeh}(2019)}]{Defenu:2019dkd}%
  \BibitemOpen
  \bibfield  {author} {\bibinfo {author} {\bibfnamefont {N.}~\bibnamefont
  {Defenu}}, \bibinfo {author} {\bibfnamefont {T.}~\bibnamefont {Enss}}, \ and\
  \bibinfo {author} {\bibfnamefont {J.~C.}\ \bibnamefont {Halimeh}},\
  }\bibfield  {title} {\emph {\bibinfo {title} {{Dynamical criticality and
  domain-wall coupling in long-range Hamiltonians}},}\ }\href {\doibase
  10.1103/PhysRevB.100.014434} {\bibfield  {journal} {\bibinfo  {journal}
  {Phys. Rev. B}\ }\textbf {\bibinfo {volume} {100}},\ \bibinfo {pages}
  {014434} (\bibinfo {year} {2019})},\ \Eprint
  {http://arxiv.org/abs/1902.08621} {arXiv:1902.08621 [cond-mat.stat-mech]}
\bibitem [{\citenamefont {Halimeh}\ \emph {et~al.}(2021)\citenamefont
  {Halimeh}, \citenamefont {Damme}, \citenamefont {Guo}, \citenamefont {Lang},\
  and\ \citenamefont {Hauke}}]{halimeh2021dynamical}%
  \BibitemOpen
  \bibfield  {author} {\bibinfo {author} {\bibfnamefont {J.~C.}\ \bibnamefont
  {Halimeh}}, \bibinfo {author} {\bibfnamefont {M.~V.}\ \bibnamefont {Damme}},
  \bibinfo {author} {\bibfnamefont {L.}~\bibnamefont {Guo}}, \bibinfo {author}
  {\bibfnamefont {J.}~\bibnamefont {Lang}}, \ and\ \bibinfo {author}
  {\bibfnamefont {P.}~\bibnamefont {Hauke}},\ }\bibfield  {title} {\emph
  {\bibinfo {title} {Dynamical phase transitions in quantum spin models with
  antiferromagnetic long-range interactions},}\ }\href@noop {} {\  (\bibinfo
  {year} {2021})},\ \Eprint {http://arxiv.org/abs/2106.05282} {arXiv:2106.05282
  [cond-mat.quant-gas]}
\bibitem [{\citenamefont {Hashizume}, \citenamefont {McCulloch},\ and\
  \citenamefont {Halimeh}(2020)}]{hashizume2020dynamical}%
  \BibitemOpen
  \bibfield  {author} {\bibinfo {author} {\bibfnamefont {T.}~\bibnamefont
  {Hashizume}}, \bibinfo {author} {\bibfnamefont {I.~P.}\ \bibnamefont
  {McCulloch}}, \ and\ \bibinfo {author} {\bibfnamefont {J.~C.}\ \bibnamefont
  {Halimeh}},\ }\bibfield  {title} {\emph {\bibinfo {title} {Dynamical phase
  transitions in the two-dimensional transverse-field ising model},}\
  }\href@noop {} {\  (\bibinfo {year} {2020})},\ \Eprint
  {http://arxiv.org/abs/1811.09275} {arXiv:1811.09275 [cond-mat.str-el]}
\bibitem [{\citenamefont {Hashizume}, \citenamefont {Halimeh},\ and\
  \citenamefont {McCulloch}(2020)}]{PhysRevB.102.035115}%
  \BibitemOpen
  \bibfield  {author} {\bibinfo {author} {\bibfnamefont {T.}~\bibnamefont
  {Hashizume}}, \bibinfo {author} {\bibfnamefont {J.~C.}\ \bibnamefont
  {Halimeh}}, \ and\ \bibinfo {author} {\bibfnamefont {I.~P.}\ \bibnamefont
  {McCulloch}},\ }\bibfield  {title} {\emph {\bibinfo {title} {Hybrid infinite
  time-evolving block decimation algorithm for long-range multidimensional
  quantum many-body systems},}\ }\href {\doibase 10.1103/PhysRevB.102.035115}
  {\bibfield  {journal} {\bibinfo  {journal} {Phys. Rev. B}\ }\textbf {\bibinfo
  {volume} {102}},\ \bibinfo {pages} {035115} (\bibinfo {year} {2020})},\
  \Eprint {http://arxiv.org/abs/1910.10726} {arXiv:1910.10726
  [cond-mat.str-el]}
\bibitem [{\citenamefont {Tan}\ \emph {et~al.}(2021)\citenamefont {Tan} \emph
  {et~al.}}]{Tan:2019kya}%
  \BibitemOpen
  \bibfield  {author} {\bibinfo {author} {\bibfnamefont {W.~L.}\ \bibnamefont
  {Tan}} \emph {et~al.},\ }\bibfield  {title} {\emph {\bibinfo {title}
  {{Domain-wall confinement and dynamics in a quantum simulator}},}\ }\href
  {\doibase 10.1038/s41567-021-01194-3} {\bibfield  {journal} {\bibinfo
  {journal} {Nature Phys.}\ }\textbf {\bibinfo {volume} {17}},\ \bibinfo
  {pages} {742} (\bibinfo {year} {2021})},\ \Eprint
  {http://arxiv.org/abs/1912.11117} {arXiv:1912.11117 [quant-ph]}
\bibitem [{\citenamefont {Vovrosh}\ and\ \citenamefont
  {Knolle}(2021)}]{Vovrosh_2021}%
  \BibitemOpen
  \bibfield  {author} {\bibinfo {author} {\bibfnamefont {J.}~\bibnamefont
  {Vovrosh}}\ and\ \bibinfo {author} {\bibfnamefont {J.}~\bibnamefont
  {Knolle}},\ }\bibfield  {title} {\emph {\bibinfo {title} {Confinement and
  entanglement dynamics on a digital quantum computer},}\ }\href {\doibase
  10.1038/s41598-021-90849-5} {\bibfield  {journal} {\bibinfo  {journal}
  {Scientific Reports}\ }\textbf {\bibinfo {volume} {11}},\ \bibinfo {pages}
  {11577} (\bibinfo {year} {2021})},\ \Eprint {http://arxiv.org/abs/2001.03044}
  {arXiv:2001.03044 [cond-mat.str-el]}
\bibitem [{\citenamefont {Schuckert}\ and\ \citenamefont
  {Knap}(2020)}]{Schuckert:2020qeo}%
  \BibitemOpen
  \bibfield  {author} {\bibinfo {author} {\bibfnamefont {A.}~\bibnamefont
  {Schuckert}}\ and\ \bibinfo {author} {\bibfnamefont {M.}~\bibnamefont
  {Knap}},\ }\bibfield  {title} {\emph {\bibinfo {title} {{Probing eigenstate
  thermalization in quantum simulators via fluctuation-dissipation
  relations}},}\ }\href {\doibase 10.1103/PhysRevResearch.2.043315} {\bibfield
  {journal} {\bibinfo  {journal} {Phys. Rev. Res.}\ }\textbf {\bibinfo {volume}
  {2}},\ \bibinfo {pages} {043315} (\bibinfo {year} {2020})},\ \Eprint
  {http://arxiv.org/abs/2007.10347} {arXiv:2007.10347 [cond-mat.quant-gas]}
\bibitem [{\citenamefont {Geier}\ and\ \citenamefont
  {Hauke}(2021)}]{Geier2021}%
  \BibitemOpen
  \bibfield  {author} {\bibinfo {author} {\bibfnamefont {K.~T.}\ \bibnamefont
  {Geier}}\ and\ \bibinfo {author} {\bibfnamefont {P.}~\bibnamefont {Hauke}},\
  }\bibfield  {title} {\emph {\bibinfo {title} {{From non-Hermitian linear
  response to dynamical correlations and fluctuation-dissipation relations in
  quantum many-body systems}},}\ }\href@noop {} {\  (\bibinfo {year} {2021})},\
  \Eprint {http://arxiv.org/abs/2104.03983} {arXiv:2104.03983
  [cond-mat.quant-gas]}
\bibitem [{\citenamefont {Vovrosh}\ \emph {et~al.}(2021)\citenamefont
  {Vovrosh}, \citenamefont {Khosla}, \citenamefont {Greenaway}, \citenamefont
  {Self}, \citenamefont {Kim},\ and\ \citenamefont {Knolle}}]{Vovrosh:2021ocf}%
  \BibitemOpen
  \bibfield  {author} {\bibinfo {author} {\bibfnamefont {J.}~\bibnamefont
  {Vovrosh}}, \bibinfo {author} {\bibfnamefont {K.~E.}\ \bibnamefont {Khosla}},
  \bibinfo {author} {\bibfnamefont {S.}~\bibnamefont {Greenaway}}, \bibinfo
  {author} {\bibfnamefont {C.}~\bibnamefont {Self}}, \bibinfo {author}
  {\bibfnamefont {M.}~\bibnamefont {Kim}}, \ and\ \bibinfo {author}
  {\bibfnamefont {J.}~\bibnamefont {Knolle}},\ }\bibfield  {title} {\emph
  {\bibinfo {title} {{Simple mitigation of global depolarizing errors in
  quantum simulations}},}\ }\href {\doibase 10.1103/PhysRevE.104.035309}
  {\bibfield  {journal} {\bibinfo  {journal} {Phys. Rev. E}\ }\textbf {\bibinfo
  {volume} {104}},\ \bibinfo {pages} {035309} (\bibinfo {year} {2021})},\
  \Eprint {http://arxiv.org/abs/2101.01690} {arXiv:2101.01690 [quant-ph]}
\bibitem [{\citenamefont {Verdel}\ \emph {et~al.}(2020)\citenamefont {Verdel},
  \citenamefont {Liu}, \citenamefont {Whitsitt}, \citenamefont {Gorshkov},\
  and\ \citenamefont {Heyl}}]{Verdel:2019chj}%
  \BibitemOpen
  \bibfield  {author} {\bibinfo {author} {\bibfnamefont {R.}~\bibnamefont
  {Verdel}}, \bibinfo {author} {\bibfnamefont {F.}~\bibnamefont {Liu}},
  \bibinfo {author} {\bibfnamefont {S.}~\bibnamefont {Whitsitt}}, \bibinfo
  {author} {\bibfnamefont {A.~V.}\ \bibnamefont {Gorshkov}}, \ and\ \bibinfo
  {author} {\bibfnamefont {M.}~\bibnamefont {Heyl}},\ }\bibfield  {title}
  {\emph {\bibinfo {title} {{Real-time dynamics of string breaking in quantum
  spin chains}},}\ }\href {\doibase 10.1103/PhysRevB.102.014308} {\bibfield
  {journal} {\bibinfo  {journal} {Phys. Rev. B}\ }\textbf {\bibinfo {volume}
  {102}},\ \bibinfo {pages} {014308} (\bibinfo {year} {2020})},\ \Eprint
  {http://arxiv.org/abs/1911.11382} {arXiv:1911.11382 [cond-mat.stat-mech]}
\bibitem [{\citenamefont {Surace}\ and\ \citenamefont
  {Lerose}(2020)}]{Surace:2020ycc}%
  \BibitemOpen
  \bibfield  {author} {\bibinfo {author} {\bibfnamefont {F.~M.}\ \bibnamefont
  {Surace}}\ and\ \bibinfo {author} {\bibfnamefont {A.}~\bibnamefont
  {Lerose}},\ }\bibfield  {title} {\emph {\bibinfo {title} {{Scattering of
  mesons in quantum simulators}},}\ }\href@noop {} {\  (\bibinfo {year}
  {2020})},\ \Eprint {http://arxiv.org/abs/2011.10583} {arXiv:2011.10583
  [cond-mat.quant-gas]}
\bibitem [{\citenamefont {Karpov}\ \emph {et~al.}(2020)\citenamefont {Karpov},
  \citenamefont {Zhu}, \citenamefont {Heller},\ and\ \citenamefont
  {Heyl}}]{Karpov:2020pqe}%
  \BibitemOpen
  \bibfield  {author} {\bibinfo {author} {\bibfnamefont {P.~I.}\ \bibnamefont
  {Karpov}}, \bibinfo {author} {\bibfnamefont {G.~Y.}\ \bibnamefont {Zhu}},
  \bibinfo {author} {\bibfnamefont {M.~P.}\ \bibnamefont {Heller}}, \ and\
  \bibinfo {author} {\bibfnamefont {M.}~\bibnamefont {Heyl}},\ }\bibfield
  {title} {\emph {\bibinfo {title} {{Spatiotemporal dynamics of particle
  collisions in quantum spin chains}},}\ }\href@noop {} {\  (\bibinfo {year}
  {2020})},\ \Eprint {http://arxiv.org/abs/2011.11624} {arXiv:2011.11624
  [cond-mat.quant-gas]}
\bibitem [{Note1()}]{Note1}%
  \BibitemOpen
  \bibinfo {note} {See also \cite {Rigobello:2021fxw} for a QED study of meson
  scattering with tensor network simulations.}
\bibitem [{Note2()}]{Note2}%
  \BibitemOpen
  \bibinfo {note} {As discussed in detail in \cite {Kormos2017}, the
  semiclassical regime is valid deep in the ferromagnetic phase, i.e.\ $h<J$ in
  \protect \textup {\hbox {\mathsurround \z@ \protect \normalfont
  (\ignorespaces \ref {eq:H_NN}\unskip \@@italiccorr )}} and \protect \textup
  {\hbox {\mathsurround \z@ \protect \normalfont (\ignorespaces \ref
  {eq:H_LR}\unskip \@@italiccorr )}}, when the system is not close to
  criticality. In the case of nearest-neighbor interactions, isolated spin
  domains of magnetization opposite to the longitudinal field give rise to a
  confining potential linear in its spatial extension. Mesonic excitations and
  their properties can then be estimated from the Bohr--Sommerfeld quantization
  condition. In a similar way, using a two-kink model, it was shown in \cite
  {Liu:2018fza} that long-range interactions give rise to an effective
  confining potential when restricting to the Hilbert space of two domain-wall
  states. Details on a reformulation of the nearest-neighbor Ising model at
  very weak transverse field as a nonrelativistic $\protect \mathds {Z}_2$
  gauge theory are discussed, e.g., in \cite {Karpov:2020pqe}.}
\bibitem [{\citenamefont {Zamolodchikov}(1989)}]{Zamolodchikov:1989fp}%
  \BibitemOpen
  \bibfield  {author} {\bibinfo {author} {\bibfnamefont {A.~B.}\ \bibnamefont
  {Zamolodchikov}},\ }\bibfield  {title} {\emph {\bibinfo {title} {{Integrals
  of Motion and S-Matrix of the (Scaled) T=T{$_c$} Ising Model with Magnetic
  Field}},}\ }\href {\doibase 10.1142/S0217751X8900176X} {\bibfield  {journal}
  {\bibinfo  {journal} {Int. J. Mod. Phys.}\ }\textbf {\bibinfo {volume}
  {A4}},\ \bibinfo {pages} {4235} (\bibinfo {year} {1989})}
\bibitem [{\citenamefont {Coldea}\ \emph {et~al.}(2010)\citenamefont {Coldea},
  \citenamefont {Tennant}, \citenamefont {Wheeler}, \citenamefont {Wawrzynska},
  \citenamefont {Prabhakaran}, \citenamefont {Telling}, \citenamefont
  {Habicht}, \citenamefont {Smeibidl},\ and\ \citenamefont
  {Kiefer}}]{Coldea_2010}%
  \BibitemOpen
  \bibfield  {author} {\bibinfo {author} {\bibfnamefont {R.}~\bibnamefont
  {Coldea}}, \bibinfo {author} {\bibfnamefont {D.~A.}\ \bibnamefont {Tennant}},
  \bibinfo {author} {\bibfnamefont {E.~M.}\ \bibnamefont {Wheeler}}, \bibinfo
  {author} {\bibfnamefont {E.}~\bibnamefont {Wawrzynska}}, \bibinfo {author}
  {\bibfnamefont {D.}~\bibnamefont {Prabhakaran}}, \bibinfo {author}
  {\bibfnamefont {M.}~\bibnamefont {Telling}}, \bibinfo {author} {\bibfnamefont
  {K.}~\bibnamefont {Habicht}}, \bibinfo {author} {\bibfnamefont
  {P.}~\bibnamefont {Smeibidl}}, \ and\ \bibinfo {author} {\bibfnamefont
  {K.}~\bibnamefont {Kiefer}},\ }\bibfield  {title} {\emph {\bibinfo {title}
  {{Quantum Criticality in an Ising Chain: Experimental Evidence for Emergent
  E{$_8$} Symmetry}},}\ }\href {\doibase 10.1126/science.1180085} {\bibfield
  {journal} {\bibinfo  {journal} {Science}\ }\textbf {\bibinfo {volume}
  {327}},\ \bibinfo {pages} {177–180} (\bibinfo {year} {2010})},\ \Eprint
  {http://arxiv.org/abs/1103.3694} {arXiv:1103.3694 [cond-mat.str-el]}
\bibitem [{\citenamefont {Zou}\ \emph {et~al.}(2020)\citenamefont {Zou} \emph
  {et~al.}}]{Zou:2020ouw}%
  \BibitemOpen
  \bibfield  {author} {\bibinfo {author} {\bibfnamefont {H.}~\bibnamefont
  {Zou}} \emph {et~al.},\ }\bibfield  {title} {\emph {\bibinfo {title} {{$E_8$
  Spectra of Quasi-one-dimensional Antiferromagnet BaCo$_2$V$_2$O$_8$ under
  Transverse Field}},}\ }\href@noop {} {\  (\bibinfo {year} {2020})},\ \Eprint
  {http://arxiv.org/abs/2005.13302} {arXiv:2005.13302 [cond-mat.str-el]}
\bibitem [{\citenamefont {Zhang}\ \emph {et~al.}(2020)\citenamefont {Zhang}
  \emph {et~al.}}]{Zhang:2020enf}%
  \BibitemOpen
  \bibfield  {author} {\bibinfo {author} {\bibfnamefont {Z.}~\bibnamefont
  {Zhang}} \emph {et~al.},\ }\bibfield  {title} {\emph {\bibinfo {title}
  {{Observation of E8 Particles in an Ising Chain Antiferromagnet}},}\ }\href
  {\doibase 10.1103/PhysRevB.101.220411} {\bibfield  {journal} {\bibinfo
  {journal} {Phys. Rev. B}\ }\textbf {\bibinfo {volume} {101}},\ \bibinfo
  {pages} {220411} (\bibinfo {year} {2020})},\ \Eprint
  {http://arxiv.org/abs/2005.13772} {arXiv:2005.13772 [cond-mat.str-el]}
\bibitem [{\citenamefont {Amelin}\ \emph {et~al.}(2020)\citenamefont {Amelin},
  \citenamefont {Engelmayer}, \citenamefont {Viirok}, \citenamefont {Nagel},
  \citenamefont {R\~o\ om}, \citenamefont {Lorenz},\ and\ \citenamefont
  {Wang}}]{Amelin:2020mif}%
  \BibitemOpen
  \bibfield  {author} {\bibinfo {author} {\bibfnamefont {K.}~\bibnamefont
  {Amelin}}, \bibinfo {author} {\bibfnamefont {J.}~\bibnamefont {Engelmayer}},
  \bibinfo {author} {\bibfnamefont {J.}~\bibnamefont {Viirok}}, \bibinfo
  {author} {\bibfnamefont {U.}~\bibnamefont {Nagel}}, \bibinfo {author}
  {\bibfnamefont {T.}~\bibnamefont {R\~o\ om}}, \bibinfo {author}
  {\bibfnamefont {T.}~\bibnamefont {Lorenz}}, \ and\ \bibinfo {author}
  {\bibfnamefont {Z.}~\bibnamefont {Wang}},\ }\bibfield  {title} {\emph
  {\bibinfo {title} {{Experimental Observation of Quantum Many-Body Excitations
  of $E_8$ Symmetry in the Ising Chain Ferromagnet CoNb$_2$O$_6$}},}\ }\href
  {\doibase 10.1103/PhysRevB.102.104431} {\bibfield  {journal} {\bibinfo
  {journal} {Phys. Rev. B}\ }\textbf {\bibinfo {volume} {102}},\ \bibinfo
  {pages} {104431} (\bibinfo {year} {2020})},\ \Eprint
  {http://arxiv.org/abs/2006.12956} {arXiv:2006.12956 [cond-mat.str-el]}
\bibitem [{\citenamefont {Hauke}\ \emph {et~al.}(2013)\citenamefont {Hauke},
  \citenamefont {Marcos}, \citenamefont {Dalmonte},\ and\ \citenamefont
  {Zoller}}]{Hauke2013b}%
  \BibitemOpen
  \bibfield  {author} {\bibinfo {author} {\bibfnamefont {P.}~\bibnamefont
  {Hauke}}, \bibinfo {author} {\bibfnamefont {D.}~\bibnamefont {Marcos}},
  \bibinfo {author} {\bibfnamefont {M.}~\bibnamefont {Dalmonte}}, \ and\
  \bibinfo {author} {\bibfnamefont {P.}~\bibnamefont {Zoller}},\ }\bibfield
  {title} {\emph {\bibinfo {title} {{Quantum simulation of a lattice
  {S}chwinger model in a chain of trapped ions}},}\ }\href@noop {} {\bibfield
  {journal} {\bibinfo  {journal} {Phys. Rev. X}\ }\textbf {\bibinfo {volume}
  {3}},\ \bibinfo {pages} {041018} (\bibinfo {year} {2013})}
\bibitem [{\citenamefont {Yang}\ \emph {et~al.}(2016)\citenamefont {Yang},
  \citenamefont {Giri}, \citenamefont {Johanning}, \citenamefont {Wunderlich},
  \citenamefont {Zoller},\ and\ \citenamefont {Hauke}}]{Yang:2016hjn}%
  \BibitemOpen
  \bibfield  {author} {\bibinfo {author} {\bibfnamefont {D.}~\bibnamefont
  {Yang}}, \bibinfo {author} {\bibfnamefont {G.~S.}\ \bibnamefont {Giri}},
  \bibinfo {author} {\bibfnamefont {M.}~\bibnamefont {Johanning}}, \bibinfo
  {author} {\bibfnamefont {C.}~\bibnamefont {Wunderlich}}, \bibinfo {author}
  {\bibfnamefont {P.}~\bibnamefont {Zoller}}, \ and\ \bibinfo {author}
  {\bibfnamefont {P.}~\bibnamefont {Hauke}},\ }\bibfield  {title} {\emph
  {\bibinfo {title} {{Analog quantum simulation of (1+1)-dimensional lattice
  QED with trapped ions}},}\ }\href {\doibase 10.1103/PhysRevA.94.052321}
  {\bibfield  {journal} {\bibinfo  {journal} {Phys. Rev. A}\ }\textbf {\bibinfo
  {volume} {94}},\ \bibinfo {pages} {052321} (\bibinfo {year} {2016})},\
  \Eprint {http://arxiv.org/abs/1604.03124} {arXiv:1604.03124 [quant-ph]}
\bibitem [{\citenamefont {Muschik}\ \emph {et~al.}(2017)\citenamefont
  {Muschik}, \citenamefont {Heyl}, \citenamefont {Martinez}, \citenamefont
  {Monz}, \citenamefont {Schindler}, \citenamefont {Vogell}, \citenamefont
  {Dalmonte}, \citenamefont {Hauke}, \citenamefont {Blatt},\ and\ \citenamefont
  {Zoller}}]{Muschik_2017}%
  \BibitemOpen
  \bibfield  {author} {\bibinfo {author} {\bibfnamefont {C.}~\bibnamefont
  {Muschik}}, \bibinfo {author} {\bibfnamefont {M.}~\bibnamefont {Heyl}},
  \bibinfo {author} {\bibfnamefont {E.}~\bibnamefont {Martinez}}, \bibinfo
  {author} {\bibfnamefont {T.}~\bibnamefont {Monz}}, \bibinfo {author}
  {\bibfnamefont {P.}~\bibnamefont {Schindler}}, \bibinfo {author}
  {\bibfnamefont {B.}~\bibnamefont {Vogell}}, \bibinfo {author} {\bibfnamefont
  {M.}~\bibnamefont {Dalmonte}}, \bibinfo {author} {\bibfnamefont
  {P.}~\bibnamefont {Hauke}}, \bibinfo {author} {\bibfnamefont
  {R.}~\bibnamefont {Blatt}}, \ and\ \bibinfo {author} {\bibfnamefont
  {P.}~\bibnamefont {Zoller}},\ }\bibfield  {title} {\emph {\bibinfo {title}
  {U(1) wilson lattice gauge theories in digital quantum simulators},}\ }\href
  {\doibase 10.1088/1367-2630/aa89ab} {\bibfield  {journal} {\bibinfo
  {journal} {New Journal of Physics}\ }\textbf {\bibinfo {volume} {19}},\
  \bibinfo {pages} {103020} (\bibinfo {year} {2017})}
\bibitem [{\citenamefont {Davoudi}\ \emph {et~al.}(2020)\citenamefont
  {Davoudi}, \citenamefont {Hafezi}, \citenamefont {Monroe}, \citenamefont
  {Pagano}, \citenamefont {Seif},\ and\ \citenamefont
  {Shaw}}]{Davoudi:2019bhy}%
  \BibitemOpen
  \bibfield  {author} {\bibinfo {author} {\bibfnamefont {Z.}~\bibnamefont
  {Davoudi}}, \bibinfo {author} {\bibfnamefont {M.}~\bibnamefont {Hafezi}},
  \bibinfo {author} {\bibfnamefont {C.}~\bibnamefont {Monroe}}, \bibinfo
  {author} {\bibfnamefont {G.}~\bibnamefont {Pagano}}, \bibinfo {author}
  {\bibfnamefont {A.}~\bibnamefont {Seif}}, \ and\ \bibinfo {author}
  {\bibfnamefont {A.}~\bibnamefont {Shaw}},\ }\bibfield  {title} {\emph
  {\bibinfo {title} {{Towards analog quantum simulations of lattice gauge
  theories with trapped ions}},}\ }\href {\doibase
  10.1103/PhysRevResearch.2.023015} {\bibfield  {journal} {\bibinfo  {journal}
  {Phys. Rev. Res.}\ }\textbf {\bibinfo {volume} {2}},\ \bibinfo {pages}
  {023015} (\bibinfo {year} {2020})},\ \Eprint
  {http://arxiv.org/abs/1908.03210} {arXiv:1908.03210 [quant-ph]}
\bibitem [{\citenamefont {Paulson}\ \emph {et~al.}(2020)\citenamefont {Paulson}
  \emph {et~al.}}]{Paulson:2020zjd}%
  \BibitemOpen
  \bibfield  {author} {\bibinfo {author} {\bibfnamefont {D.}~\bibnamefont
  {Paulson}} \emph {et~al.},\ }\bibfield  {title} {\emph {\bibinfo {title}
  {{Towards simulating 2D effects in lattice gauge theories on a quantum
  computer}},}\ }\href@noop {} {\  (\bibinfo {year} {2020})},\ \Eprint
  {http://arxiv.org/abs/2008.09252} {arXiv:2008.09252 [quant-ph]}
\bibitem [{\citenamefont {Davoudi}, \citenamefont {Linke},\ and\ \citenamefont
  {Pagano}(2021)}]{Davoudi:2021ney}%
  \BibitemOpen
  \bibfield  {author} {\bibinfo {author} {\bibfnamefont {Z.}~\bibnamefont
  {Davoudi}}, \bibinfo {author} {\bibfnamefont {N.~M.}\ \bibnamefont {Linke}},
  \ and\ \bibinfo {author} {\bibfnamefont {G.}~\bibnamefont {Pagano}},\
  }\bibfield  {title} {\emph {\bibinfo {title} {{Toward simulating quantum
  field theories with controlled phonon-ion dynamics: A hybrid analog-digital
  approach}},}\ }\href@noop {} {\  (\bibinfo {year} {2021})},\ \Eprint
  {http://arxiv.org/abs/2104.09346} {arXiv:2104.09346 [quant-ph]}
\bibitem [{\citenamefont {Zohar}(2021)}]{Zohar:2021nyc}%
  \BibitemOpen
  \bibfield  {author} {\bibinfo {author} {\bibfnamefont {E.}~\bibnamefont
  {Zohar}},\ }\bibfield  {title} {\emph {\bibinfo {title} {{Quantum Simulation
  of Lattice Gauge Theories in more than One Space Dimension -- Requirements,
  Challenges, Methods}},}\ }\href@noop {} {\  (\bibinfo {year} {2021})},\
  \Eprint {http://arxiv.org/abs/2106.04609} {arXiv:2106.04609 [quant-ph]}
\bibitem [{\citenamefont {Sachdev}(2011)}]{sachdev_2011}%
  \BibitemOpen
  \bibfield  {author} {\bibinfo {author} {\bibfnamefont {S.}~\bibnamefont
  {Sachdev}},\ }\href {\doibase 10.1017/CBO9780511973765} {\emph {\bibinfo
  {title} {Quantum Phase Transitions}}},\ \bibinfo {edition} {2nd}\ ed.\
  (\bibinfo  {publisher} {Cambridge University Press},\ \bibinfo {year} {2011})
\bibitem [{\citenamefont {Rakovszky}\ \emph {et~al.}(2016)\citenamefont
  {Rakovszky}, \citenamefont {Mesty{\'a}n}, \citenamefont {Collura},
  \citenamefont {Kormos},\ and\ \citenamefont
  {Tak{\'a}cs}}]{Rakovszky:2016ugs}%
  \BibitemOpen
  \bibfield  {author} {\bibinfo {author} {\bibfnamefont {T.}~\bibnamefont
  {Rakovszky}}, \bibinfo {author} {\bibfnamefont {M.}~\bibnamefont
  {Mesty{\'a}n}}, \bibinfo {author} {\bibfnamefont {M.}~\bibnamefont
  {Collura}}, \bibinfo {author} {\bibfnamefont {M.}~\bibnamefont {Kormos}}, \
  and\ \bibinfo {author} {\bibfnamefont {G.}~\bibnamefont {Tak{\'a}cs}},\
  }\bibfield  {title} {\emph {\bibinfo {title} {{Hamiltonian truncation
  approach to quenches in the Ising field theory}},}\ }\href {\doibase
  10.1016/j.nuclphysb.2016.08.024} {\bibfield  {journal} {\bibinfo  {journal}
  {Nucl. Phys.}\ }\textbf {\bibinfo {volume} {B911}},\ \bibinfo {pages} {805}
  (\bibinfo {year} {2016})},\ \Eprint {http://arxiv.org/abs/1607.01068}
  {arXiv:1607.01068 [cond-mat.stat-mech]}
\bibitem [{\citenamefont {H{\'o}ds{\'a}gi}, \citenamefont {Kormos},\ and\
  \citenamefont {Tak{\'a}cs}(2018)}]{Hodsagi:2018sul}%
  \BibitemOpen
  \bibfield  {author} {\bibinfo {author} {\bibfnamefont {K.}~\bibnamefont
  {H{\'o}ds{\'a}gi}}, \bibinfo {author} {\bibfnamefont {M.}~\bibnamefont
  {Kormos}}, \ and\ \bibinfo {author} {\bibfnamefont {G.}~\bibnamefont
  {Tak{\'a}cs}},\ }\bibfield  {title} {\emph {\bibinfo {title} {{Quench
  dynamics of the Ising field theory in a magnetic field}},}\ }\href {\doibase
  10.21468/SciPostPhys.5.3.027} {\bibfield  {journal} {\bibinfo  {journal}
  {SciPost Phys.}\ }\textbf {\bibinfo {volume} {5}},\ \bibinfo {pages} {027}
  (\bibinfo {year} {2018})},\ \Eprint {http://arxiv.org/abs/1803.01158}
  {arXiv:1803.01158 [cond-mat.stat-mech]}
\bibitem [{Note3()}]{Note3}%
  \BibitemOpen
  \bibinfo {note} {Physical consequences of algebraic long-range interactions
  in QMB systems for eigenstate thermalization and symmetry properties are
  studied, e.g., in \cite {russomanno2020longrange}.}
\bibitem [{\citenamefont {Porras}\ and\ \citenamefont
  {Cirac}(2004)}]{Porras2004a}%
  \BibitemOpen
  \bibfield  {author} {\bibinfo {author} {\bibfnamefont {D.}~\bibnamefont
  {Porras}}\ and\ \bibinfo {author} {\bibfnamefont {J.~I.}\ \bibnamefont
  {Cirac}},\ }\bibfield  {title} {\emph {\bibinfo {title} {Effective quantum
  spin systems with trapped ions},}\ }\href@noop {} {\bibfield  {journal}
  {\bibinfo  {journal} {Phys. Rev. Lett.}\ }\textbf {\bibinfo {volume} {92}},\
  \bibinfo {pages} {207901} (\bibinfo {year} {2004})}
\bibitem [{\citenamefont {Trautmann}\ and\ \citenamefont
  {Hauke}(2018)}]{Trautmann_2018}%
  \BibitemOpen
  \bibfield  {author} {\bibinfo {author} {\bibfnamefont {N.}~\bibnamefont
  {Trautmann}}\ and\ \bibinfo {author} {\bibfnamefont {P.}~\bibnamefont
  {Hauke}},\ }\bibfield  {title} {\emph {\bibinfo {title} {Trapped-ion quantum
  simulation of excitation transport: Disordered, noisy, and long-range
  connected quantum networks},}\ }\href {\doibase 10.1103/physreva.97.023606}
  {\bibfield  {journal} {\bibinfo  {journal} {Physical Review A}\ }\textbf
  {\bibinfo {volume} {97}} (\bibinfo {year} {2018}),\
  10.1103/physreva.97.023606}
\bibitem [{\citenamefont {Hauke}\ and\ \citenamefont
  {Tagliacozzo}(2013)}]{Hauke_2013}%
  \BibitemOpen
  \bibfield  {author} {\bibinfo {author} {\bibfnamefont {P.}~\bibnamefont
  {Hauke}}\ and\ \bibinfo {author} {\bibfnamefont {L.}~\bibnamefont
  {Tagliacozzo}},\ }\bibfield  {title} {\emph {\bibinfo {title} {Spread of
  correlations in long-range interacting quantum systems},}\ }\href {\doibase
  10.1103/physrevlett.111.207202} {\bibfield  {journal} {\bibinfo  {journal}
  {Physical Review Letters}\ }\textbf {\bibinfo {volume} {111}} (\bibinfo
  {year} {2013}),\ 10.1103/physrevlett.111.207202},\ \Eprint
  {http://arxiv.org/abs/1304.7725} {arXiv:1304.7725 [quant-ph]}
\bibitem [{\citenamefont {Kj{\"a}ll}, \citenamefont {Pollmann},\ and\
  \citenamefont {Moore}(2011)}]{Kjall_2011}%
  \BibitemOpen
  \bibfield  {author} {\bibinfo {author} {\bibfnamefont {J.~A.}\ \bibnamefont
  {Kj{\"a}ll}}, \bibinfo {author} {\bibfnamefont {F.}~\bibnamefont {Pollmann}},
  \ and\ \bibinfo {author} {\bibfnamefont {J.~E.}\ \bibnamefont {Moore}},\
  }\bibfield  {title} {\emph {\bibinfo {title} {{Bound states and E{$_8$}
  symmetry effects in perturbed quantum Ising chains}},}\ }\href {\doibase
  10.1103/PhysRevB.83.020407} {\bibfield  {journal} {\bibinfo  {journal} {Phys.
  Rev. B}\ }\textbf {\bibinfo {volume} {83}},\ \bibinfo {pages} {020407}
  (\bibinfo {year} {2011})},\ \Eprint {http://arxiv.org/abs/1008.3534}
  {arXiv:1008.3534 [cond-mat.str-el]}
\bibitem [{Note4()}]{Note4}%
  \BibitemOpen
  \bibinfo {note} {The mass gap $m_n$ of level $n$ is defined as the energy
  difference to the groundstate, i.e.\ $m_n=E_n-E_0$.}
\bibitem [{\citenamefont {L{\"u}scher}(1986)}]{Luscher:1986}%
  \BibitemOpen
  \bibfield  {author} {\bibinfo {author} {\bibfnamefont {M.}~\bibnamefont
  {L{\"u}scher}},\ }\bibfield  {title} {\emph {\bibinfo {title} {{Volume
  dependence of the energy spectrum in massive quantum field theories}},}\
  }\href {\doibase 10.1007/BF01211589} {\bibfield  {journal} {\bibinfo
  {journal} {Commun. Math. Phys.}\ }\textbf {\bibinfo {volume} {104}},\
  \bibinfo {pages} {177} (\bibinfo {year} {1986})}
\bibitem [{Note5()}]{Note5}%
  \BibitemOpen
  \bibinfo {note} {See e.g.\ \cite {jensen1991rare} for an introduction.}
\bibitem [{\citenamefont {Wiener}(1930)}]{Wiener1930}%
  \BibitemOpen
  \bibfield  {author} {\bibinfo {author} {\bibfnamefont {N.}~\bibnamefont
  {Wiener}},\ }\bibfield  {title} {\emph {\bibinfo {title} {{Generalized
  harmonic analysis}},}\ }\href {\doibase 10.1007/BF02546511} {\bibfield
  {journal} {\bibinfo  {journal} {Acta Mathematica}\ }\textbf {\bibinfo
  {volume} {55}},\ \bibinfo {pages} {117 } (\bibinfo {year} {1930})}
\bibitem [{\citenamefont {Khintchine}(1934)}]{Khintchine1934}%
  \BibitemOpen
  \bibfield  {author} {\bibinfo {author} {\bibfnamefont {A.}~\bibnamefont
  {Khintchine}},\ }\bibfield  {title} {\emph {\bibinfo {title}
  {{Korrelationstheorie der station\"aren stochastischen Prozesse}},}\ }\href
  {\doibase 10.1007/BF01449156} {\bibfield  {journal} {\bibinfo  {journal}
  {Mathematische Annalen}\ }\textbf {\bibinfo {volume} {109}},\ \bibinfo
  {pages} {604 } (\bibinfo {year} {1934})}
\bibitem [{Note6()}]{Note6}%
  \BibitemOpen
  \bibinfo {note} {We refer to the appendices for further details on the finite
  size, long-range and longitudinal field dependence of the energy and
  absorption spectra discussed in this section.}
\bibitem [{\citenamefont {Wang}\ \emph {et~al.}(2021)\citenamefont {Wang},
  \citenamefont {Zou}, \citenamefont {Hodsagi}, \citenamefont {Kormos},
  \citenamefont {Takacs},\ and\ \citenamefont {Wu}}]{wang2021spin}%
  \BibitemOpen
  \bibfield  {author} {\bibinfo {author} {\bibfnamefont {X.}~\bibnamefont
  {Wang}}, \bibinfo {author} {\bibfnamefont {H.}~\bibnamefont {Zou}}, \bibinfo
  {author} {\bibfnamefont {K.}~\bibnamefont {Hodsagi}}, \bibinfo {author}
  {\bibfnamefont {M.}~\bibnamefont {Kormos}}, \bibinfo {author} {\bibfnamefont
  {G.}~\bibnamefont {Takacs}}, \ and\ \bibinfo {author} {\bibfnamefont
  {J.}~\bibnamefont {Wu}},\ }\bibfield  {title} {\emph {\bibinfo {title} {Spin
  dynamics of a perturbed quantum critical ising chain},}\ }\href@noop {} {\
  (\bibinfo {year} {2021})},\ \Eprint {http://arxiv.org/abs/2103.09128}
  {arXiv:2103.09128 [cond-mat.str-el]}
\bibitem [{\citenamefont {Gu}\ \emph {et~al.}(2008)\citenamefont {Gu},
  \citenamefont {Kwok}, \citenamefont {Ning},\ and\ \citenamefont
  {Lin}}]{Gu_2008}%
  \BibitemOpen
  \bibfield  {author} {\bibinfo {author} {\bibfnamefont {S.-J.}\ \bibnamefont
  {Gu}}, \bibinfo {author} {\bibfnamefont {H.-M.}\ \bibnamefont {Kwok}},
  \bibinfo {author} {\bibfnamefont {W.-Q.}\ \bibnamefont {Ning}}, \ and\
  \bibinfo {author} {\bibfnamefont {H.-Q.}\ \bibnamefont {Lin}},\ }\bibfield
  {title} {\emph {\bibinfo {title} {Fidelity susceptibility, scaling, and
  universality in quantum critical phenomena},}\ }\href {\doibase
  10.1103/PhysRevB.77.245109} {\bibfield  {journal} {\bibinfo  {journal} {Phys.
  Rev. B}\ }\textbf {\bibinfo {volume} {77}},\ \bibinfo {pages} {245109}
  (\bibinfo {year} {2008})},\ \Eprint {http://arxiv.org/abs/0706.2495}
  {arXiv:0706.2495 [quant-ph]}
\bibitem [{\citenamefont {Gu}(2010)}]{GU_2010}%
  \BibitemOpen
  \bibfield  {author} {\bibinfo {author} {\bibfnamefont {S.-J.}\ \bibnamefont
  {Gu}},\ }\bibfield  {title} {\emph {\bibinfo {title} {{Fidelity approach to
  quantum phase transitions}},}\ }\href {\doibase 10.1142/s0217979210056335}
  {\bibfield  {journal} {\bibinfo  {journal} {International Journal of Modern
  Physics B}\ }\textbf {\bibinfo {volume} {24}},\ \bibinfo {pages}
  {4371–4458} (\bibinfo {year} {2010})},\ \Eprint
  {http://arxiv.org/abs/0811.3127} {arXiv:0811.3127 [quant-ph]}
\bibitem [{\citenamefont {Khalouf-Rivera}, \citenamefont {Carvajal},\ and\
  \citenamefont {Perez-Bernal}(2021)}]{khaloufrivera2021quantum}%
  \BibitemOpen
  \bibfield  {author} {\bibinfo {author} {\bibfnamefont {J.}~\bibnamefont
  {Khalouf-Rivera}}, \bibinfo {author} {\bibfnamefont {M.}~\bibnamefont
  {Carvajal}}, \ and\ \bibinfo {author} {\bibfnamefont {F.}~\bibnamefont
  {Perez-Bernal}},\ }\href@noop {} {\emph {\bibinfo {title} {Quantum fidelity
  susceptibility in excited state quantum phase transitions: application to the
  bending spectra of nonrigid molecules},}\ } (\bibinfo {year} {2021}),\
  \Eprint {http://arxiv.org/abs/2102.12335} {arXiv:2102.12335 [quant-ph]}
\bibitem [{\citenamefont {Russomanno}, \citenamefont {Fava},\ and\
  \citenamefont {Heyl}(2020)}]{russomanno2020longrange}%
  \BibitemOpen
  \bibfield  {author} {\bibinfo {author} {\bibfnamefont {A.}~\bibnamefont
  {Russomanno}}, \bibinfo {author} {\bibfnamefont {M.}~\bibnamefont {Fava}}, \
  and\ \bibinfo {author} {\bibfnamefont {M.}~\bibnamefont {Heyl}},\ }\bibfield
  {title} {\emph {\bibinfo {title} {Long-range ising chains: eigenstate
  thermalization and symmetry breaking of excited states},}\ }\href@noop {} {\
  (\bibinfo {year} {2020})},\ \Eprint {http://arxiv.org/abs/2012.06505}
  {arXiv:2012.06505 [cond-mat.stat-mech]}
\bibitem [{\citenamefont {Cevolani}, \citenamefont {Carleo},\ and\
  \citenamefont {Sanchez-Palencia}(2015)}]{Cevolani2015}%
  \BibitemOpen
  \bibfield  {author} {\bibinfo {author} {\bibfnamefont {L.}~\bibnamefont
  {Cevolani}}, \bibinfo {author} {\bibfnamefont {G.}~\bibnamefont {Carleo}}, \
  and\ \bibinfo {author} {\bibfnamefont {L.}~\bibnamefont {Sanchez-Palencia}},\
  }\bibfield  {title} {\emph {\bibinfo {title} {Protected quasilocality in
  quantum systems with long-range interactions},}\ }\href {\doibase
  10.1103/PhysRevA.92.041603} {\bibfield  {journal} {\bibinfo  {journal} {Phys.
  Rev. A}\ }\textbf {\bibinfo {volume} {92}},\ \bibinfo {pages} {041603}
  (\bibinfo {year} {2015})}
\bibitem [{Note7()}]{Note7}%
  \BibitemOpen
  \bibinfo {note} {Further details on the underlying spectra are discussed in
  appendix~\ref {app:field_dependence}.}
\bibitem [{\citenamefont {S\o{}rensen}\ and\ \citenamefont
  {M\o{}lmer}(2000)}]{MoelmerSoerensen2000}%
  \BibitemOpen
  \bibfield  {author} {\bibinfo {author} {\bibfnamefont {A.}~\bibnamefont
  {S\o{}rensen}}\ and\ \bibinfo {author} {\bibfnamefont {K.}~\bibnamefont
  {M\o{}lmer}},\ }\bibfield  {title} {\emph {\bibinfo {title} {Entanglement and
  quantum computation with ions in thermal motion},}\ }\href {\doibase
  10.1103/PhysRevA.62.022311} {\bibfield  {journal} {\bibinfo  {journal} {Phys.
  Rev. A}\ }\textbf {\bibinfo {volume} {62}},\ \bibinfo {pages} {022311}
  (\bibinfo {year} {2000})}
\bibitem [{\citenamefont {Wang}\ and\ \citenamefont
  {Freericks}(2012)}]{Wang2012}%
  \BibitemOpen
  \bibfield  {author} {\bibinfo {author} {\bibfnamefont {C.-C.~J.}\
  \bibnamefont {Wang}}\ and\ \bibinfo {author} {\bibfnamefont {J.~K.}\
  \bibnamefont {Freericks}},\ }\bibfield  {title} {\emph {\bibinfo {title}
  {Intrinsic phonon effects on analog quantum simulators with ultracold trapped
  ions},}\ }\href@noop {} {\bibfield  {journal} {\bibinfo  {journal} {Phys.
  Rev. A}\ }\textbf {\bibinfo {volume} {86}},\ \bibinfo {pages} {032329}
  (\bibinfo {year} {2012})}
\bibitem [{\citenamefont {Hauke}\ \emph {et~al.}(2015)\citenamefont {Hauke},
  \citenamefont {Bonnes}, \citenamefont {Heyl},\ and\ \citenamefont
  {Lechner}}]{Hauke_2015}%
  \BibitemOpen
  \bibfield  {author} {\bibinfo {author} {\bibfnamefont {P.}~\bibnamefont
  {Hauke}}, \bibinfo {author} {\bibfnamefont {L.}~\bibnamefont {Bonnes}},
  \bibinfo {author} {\bibfnamefont {M.}~\bibnamefont {Heyl}}, \ and\ \bibinfo
  {author} {\bibfnamefont {W.}~\bibnamefont {Lechner}},\ }\bibfield  {title}
  {\emph {\bibinfo {title} {Probing entanglement in adiabatic quantum
  optimization with trapped ions},}\ }\href {\doibase 10.3389/fphy.2015.00021}
  {\bibfield  {journal} {\bibinfo  {journal} {Front. Phys.}\ }\textbf {\bibinfo
  {volume} {3}},\ \bibinfo {pages} {21} (\bibinfo {year} {2015})}
\bibitem [{\citenamefont {Schindler}\ \emph {et~al.}(2013)\citenamefont
  {Schindler}, \citenamefont {Nigg}, \citenamefont {Monz}, \citenamefont
  {Barreiro}, \citenamefont {Martinez}, \citenamefont {Wang}, \citenamefont
  {Quint}, \citenamefont {Brandl}, \citenamefont {Nebendahl}, \citenamefont
  {Roos},\ and\ \citenamefont {et~al.}}]{Schindler_2013}%
  \BibitemOpen
  \bibfield  {author} {\bibinfo {author} {\bibfnamefont {P.}~\bibnamefont
  {Schindler}}, \bibinfo {author} {\bibfnamefont {D.}~\bibnamefont {Nigg}},
  \bibinfo {author} {\bibfnamefont {T.}~\bibnamefont {Monz}}, \bibinfo {author}
  {\bibfnamefont {J.~T.}\ \bibnamefont {Barreiro}}, \bibinfo {author}
  {\bibfnamefont {E.}~\bibnamefont {Martinez}}, \bibinfo {author}
  {\bibfnamefont {S.~X.}\ \bibnamefont {Wang}}, \bibinfo {author}
  {\bibfnamefont {S.}~\bibnamefont {Quint}}, \bibinfo {author} {\bibfnamefont
  {M.~F.}\ \bibnamefont {Brandl}}, \bibinfo {author} {\bibfnamefont
  {V.}~\bibnamefont {Nebendahl}}, \bibinfo {author} {\bibfnamefont {C.~F.}\
  \bibnamefont {Roos}}, \ and\ \bibinfo {author} {\bibnamefont {et~al.}},\
  }\bibfield  {title} {\emph {\bibinfo {title} {A quantum information processor
  with trapped ions},}\ }\href {\doibase 10.1088/1367-2630/15/12/123012}
  {\bibfield  {journal} {\bibinfo  {journal} {New Journal of Physics}\ }\textbf
  {\bibinfo {volume} {15}},\ \bibinfo {pages} {123012} (\bibinfo {year}
  {2013})}
\bibitem [{\citenamefont {Monroe}\ \emph
  {et~al.}(2021{\natexlab{b}})\citenamefont {Monroe} \emph
  {et~al.}}]{Monroe:2019asq}%
  \BibitemOpen
  \bibfield  {author} {\bibinfo {author} {\bibfnamefont {C.}~\bibnamefont
  {Monroe}} \emph {et~al.},\ }\bibfield  {title} {\emph {\bibinfo {title}
  {{Programmable quantum simulations of spin systems with trapped ions}},}\
  }\href {\doibase 10.1103/RevModPhys.93.025001} {\bibfield  {journal}
  {\bibinfo  {journal} {Rev. Mod. Phys.}\ }\textbf {\bibinfo {volume} {93}},\
  \bibinfo {pages} {025001} (\bibinfo {year} {2021}{\natexlab{b}})},\ \Eprint
  {http://arxiv.org/abs/1912.07845} {arXiv:1912.07845 [quant-ph]}
\bibitem [{\citenamefont {Jensen}\ and\ \citenamefont
  {Mackintosh}(1991)}]{jensen1991rare}%
  \BibitemOpen
  \bibfield  {author} {\bibinfo {author} {\bibfnamefont {J.}~\bibnamefont
  {Jensen}}\ and\ \bibinfo {author} {\bibfnamefont {A.}~\bibnamefont
  {Mackintosh}},\ }\href {https://www.fys.ku.dk/~jjensen/Book/Ebook.pdf} {\emph
  {\bibinfo {title} {{Rare Earth Magnetism: Structures and Excitations}}}},\
  International Series of Monographs on Physics\ (\bibinfo  {publisher}
  {Clarendon Press},\ \bibinfo {year} {1991})
\bibitem [{\citenamefont {Nevado}\ and\ \citenamefont
  {Porras}(2016)}]{Nevado_2016}%
  \BibitemOpen
  \bibfield  {author} {\bibinfo {author} {\bibfnamefont {P.}~\bibnamefont
  {Nevado}}\ and\ \bibinfo {author} {\bibfnamefont {D.}~\bibnamefont
  {Porras}},\ }\bibfield  {title} {\emph {\bibinfo {title} {Hidden frustrated
  interactions and quantum annealing in trapped-ion spin-phonon chains},}\
  }\href {\doibase 10.1103/physreva.93.013625} {\bibfield  {journal} {\bibinfo
  {journal} {Physical Review A}\ }\textbf {\bibinfo {volume} {93}} (\bibinfo
  {year} {2016}),\ 10.1103/physreva.93.013625}
\bibitem [{\citenamefont {Korenblit}\ \emph {et~al.}(2012)\citenamefont
  {Korenblit}, \citenamefont {Kafri}, \citenamefont {Campbell}, \citenamefont
  {Islam}, \citenamefont {Edwards}, \citenamefont {Gong}, \citenamefont {Lin},
  \citenamefont {Duan}, \citenamefont {Kim}, \citenamefont {Kim},\ and\
  \citenamefont {Monroe}}]{Korenblit2012}%
  \BibitemOpen
  \bibfield  {author} {\bibinfo {author} {\bibfnamefont {S.}~\bibnamefont
  {Korenblit}}, \bibinfo {author} {\bibfnamefont {D.}~\bibnamefont {Kafri}},
  \bibinfo {author} {\bibfnamefont {W.~C.}\ \bibnamefont {Campbell}}, \bibinfo
  {author} {\bibfnamefont {R.}~\bibnamefont {Islam}}, \bibinfo {author}
  {\bibfnamefont {E.~E.}\ \bibnamefont {Edwards}}, \bibinfo {author}
  {\bibfnamefont {Z.-X.}\ \bibnamefont {Gong}}, \bibinfo {author}
  {\bibfnamefont {G.-D.}\ \bibnamefont {Lin}}, \bibinfo {author} {\bibfnamefont
  {L.-M.}\ \bibnamefont {Duan}}, \bibinfo {author} {\bibfnamefont
  {J.}~\bibnamefont {Kim}}, \bibinfo {author} {\bibfnamefont {K.}~\bibnamefont
  {Kim}}, \ and\ \bibinfo {author} {\bibfnamefont {C.}~\bibnamefont {Monroe}},\
  }\bibfield  {title} {\emph {\bibinfo {title} {Quantum simulation of spin
  models on an arbitrary lattice with trapped ions},}\ }\href@noop {}
  {\bibfield  {journal} {\bibinfo  {journal} {New J. Phys.}\ }\textbf {\bibinfo
  {volume} {14}},\ \bibinfo {pages} {095024} (\bibinfo {year} {2012})}
\bibitem [{\citenamefont {Nevado}, \citenamefont {Fernandez-Lorenzo},\ and\
  \citenamefont {Porras}(2017)}]{Nevado_2017}%
  \BibitemOpen
  \bibfield  {author} {\bibinfo {author} {\bibfnamefont {P.}~\bibnamefont
  {Nevado}}, \bibinfo {author} {\bibfnamefont {S.}~\bibnamefont
  {Fernandez-Lorenzo}}, \ and\ \bibinfo {author} {\bibfnamefont
  {D.}~\bibnamefont {Porras}},\ }\bibfield  {title} {\emph {\bibinfo {title}
  {Topological edge states in periodically driven trapped-ion chains},}\ }\href
  {\doibase 10.1103/physrevlett.119.210401} {\bibfield  {journal} {\bibinfo
  {journal} {Physical Review Letters}\ }\textbf {\bibinfo {volume} {119}}
  (\bibinfo {year} {2017}),\ 10.1103/physrevlett.119.210401},\ \Eprint
  {http://arxiv.org/abs/1706.04136} {arXiv:1706.04136 [quant-ph]}
\bibitem [{\citenamefont {Manovitz}\ \emph {et~al.}(2020)\citenamefont
  {Manovitz}, \citenamefont {Shapira}, \citenamefont {Akerman}, \citenamefont
  {Stern},\ and\ \citenamefont {Ozeri}}]{Manovitz2020}%
  \BibitemOpen
  \bibfield  {author} {\bibinfo {author} {\bibfnamefont {T.}~\bibnamefont
  {Manovitz}}, \bibinfo {author} {\bibfnamefont {Y.}~\bibnamefont {Shapira}},
  \bibinfo {author} {\bibfnamefont {N.}~\bibnamefont {Akerman}}, \bibinfo
  {author} {\bibfnamefont {A.}~\bibnamefont {Stern}}, \ and\ \bibinfo {author}
  {\bibfnamefont {R.}~\bibnamefont {Ozeri}},\ }\bibfield  {title} {\emph
  {\bibinfo {title} {Quantum simulations with complex geometries and synthetic
  gauge fields in a trapped ion chain},}\ }\href {\doibase
  10.1103/PRXQuantum.1.020303} {\bibfield  {journal} {\bibinfo  {journal} {PRX
  Quantum}\ }\textbf {\bibinfo {volume} {1}},\ \bibinfo {pages} {020303}
  (\bibinfo {year} {2020})}
\bibitem [{\citenamefont {Bermudez}\ \emph {et~al.}(2012)\citenamefont
  {Bermudez}, \citenamefont {Almeida}, \citenamefont {Ott}, \citenamefont
  {Kaufmann}, \citenamefont {Ulm}, \citenamefont {Poschinger}, \citenamefont
  {Schmidt-Kaler}, \citenamefont {Retzker},\ and\ \citenamefont
  {Plenio}}]{Bermudez2012b}%
  \BibitemOpen
  \bibfield  {author} {\bibinfo {author} {\bibfnamefont {A.}~\bibnamefont
  {Bermudez}}, \bibinfo {author} {\bibfnamefont {J.}~\bibnamefont {Almeida}},
  \bibinfo {author} {\bibfnamefont {K.}~\bibnamefont {Ott}}, \bibinfo {author}
  {\bibfnamefont {H.}~\bibnamefont {Kaufmann}}, \bibinfo {author}
  {\bibfnamefont {S.}~\bibnamefont {Ulm}}, \bibinfo {author} {\bibfnamefont
  {U.}~\bibnamefont {Poschinger}}, \bibinfo {author} {\bibfnamefont
  {F.}~\bibnamefont {Schmidt-Kaler}}, \bibinfo {author} {\bibfnamefont
  {A.}~\bibnamefont {Retzker}}, \ and\ \bibinfo {author} {\bibfnamefont
  {M.~B.}\ \bibnamefont {Plenio}},\ }\bibfield  {title} {\emph {\bibinfo
  {title} {Quantum magnetism of spin-ladder compounds with trapped-ion
  crystals},}\ }\href@noop {} {\bibfield  {journal} {\bibinfo  {journal} {New
  J. Phys.}\ }\textbf {\bibinfo {volume} {14}},\ \bibinfo {pages} {093042}
  (\bibinfo {year} {2012})}
\bibitem [{\citenamefont {Zippilli}\ \emph {et~al.}(2014)\citenamefont
  {Zippilli}, \citenamefont {Johanning}, \citenamefont {Giampaolo},
  \citenamefont {Wunderlich},\ and\ \citenamefont {Illuminati}}]{Zippilli2014}%
  \BibitemOpen
  \bibfield  {author} {\bibinfo {author} {\bibfnamefont {S.}~\bibnamefont
  {Zippilli}}, \bibinfo {author} {\bibfnamefont {M.}~\bibnamefont {Johanning}},
  \bibinfo {author} {\bibfnamefont {S.~M.}\ \bibnamefont {Giampaolo}}, \bibinfo
  {author} {\bibfnamefont {C.}~\bibnamefont {Wunderlich}}, \ and\ \bibinfo
  {author} {\bibfnamefont {F.}~\bibnamefont {Illuminati}},\ }\bibfield  {title}
  {\emph {\bibinfo {title} {Adiabatic quantum simulation with a segmented ion
  trap: Application to long-distance entanglement in quantum spin systems},}\
  }\href@noop {} {\bibfield  {journal} {\bibinfo  {journal} {Phys. Rev. A}\
  }\textbf {\bibinfo {volume} {89}},\ \bibinfo {pages} {042308} (\bibinfo
  {year} {2014})}
\bibitem [{\citenamefont {Guti\'errez}, \citenamefont {M\"uller},\ and\
  \citenamefont {Berm\'udez}(2019)}]{Gutierrez2019}%
  \BibitemOpen
  \bibfield  {author} {\bibinfo {author} {\bibfnamefont {M.}~\bibnamefont
  {Guti\'errez}}, \bibinfo {author} {\bibfnamefont {M.}~\bibnamefont
  {M\"uller}}, \ and\ \bibinfo {author} {\bibfnamefont {A.}~\bibnamefont
  {Berm\'udez}},\ }\bibfield  {title} {\emph {\bibinfo {title} {Transversality
  and lattice surgery: Exploring realistic routes toward coupled logical qubits
  with trapped-ion quantum processors},}\ }\href {\doibase
  10.1103/PhysRevA.99.022330} {\bibfield  {journal} {\bibinfo  {journal} {Phys.
  Rev. A}\ }\textbf {\bibinfo {volume} {99}},\ \bibinfo {pages} {022330}
  (\bibinfo {year} {2019})}
\bibitem [{\citenamefont {Horstmann}\ \emph {et~al.}(2010)\citenamefont
  {Horstmann}, \citenamefont {Reznik}, \citenamefont {Fagnocchi},\ and\
  \citenamefont {Cirac}}]{Horstmann2010}%
  \BibitemOpen
  \bibfield  {author} {\bibinfo {author} {\bibfnamefont {B.}~\bibnamefont
  {Horstmann}}, \bibinfo {author} {\bibfnamefont {B.}~\bibnamefont {Reznik}},
  \bibinfo {author} {\bibfnamefont {S.}~\bibnamefont {Fagnocchi}}, \ and\
  \bibinfo {author} {\bibfnamefont {J.}~\bibnamefont {Cirac}},\ }\bibfield
  {title} {\emph {\bibinfo {title} {Hawking radiation from an acoustic black
  hole on an ion ring},}\ }\href {\doibase 10.1103/PhysRevLett.104.250403}
  {\bibfield  {journal} {\bibinfo  {journal} {Phys. Rev. Lett.}\ }\textbf
  {\bibinfo {volume} {104}},\ \bibinfo {pages} {250403} (\bibinfo {year}
  {2010})}
\bibitem [{\citenamefont {Horstmann}\ \emph {et~al.}(2011)\citenamefont
  {Horstmann}, \citenamefont {Sch\"{u}tzhold}, \citenamefont {Reznik},
  \citenamefont {Fagnocchi},\ and\ \citenamefont {Cirac}}]{Horstmann2011}%
  \BibitemOpen
  \bibfield  {author} {\bibinfo {author} {\bibfnamefont {B.}~\bibnamefont
  {Horstmann}}, \bibinfo {author} {\bibfnamefont {R.}~\bibnamefont
  {Sch\"{u}tzhold}}, \bibinfo {author} {\bibfnamefont {B.}~\bibnamefont
  {Reznik}}, \bibinfo {author} {\bibfnamefont {S.}~\bibnamefont {Fagnocchi}}, \
  and\ \bibinfo {author} {\bibfnamefont {J.}~\bibnamefont {Cirac}},\ }\bibfield
   {title} {\emph {\bibinfo {title} {Hawking radiation on an ion ring in the
  quantum regime},}\ }\href {\doibase 10.1088/1367-2630/13/4/045008} {\bibfield
   {journal} {\bibinfo  {journal} {New J. Phys.}\ }\textbf {\bibinfo {volume}
  {13}},\ \bibinfo {pages} {045008} (\bibinfo {year} {2011})}
\bibitem [{\citenamefont {Li}\ \emph {et~al.}(2017)\citenamefont {Li},
  \citenamefont {Urban}, \citenamefont {Noel}, \citenamefont {Chuang},
  \citenamefont {Xia}, \citenamefont {Ransford}, \citenamefont {Hemmerling},
  \citenamefont {Wang}, \citenamefont {Li}, \citenamefont {H\"affner},\ and\
  \citenamefont {et~al.}}]{Li_2017}%
  \BibitemOpen
  \bibfield  {author} {\bibinfo {author} {\bibfnamefont {H.-K.}\ \bibnamefont
  {Li}}, \bibinfo {author} {\bibfnamefont {E.}~\bibnamefont {Urban}}, \bibinfo
  {author} {\bibfnamefont {C.}~\bibnamefont {Noel}}, \bibinfo {author}
  {\bibfnamefont {A.}~\bibnamefont {Chuang}}, \bibinfo {author} {\bibfnamefont
  {Y.}~\bibnamefont {Xia}}, \bibinfo {author} {\bibfnamefont {A.}~\bibnamefont
  {Ransford}}, \bibinfo {author} {\bibfnamefont {B.}~\bibnamefont
  {Hemmerling}}, \bibinfo {author} {\bibfnamefont {Y.}~\bibnamefont {Wang}},
  \bibinfo {author} {\bibfnamefont {T.}~\bibnamefont {Li}}, \bibinfo {author}
  {\bibfnamefont {H.}~\bibnamefont {H\"affner}}, \ and\ \bibinfo {author}
  {\bibnamefont {et~al.}},\ }\bibfield  {title} {\emph {\bibinfo {title}
  {Realization of translational symmetry in trapped cold ion rings},}\ }\href
  {\doibase 10.1103/physrevlett.118.053001} {\bibfield  {journal} {\bibinfo
  {journal} {Physical Review Letters}\ }\textbf {\bibinfo {volume} {118}}
  (\bibinfo {year} {2017}),\ 10.1103/physrevlett.118.053001}
\bibitem [{\citenamefont {Ba\~nuls}\ \emph {et~al.}(2020)\citenamefont
  {Ba\~nuls} \emph {et~al.}}]{Banuls:2019bmf}%
  \BibitemOpen
  \bibfield  {author} {\bibinfo {author} {\bibfnamefont {M.~C.}\ \bibnamefont
  {Ba\~nuls}} \emph {et~al.},\ }\bibfield  {title} {\emph {\bibinfo {title}
  {{Simulating Lattice Gauge Theories within Quantum Technologies}},}\ }\href
  {\doibase 10.1140/epjd/e2020-100571-8} {\bibfield  {journal} {\bibinfo
  {journal} {Eur. Phys. J. D}\ }\textbf {\bibinfo {volume} {74}},\ \bibinfo
  {pages} {165} (\bibinfo {year} {2020})},\ \Eprint
  {http://arxiv.org/abs/1911.00003} {arXiv:1911.00003 [quant-ph]}
\bibitem [{\citenamefont {Zhu}\ \emph {et~al.}(2020)\citenamefont {Zhu},
  \citenamefont {Johri}, \citenamefont {Linke}, \citenamefont {Landsman},
  \citenamefont {Nguyen}, \citenamefont {Alderete}, \citenamefont {Matsuura},
  \citenamefont {Hsieh},\ and\ \citenamefont {Monroe}}]{Zhu:2019bri}%
  \BibitemOpen
  \bibfield  {author} {\bibinfo {author} {\bibfnamefont {D.}~\bibnamefont
  {Zhu}}, \bibinfo {author} {\bibfnamefont {S.}~\bibnamefont {Johri}}, \bibinfo
  {author} {\bibfnamefont {N.~M.}\ \bibnamefont {Linke}}, \bibinfo {author}
  {\bibfnamefont {K.~A.}\ \bibnamefont {Landsman}}, \bibinfo {author}
  {\bibfnamefont {N.~H.}\ \bibnamefont {Nguyen}}, \bibinfo {author}
  {\bibfnamefont {C.~H.}\ \bibnamefont {Alderete}}, \bibinfo {author}
  {\bibfnamefont {A.~Y.}\ \bibnamefont {Matsuura}}, \bibinfo {author}
  {\bibfnamefont {T.~H.}\ \bibnamefont {Hsieh}}, \ and\ \bibinfo {author}
  {\bibfnamefont {C.}~\bibnamefont {Monroe}},\ }\bibfield  {title} {\emph
  {\bibinfo {title} {{Generation of thermofield double states and critical
  ground states with a quantum computer}},}\ }\href {\doibase
  10.1073/pnas.2006337117} {\bibfield  {journal} {\bibinfo  {journal} {Proc.
  Nat. Acad. Sci.}\ }\textbf {\bibinfo {volume} {117}},\ \bibinfo {pages}
  {25402} (\bibinfo {year} {2020})},\ \Eprint {http://arxiv.org/abs/1906.02699}
  {arXiv:1906.02699 [quant-ph]}
\bibitem [{\citenamefont {Mildenberger}(2019)}]{Mildenberger_MA}%
  \BibitemOpen
  \bibfield  {author} {\bibinfo {author} {\bibfnamefont {J.}~\bibnamefont
  {Mildenberger}},\ }\emph {\bibinfo {title} {{Trapped-Ion Quantum Simulations
  of Spin Systems at Non-Vanishing Temperature}}},\ \href
  {https://www.kip.uni-heidelberg.de/Veroeffentlichungen/details.php?id=3997}
  {\bibinfo {type} {Master thesis}},\ \bibinfo  {school} {Universit{\"at}
  Heidelberg} (\bibinfo {year} {2019}),\ \bibinfo {note}
  {\url{https://www.kip.uni-heidelberg.de/Veroeffentlichungen/details.php?id=3997}}
\bibitem [{\citenamefont {Rothkopf}(2020)}]{Rothkopf:2019ipj}%
  \BibitemOpen
  \bibfield  {author} {\bibinfo {author} {\bibfnamefont {A.}~\bibnamefont
  {Rothkopf}},\ }\bibfield  {title} {\emph {\bibinfo {title} {{Heavy Quarkonium
  in Extreme Conditions}},}\ }\href {\doibase 10.1016/j.physrep.2020.02.006}
  {\bibfield  {journal} {\bibinfo  {journal} {Phys. Rept.}\ }\textbf {\bibinfo
  {volume} {858}},\ \bibinfo {pages} {1} (\bibinfo {year} {2020})},\ \Eprint
  {http://arxiv.org/abs/1912.02253} {arXiv:1912.02253 [hep-ph]}
\bibitem [{Note8()}]{Note8}%
  \BibitemOpen
  \bibinfo {note} {See \cite {meson:project} for a related study of this
  phenomenon from a tensor network perspective in quantum spin chains.}
\bibitem [{\citenamefont {Rigobello}\ \emph {et~al.}(2021)\citenamefont
  {Rigobello}, \citenamefont {Notarnicola}, \citenamefont {Magnifico},\ and\
  \citenamefont {Montangero}}]{Rigobello:2021fxw}%
  \BibitemOpen
  \bibfield  {author} {\bibinfo {author} {\bibfnamefont {M.}~\bibnamefont
  {Rigobello}}, \bibinfo {author} {\bibfnamefont {S.}~\bibnamefont
  {Notarnicola}}, \bibinfo {author} {\bibfnamefont {G.}~\bibnamefont
  {Magnifico}}, \ and\ \bibinfo {author} {\bibfnamefont {S.}~\bibnamefont
  {Montangero}},\ }\bibfield  {title} {\emph {\bibinfo {title} {{Entanglement
  generation in QED scattering processes}},}\ }\href@noop {} {\  (\bibinfo
  {year} {2021})},\ \Eprint {http://arxiv.org/abs/2105.03445} {arXiv:2105.03445
  [hep-lat]}
\bibitem [{\citenamefont {Ba\~nuls}\ \emph {et~al.}(2022)\citenamefont
  {Ba\~nuls}, \citenamefont {Heller}, \citenamefont {Jansen}, \citenamefont
  {Knaute},\ and\ \citenamefont {Svensson}}]{meson:project}%
  \BibitemOpen
  \bibfield  {author} {\bibinfo {author} {\bibfnamefont {M.~C.}\ \bibnamefont
  {Ba\~nuls}}, \bibinfo {author} {\bibfnamefont {M.~P.}\ \bibnamefont
  {Heller}}, \bibinfo {author} {\bibfnamefont {K.}~\bibnamefont {Jansen}},
  \bibinfo {author} {\bibfnamefont {J.}~\bibnamefont {Knaute}}, \ and\ \bibinfo
  {author} {\bibfnamefont {V.}~\bibnamefont {Svensson}},\ }\href@noop {} {\emph
  {\bibinfo {title} {{to appear}},}\ } (\bibinfo {year} {2022})
\end{thebibliography}%

\end{document}